\magnification=\magstephalf
\baselineskip=14pt plus0.1pt minus0.1pt 
\parindent=25pt
\lineskip=4pt\lineskiplimit=0.1pt      
\parskip=0.1pt plus1pt
%

%

%

%

%

%
\font\sixrm=cmr6

%

%
%
\let\a=\alpha \let\b=\beta   \let\d=\delta  \let\e=\varepsilon
\let\f=\varphi \let\g=\gamma \let\h=\eta    \let\k=\kappa  \let\l=\lambda
\let\m=\mu      \let\o=\omega    \let\p=\pi  
\let\r=\rho  \let\s=\sigma \let\t=\tau   \let\th=\vartheta
  \let\z=\zeta
\let\D=\Delta \let\F=\Phi  \let\G=\Gamma  \let\L=\Lambda 
\let\O=\Omega      \let\X=\Xi

%
%
\def\cA{{\cal A}} \def\cB{{\cal B}} \def\cC{{\cal C}} \def\cD{{\cal D}}
 \def\cF{{\cal F}}  
   \def\cL{{\cal L}}
\def\cM{{\cal M}} \def\cN{{\cal N}} \def\cO{{\cal O}} 
 \def\cR{{\cal R}}

%
%
%
\newdimen\xshift 
\newdimen\yshift
\newdimen\xwidth 
\def\eqfig#1#2#3#4#5#6{
  \par\xwidth=#1 \xshift=\hsize \advance\xshift 
  by-\xwidth \divide\xshift by 2
  \yshift=#2 \divide\yshift by 2
  \vbox{
  \line{\hglue\xshift \vbox to #2{
    \smallskip
    \vfil#3 
    \includegraphics{#4.ps}
    }
    \hfill\raise\yshift\hbox{#5}
  }
  \smallskip
  \centerline{#6}
  }
  \smallskip}
%

%

%
%
%
\def\data{\number\day/\ifcase\month\or gennaio \or febbraio \or marzo \or
aprile \or maggio \or giugno \or luglio \or agosto \or settembre
\or ottobre \or novembre \or dicembre \fi/\number\year}
%
%
%
\newcount\pgn 
\pgn=1

%
%
%
\def\begintex{%
  \openin14=\jobname.aux 
  \ifeof14 
    \relax 
  \else
    \input \jobname.aux \closein14 
  \fi 
  \openout15=\jobname.aux
}

\global\newcount\numsec
\global\newcount\numfor
\global\newcount\numfig
\global\newcount\numtheo
\gdef\profonditastruttura{\dp\strutbox}
\def\senondefinito#1{\expandafter\ifx\csname#1\endcsname\relax}
\def\SIA #1,#2,#3 {\senondefinito{#1#2}%
   \expandafter\xdef\csname #1#2\endcsname{#3}\else
   \write16{???? ma #1,#2 e' gia' stato definito !!!!}\fi}
\def\etichetta(#1){(\numsection\numfor)
   \SIA e,#1,(\numsection\numfor)
   \global\advance\numfor by 1
   \write15{\string\FU (#1){\equ(#1)}}
   \write16{ EQ \equ(#1) == #1  }}
\def\oldetichetta(#1){
   \senondefinito{fu#1}\clubsuit(#1)\else
   \csname fu#1\endcsname\fi}
\def\FU(#1)#2{\SIA fu,#1,#2 }
\def\tetichetta(#1){{\numsection\numtheo}%
   \SIA theo,#1,{\numsection\numtheo}
   \global\advance\numtheo by 1%
   \write15{\string\FUth (#1){\thm[#1]}}%
   \write16{ TH \thm[#1] == #1  }}

\def\oldtetichetta(#1){
   \senondefinito{futh#1}\clubsuit(#1)\else
   \csname futh#1\endcsname\fi}
\def\FUth(#1)#2{\SIA futh,#1,#2 }
\def\getichetta(#1){Fig. \number\numfig
 \SIA e,#1,{\number\numfig}
 \global\advance\numfig by 1
 \write15{\string\FU (#1){\equ(#1)}}
 \write16{ Fig. \equ(#1) ha simbolo  #1  }}
\newdimen\gwidth
\def\BOZZA{
 \def\alato(##1){
 {\vtop to \profonditastruttura{\baselineskip
 \profonditastruttura\vss
 \rlap{\kern-\hsize\kern-1.3truecm{$\scriptstyle##1$}}}}}
 \def\galato(##1){ \gwidth=\hsize \divide\gwidth by 2
 {\vtop to \profonditastruttura{\baselineskip
 \profonditastruttura\vss
 \rlap{\kern-\gwidth\kern-1.3truecm{$\scriptstyle##1$}}}}}
 \def\talato(##1){\rlap{\sixrm\kern -1.3truecm ##1}}
 \def\thm{\teo}\def\thf{\teo}
}
\def\alato(#1){}
\def\galato(#1){}
\def\talato(#1){}
\def\numsection#1{%
  \ifnum\numsec=-100%
    \relax\number#1%
  \else
  \ifnum\numsec=-101%
    \relax A\number#1%
  \else
  \ifnum\numsec<0%
    A\number-\numsec.\number#1%
  \else
    \number\numsec.\number#1%
\fi
\fi
\fi
}
\def\Thm[#1]{\tetichetta(#1)}
\def\thf[#1]{\senondefinito{futh#1}$\clubsuit$[#1]\else
   \csname futh#1\endcsname\fi}
\def\thm[#1]{\senondefinito{theo#1}$\spadesuit$[#1]\else
   \csname theo#1\endcsname\fi}
\def\Eq(#1){\eqno{\etichetta(#1)\alato(#1)}}
\def\eq(#1){\etichetta(#1)\alato(#1)}
\def\eqv(#1){\senondefinito{fu#1}$\clubsuit$(#1)\else
   \csname fu#1\endcsname\fi}
\def\equ(#1){\senondefinito{e#1}$\spadesuit$(#1)\else
   \csname e#1\endcsname\fi}
\let\eqf=\eqv
\def\nonumeration{
  \let\etichetta=\oldetichetta
  \let\tetichetta=\oldtetichetta
  \let\equ=\eqf
  \let\thm=\thf
}
%
%
%
%
\def\include#1{%
\openin13=#1.aux \ifeof13 \relax \else
\input #1.aux \closein13 \fi
}
%
%
%
%

%
%

%
%

\def\\{\hfill\break}

\def\tthsp{\kern .083333 em}

\def\?{\mskip -10mu}
%
%

\def\indbox#1{\hbox to \parindent{\hfil\ #1\hfil} }

\def\ref[#1]{[#1]}
\def\beginsubsection#1\par{\bigskip\leftline{\it #1}\nobreak\smallskip
	    \noindent}
\newfam\msafam
\newfam\msbfam
\newfam\eufmfam
%
%
%
\def\hexnumber#1{%
  \ifcase#1 0\or 1\or 2\or 3\or 4\or 5\or 6\or 7\or 8\or
  9\or A\or B\or C\or D\or E\or F\fi}
\font\tenmsa=msam10
\font\sevenmsa=msam7
\font\fivemsa=msam5
\textfont\msafam=\tenmsa
\scriptfont\msafam=\sevenmsa
\scriptscriptfont\msafam=\fivemsa        
\edef\msafamhexnumber{\hexnumber\msafam}%
%
%
\mathchardef\restriction"1\msafamhexnumber16
\mathchardef\ssim"0218
\mathchardef\square"0\msafamhexnumber03
\mathchardef\eqd"3\msafamhexnumber2C
\def\QED{\ifhmode\unskip\nobreak\fi\quad
  \ifmmode\square\else$\square$\fi}            
\font\tenmsb=msbm10
\font\sevenmsb=msbm7
\font\fivemsb=msbm5
\textfont\msbfam=\tenmsb
\scriptfont\msbfam=\sevenmsb
\scriptscriptfont\msbfam=\fivemsb
\def\Bbb#1{\fam\msbfam\relax#1}    
\font\teneufm=eufm10
\font\seveneufm=eufm7
\font\fiveeufm=eufm5
\textfont\eufmfam=\teneufm
\scriptfont\eufmfam=\seveneufm
\scriptscriptfont\eufmfam=\fiveeufm

\def\bZ{{\Bbb Z}}

\def\bR{{\Bbb R}}

\def\bN{{\Bbb N}}

\def\({\left(}
\def\){\right)}
%
%
%

\let\Z=\integer

\let\neper=e
\let\ii=i

\let\id=\identity

\let\de=\partial

\let\<=\langle
\let\>=\rangle

\def\Var{ \mathop{\rm Var}\nolimits }

\def\diam{\mathop{\rm diam}\nolimits}

\def\supp{\mathop{\rm supp}\nolimits}

\outer\def\nproclaim#1 [#2]#3. #4\par{\medbreak \noindent
   \talato(#2){\bf #1 \Thm[#2]#3.\enspace }%
   {\sl #4\par }\ifdim \lastskip <\medskipamount 
   \removelastskip \penalty 55\medskip \fi}
\def\thmm[#1]{#1}
\def\teo[#1]{#1}
%
%
\def\sttilde#1{%
\dimen2=\fontdimen5\textfont0
\setbox0=\hbox{$\mathchar"7E$}
\setbox1=\hbox{$\scriptstyle #1$}
\dimen0=\wd0
\dimen1=\wd1
\advance\dimen1 by -\dimen0
\divide\dimen1 by 2
\vbox{\offinterlineskip%
   \moveright\dimen1 \box0 \kern - \dimen2\box1}
}
%

%
%
%

\def\bye{%
\par\vfill\supereject
\message{******** Run TeX twice to resolve cross-references *****************}%
\message{******** Run TeX twice to resolve cross-references *****************}%
\message{******** Run TeX twice to resolve cross-references *****************}%
\message{******** Run TeX twice to resolve cross-references *****************}%
\end}

\long\def\newsection#1\par{%
\advance\numsec by 1%
\numfor=1%
\numtheo=1%
\pgn=1%
\vskip 0pt plus.3\vsize \penalty -250 \vskip 0pt plus-.3\vsize \bigskip 
\vskip \parskip \message {#1}\leftline {\bf \number\numsec. #1}%
\nobreak \smallskip \noindent}

\long\def\appendix#1\par{%
\numsec=-101%
\numfor=1%
\numtheo=1%
\pgn=1%
\vskip 0pt plus.3\vsize \penalty -250 \vskip 0pt plus-.3\vsize \bigskip 
\vskip \parskip \message {#1}\leftline {\bf Appendix. #1}%
\nobreak \smallskip \noindent}

\def\frac#1#2{{#1\over#2}}

\let\noi=\noindent
\def\12{{1\over 2}}
\let\mi=\wedge
\def\qed{\hfill$\square$} 
\let\p=\partial
\let\ten=\rightarrow
\let\su=\subset
\def\ssu{\subset\subset}
\def\bz{{\underline z}}

\def\bt{{\underline t}}
\def\bs{{\underline s}}
\def\br{{\underline \rho}}
\def\bn{{\underline n}}
\def\bm{{\underline m}}
\def\bN{{\underline N}}
\def\bM{{\underline M}}
\def\bJ{{\Bbb J}}
\def\bA{{\Bbb A}}
\def\bD{{\Bbb D}}
\def\diam{\mathop{\rm diam}\nolimits}

\let\tilde=\widetilde

\def\bzi{{\underline \zeta}}
\def\<{\left\langle}
\def\>{\right\rangle}

\newdimen\xshift \newdimen\xwidth \newdimen\yshift \newdimen\ywidth
\def\ins#1#2#3{\vbox to0pt{\kern-#2 \hbox{\kern#1 #3}\vss}\nointerlineskip}
\def\eqfig#1#2#3#4#5{
\par\xwidth=#1 \xshift=\hsize \advance\xshift 
by-\xwidth \divide\xshift by 2
\yshift=#2 \divide\yshift by 2
\line{\hglue\xshift \vbox to #2{\vfil
#3 \includegraphics{#4.ps}}\hfill\raise\yshift\hbox{#5}}}

\expandafter\ifx\csname sezioniseparate\endcsname\relax%
\input macro \fi
\numsec=0
\numfor=1\numtheo=1\pgn=1
\noindent
\centerline {\bf RENORMALIZATION--GROUP TRANSFORMATIONS}
\centerline {\bf UNDER STRONG MIXING CONDITIONS:}
\centerline {\bf  GIBBSIANNESS AND CONVERGENCE OF RENORMALIZED INTERACTIONS.}
\vskip 2 truecm
\centerline {Lorenzo Bertini}
\vskip 0.25 truecm
\centerline {\it Dipartimento di Matematica -- Universit\`a di Roma 
``La Sapienza''}
\par\noindent
\centerline{\it P.le A. Moro 2, 00185 Roma, Italy.}
\par\noindent
\centerline {\rm E\_mail: lorenzo@carpenter.mat.uniroma1.it}
\vskip 0.5 truecm
\centerline {Emilio N.M. Cirillo}
\vskip 0.25 truecm
\centerline {\it CMI -- Universit\'e de Provence,}
\par\noindent
\centerline{\it 39 rue Joliot Curie, 13453 Marseille, France.}
\par\noindent
\centerline {\rm E\_mail: cirillo@gyptis.univ-mrs.fr}
\vskip 0.5 truecm
\centerline {Enzo Olivieri}
\vskip 0.25 truecm
\centerline {\it Dipartimento di Matematica -- II Universit\`a di Roma 
Tor Vergata}
\par\noindent
\centerline {\it Via della Ricerca Scientifica, 00133 Roma, 
Italy.}
\par\noindent
\centerline {\rm E\_mail: olivieri@mat.uniroma2.it}
\vskip 1.5 truecm
\centerline {\bf Abstract}
\vskip 0.5 truecm
\centerline{
\vbox{
\hsize=14truecm
\baselineskip 0.4cm
In this paper we study a renormalization-group map: the block
averaging transformation applied to Gibbs measures relative to a class
of finite range lattice gases, when  suitable strong mixing
conditions are satisfied. Using block decimation procedure, cluster
expansion (like in [HK]) and detailed comparison between statistical
ensembles, we are able to prove Gibbsianess and convergence to a
trivial (i.e. Gaussian and product) fixed point. Our results apply to
2D standard Ising model {\it at any} temperature above the critical
one  and arbitrary magnetic field. 
}}

\vfill\eject

\begintex 
\centerline{\bf Table of Contents}
\smallskip
\noi
{\bf 1. Introduction\dotfill 3} \break\par
\smallskip
\noi
{\bf 2. Notation and results \dotfill 9} \break
\indent
2.1. The lattice \dotfill 9 \break
\indent
2.2. The configuration space \dotfill 9 \break
\indent
2.3. The potential \dotfill 10 \break
\indent
2.4. The Gibbs measures \dotfill 10 \break
\indent
2.5. Strong Mixing Conditions \dotfill 11 \break
\indent
2.6. Lattice gases: Uniform Strong Mixing Conditions \dotfill 14 \break
\indent
2.7. The multi--grancanonical state\dotfill 15 \break
\indent
2.8. Block Averaging Transformation (BAT)\dotfill 16 \break
\indent
2.9. Main results \dotfill 17 \break \par
\smallskip
\noi
{\bf 3. Computing the renormalized potential via cluster expansion
\dotfill 19} \break 
\par
\smallskip
\noi
{\bf 4. The multi--canonical measure\dotfill 39}  \break 
\indent
4.1. Thermodynamic relationship \dotfill 39 \break 
\indent
4.2. Comparison of ensembles in finite volume \dotfill 43 \break 
\indent
4.3. Local Central Limit Theorem with multiplicative 
error \dotfill 50 \break  
\par 
\smallskip
\noi
{\bf 5. Gibbsianess and convergence\dotfill 52} \break 
\indent
5.1. Finite size condition for the constrained models\dotfill 52 \break 
\indent
5.2. Short range renormalized potential \dotfill 54 \break 
\indent
5.3. Gibbsianess of renormalized potential\dotfill 55 \break
\par
\smallskip
\noi
{\bf A.1. Proof of USM(${\cal A}$) $\Longrightarrow$ MUSM(${\cal A}$)
in dimension 2\dotfill 56}  \break \par
\smallskip
\noi
{\bf A.2. A counterexample to USM $\Longrightarrow$ MUSM in
dimension 3\dotfill 63}  \break \par
\smallskip
\noi
{\bf References\dotfill 66} \break \par
\vfill\eject

       
\newsection Introduction

This paper concerns the rigorous analysis of some 
real--space renormalization group
transformations (RGT) (see, for instance, ref. [NL] for a general
introduction to this subject). 
In the recent years many works  have been devoted to the
question of well--definedness of RGT. We refer to ref. 
[EFS] for a clear and
complete discussion of the problematic as well as
for an exhaustive description of 
the general setup of renormalization maps from the point of view of rigorous
statistical mechanics.
Already in the seventies (see [GaK], [CG], [GP], [I]) 
the question was raised of whether or not some typical RGT give
rise to a well defined renormalized interaction.
In other words, calling  $\m^{(\ell)}$ 
 the renormalized measure arising from the application of a 
RGT on scale $\ell$ to the Gibbs measure
$\m$, we pose the question of Gibbsianess of $\m^{(\ell)}$, 
namely we ask ourselves whether $\m^{(\ell)}$ is a Gibbs measure
corresponding to a finite--norm translationally invariant potential so
that the ``renormalized Hamiltonian" is well defined. 
 
More explicitly: let us assume that our RGT can be expressed as
$$
\m^{(\ell)}(\s') = \sum _{\s} T^{(\ell)} (\s',\s) \m (\s) \Eq (1.2')
$$
where $T^{(\ell)} (\s',\s)$ is a normalized non--negative kernel. 
The system described in terms of the $\s$ variables by the original measure 
$\m$ is called {\it object system}. The $\s'$'s are the
{\it renormalized variables} of the 
{\it image system} described by the renormalized measure $\mu^{(\ell)}$.
We can think of the transformation $T_\ell$ as  directly acting
at infinite volume or we can consider a finite volume version and
subsequently try to perform the thermodynamic limit (see [EFS] ).\par

The above mentioned pathological behavior (non--Gibbsianess of 
$\mu^{(\ell)}$) can be a consequence of the violation of a necessary
condition for Gibbsianess called {\it quasi--locality} (see [Ko], [EFS]).
The latter is a continuity property of the finite volume conditional
probabilities of $\mu^{(\ell)}$ which, roughly speaking, says that they 
are almost independent of very far away conditioning spins.
In many interesting examples (see [E1], [E2], [EFS], [EFK]) violation of
quasi--locality and consequently non--Gibbsianess of the renormalized
measure $\mu^{(\ell)}$  is a direct consequence of 
the appearance of a first order phase transition for the original (object) 
system  {\it conditioned} to some particular configuration of the 
image system. More precisely, given a configuration $\s'$, 
let us consider the probability measure on the original spin variables 
given by
$$\m_{\s'} (\s) \; = \; { T^{(\ell)} (\s',\s) \m(\s) 
\over \sum _{\eta}  T^{(\ell)} (\s',\eta) \m(\eta)}$$
It defines the {\it constrained} model corresponding to $\s'$
(which here plays the role of an external parameter).
For some particular $\s'$ it may happen that the corresponding measure 
$\m_{\s'}(\s)$ exhibits long range order inducing violation of quasi--locality 
  and then 
non--Gibbsianess for the image system. See [EFS] and 
also [GP], [I] where this  mechanism was first pointed out. 

This pathological behavior is often induced by configurations $\s'$ highly
non--typical with respect to the measure $\mu^{(\ell)}$.
This suggests the introduction of a weaker notion of Gibbsianess requiring
well--definedness of renormalized interactions not {\it for all} renormalized
configurations $\s'$ but, rather, 
{\it for $\mu^{(\ell)}$--almost all} $\s'$ (see [D2], [BKL], [LM]). 
However also this point of view poses
various other problems (see [ES] and references therein).
\par
It is also natural to ask ourselves about {\it robustness} of this pathology.
Sometimes it can be shown that non--Gibbsianess is 
an artifact due to a wrong choice of the scale
$\ell$ of the transformation in terms of the thermodynamic 
parameters of the object system. For instance in [MO4] 
has been considered the case of the measure 
$\mu^{(\ell)} = T_{d}^{(\ell)} \m_{\b,h}$, where
$T_{d}^{(\ell)}$ is the so--called decimation transformation on 
scale $\ell$ (see [EFS]) and
$\m_{\b,h}$  is the Gibbs measure for the standard 2D Ising model, 
$h$ and $\b$
being, respectively, the external field and the inverse temperature. 
In [EFS] the authors show non--Gibbsianess for 
some choices of $h, \b, \ell$. 
On the contrary, in [MO4] it is shown that, for the same values of
$h,\b$ for which, for suitable $\ell$, in [EFS]  the authors got
non--Gibbsianess, by changing $\ell$ into a sufficiently large $\ell'$,
depending on $\b,h$, one gets again Gibbsianess. We also mention 
it is possible to show that, by iterating the transformation, one
has convergence to a (trivial) fixed point, see [MO4] and also
[Ka] for the high temperature case.  
The above behavior is  related to the fact that, given suitable 
values of the parameters $\b, h$ (close to the coexistence line 
$h = 0, \b > \b_c$), on a
suitable scale $\ell$ some constrained models can undergo a phase transition 
(somehow related to the phase transition of the object system); whereas, given 
the same $h,\b$, for sufficiently large scale 
$\ell$ any constrained model is in the weak coupling region.
Another notion of robustness of the pathology  
is related to the  application of decimation transformations, see [LV], [MO5].

Let us stress that the fact that the object system is very well behaved in the 
sense that it is in
the unique phase region (in the strongest possible sense) does not preclude the
possibility that some constrained model undergoes a dangerous phase transition
inducing the pathology.\par

On the positive side, since the pioneering paper [CG], there are many
indications  that if the  constrained models are in the weak coupling
regime, then Gibbsianess of the renormalized measure follows.  
Recently Haller and Kennedy gave very interesting new rigorous results
in this direction. They proved, under very  general hypotheses, that
if  {\it all}  constrained models satisfy a uniform (in the
constraint)  version of the Dobrushin--Shlosman complete  analyticity
condition (see [DS2], [DS3]) then the renormalized measure is Gibbsian
with a finite norm  potential which can be computed via a convergent
cluster expansion. 

Another interesting question, which, in a sense, 
is the main object of the present
paper, is the convergence of the iterates of RGT or, 
in other words, the behavior of
the transformation $T^{(\ell)}$ for large $\ell$.
This problem has not been, up to now,  studied  very much from a 
point of view of rigorous statistical mechanics.
Here we present results referring to non--critical systems 
and so we have convergence to a trivial fixed point, i.e. Gaussian {\it
and} product (which correspond to infinite temperature). 
Indeed most of the recent papers concerning 
rigorous results on RGT refer to the non--critical region with some
exceptions, see [BMO], [CiO], [HK], where the authors consider 2D critical
Ising system but only for one step of RGT. 

Let us now introduce, for the standard
2D Ising model, the Block--Averaging Transformation (BAT). 
It is convenient to use the lattice gas variables. 
For a standard Ising system enclosed in a
finite volume $\L \subset \bZ ^2$ the configuration space 
is therefore $\{0,1\}^\L$; given a
configuration $\h \in \{0,1\}^\L$ and a site 
$x\in \L$, $\h_x \, \in \, \{0,1\} $ represents
the occupation number at $x$. For free or periodic boundary 
conditions the energy associated
to a configuration $\h \in \{0,1\}^\L$ is:
$$ 
E_{\L}(\h) := - \b   \sum _{ \<x,y\> \subset \L } 
\h_x\h_y - \l \sum _{x\in \L} \h_x 
\Eq (Ising)
$$
where $\<x,y\>$ is a pair of nearest neighbor sites, 
$\b$ is the inverse temperature and $\l $
is $\b$ times the chemical potential so that the Boltzmann 
factor is $\sim \exp\{-E_\L(\h)\}$.
Given $\b$ let $\l^*= \l^*(\b)$ be the value of $\l$ 
corresponding to the value zero of
the magnetic field  $h$ appearing in the expression of $E$ in terms of
spin variables $\s_x = 2\h_x -1$. For $\b,\l$ in the uniqueness region: 
$(\b,\l) \in \{ \b < \b_c\}  \cup \{ \b \geq \b_c, \l \neq \l^*\}$ 
($\b_c$ is the inverse critical temperature), let $\m_{\b,\l}$ be
the unique infinite volume Gibbs measure.
We partition $\bZ^2$ into square blocks $Q_{\ell}(i)$ 
of side $\ell$ and centers at the points $i$ belonging to the rescaled 
lattice $(\ell \bZ)^2$. 
Let $N_i = N_i(\h) := \sum _{x \in Q_\ell (i)} \h_x$ 
be the number of particles in the block $Q_\ell(i)$ in the $\h$
configuration, $\r =\r(\b,\l) = \m_{\b,\l}(\h_0)$ be the equilibrium density, 
$\chi =\chi (\b,\l) := \sum_{x\in\bZ ^2}
[\mu_{\b,\l} (\h_0 \h_x)-   \mu_{\b,\l} (\h_0 )\mu_{\b,\l} (\h_x)]$ be the
susceptibility; we then set:
$$
M_i := { N_i - \r |Q_\ell| \over \sqrt{|Q_\ell| \chi}}
\Eq (Centr)
$$
the random variables $M_i$ are centered and normalized; 
they take values in 
$$
\bar \O^{(\ell)}_i := \left \{{  - \r |Q_\ell| \over \sqrt{|Q_\ell| \chi}}, { 1
-
\r  |Q_\ell| \over \sqrt{|Q_\ell| \chi}}, \dots, { |Q_\ell| (1 - \r)
\over \sqrt{|Q_\ell| \chi}}\right \}
\Eq (Ombi)
$$
We expect $M_i$, $i\in (\ell \bZ)^2$  to have a product
(Gaussian) limiting distribution as $\ell \to \infty$.

The renormalized measure
$\mu^{(\ell)} = \m^{(\ell)}_{\b,\l}$  (arising from the application of  
the BAT transformation on scale $\ell$ to $\m_{\b,\l}$) 
is the joint distribution of the random variables $M_i$'s 
under  $\m_{\b,\l}$; i.e. it is  obtained by assigning to each
block $Q_{\ell}(i)$ a value $m_i \in\bar \O^{(\ell)}_i$ 
and by computing the probability, w.r.t. the original Gibbs measure 
$\m_{\b,\l}$ of the event: $M_i(\h) = m_i$. 
In other words, in the notation of \equ(1.2') in the 
case of BAT we have: $\s= \{\h_x\}, \s' = \{m_i\}$ and 
$$
T^{(\ell)}_{BAT}(m,\h) \; =\left\{\eqalign{ 
1& \;\;\;\;\hbox {if}\;\; M_i(\h) = m_i\;\;\forall i\cr
0& \;\;\;\;\hbox {otherwise}\cr}\right.
\;\; .
$$
In this case a constrained model is a {\it multi--canonical}\/
Ising model; namely an Ising
model subject to the constraint of having a fixed number of particles    
in each block $Q_{\ell}(i)$.

\nproclaim Theorem [Results].
Consider a 2D Ising system with $\b <\b_c$ and $\l \in \bR$ given. 
Then there exists $\ell_0 \in {\Bbb N}$ such that 
$\forall \, \ell >\ell_0\; \m^{(\ell)}_{\b,\l} $ is Gibbsian
with a finite norm translationally invariant potential 
$\F^{(\ell)} = \{\F^{(\ell)}_X,\; X \su  (\ell\Z)^d \}$.
\hfill\break\indent
Furthermore it is possible to 
decompose the potential into a short and a long range part,
$\F^{(\ell)}=\F^{(\ell),sr}+\F^{(\ell),lr}$, where
$\exists\kappa\in{\Bbb N}$: $\Phi_{X}^{(\ell),sr}\equiv 0$ if
$\diam(X)\ge\kappa$ and we have the following:
\item{{\it (i)}}
there is $\a>0$ such that
$$
\lim_{\ell\ten\infty} 
\sum_{X \ni 0} e^{\a |X|} \sup_{m_i \in \bar \O^{(\ell)}_i}
\left| \Phi^{(\ell),lr}_X (m_i, \, i\in X) \right| =0
$$
\item{{\it (ii)}}
there exist $a>0$ such that
$$
\eqalign{
& 
\lim_{\ell\ten\infty} \sup_{m_i\in \bar \O^{(\ell)}_i \atop
|m_i|\le \ell^a}
\left| \Phi^{(\ell),sr}_X (m_i, i\in X) \right| = 0
\quad \quad {\rm for }~~ |X|\ge 2
\cr
&
\lim_{\ell\ten\infty} \sup_{m_i\in \bar \O^{(\ell)}_i \atop |m_i|\le \ell^a}
\left| \Phi^{(\ell),sr}_{\{i\}} (m_i)  + \12 m_i^2 \right| = 0
\quad\quad  {\rm for }~~ i \in  (\ell\Z)^d 
}
$$

We want to stress that the results hold for $\ell $ sufficiently
large. Certainly, in particular, we cannot exclude that, very near to
$T_c$, for some, not sufficiently large $\ell$, the renormalized
measure is not Gibbsian. On the other side, it is easily seen that
taking the limit $\ell \to \infty$ is equivalent  
to iterate BAT transformation on a given scale $\ell_0$; 
to show this it is sufficient to take $\ell = \ell_0^n$ with $n\in{\Bbb N}$.
Theorem \thm[Results] above says that not only the renormalized measure 
$\m^{(\ell)}_{\b,\l}$, for any
sufficiently large $\ell$ is Gibbsian but the corresponding 
renormalized potential $\F^{(\ell)}$  actually converges, as $\ell \to
\infty$, to the one of a system of non-interacting harmonic
oscillators.

We notice that the limiting image system as $\ell \to \infty$ becomes an 
unbounded spin system and  the usual setup of Gibbsianess does
not apply to it (see [EFS]).  
It is therefore clear that we have
to introduce a large field cutoff. Indeed our result is almost optimal as 
we introduce this
cutoff only for the short range part of the interaction  and, moreover, 
the cutoff diverges as a
power law in $\ell$. On the other side it is not difficult to convince 
ourselves that the convergence result, at least in the form given
above, cannot hold without any restriction on the large fields.

This paper contains also other, much weaker, results that apply to
Ising model below $T_c$ at $\l \neq \l^*$, see Theorem \thf[teol] below. 
In that case we are forced to restrict the possible values of $m_i$ 
also in the computation of long--range part of the renormalized potential;
indeed we have of course to forbid that $m_i$ lies in the phase coexistence
interval.

Results in the same direction as Theorem \thm[Results]
were obtained by Cammarota [C]; the main differences w.r.t. the
present paper are that Cammarota considers a high temperature (much
higher than $T_c=\b_c^{-1}$) situation and that he introduces a finite
(not growing to infinity as $\ell\to \infty$) field cutoff.  
The approach of [C] is substantially different w.r.t. ours; [C] uses a
high temperature expansion: the small parameter is $\b$ and the system
is supposed to be weakly coupled on scale one; whereas  since we want to treat
a system with $T= {\b^{-1}}$ higher but arbitrarily close to $T_c$, we have to
use an approach  supposing weak coupling only on a
sufficiently large scale depending on the temperature $T > T_c$ that we have
chosen;  indeed we are forced to act in the scenario of the so called
restricted complete analyticity. 
Let us try to clarify this point.
Exactly in the spirit of renormalization group theory we can say that
a system above its critical point is very weakly coupled on a scale
large compared to the correlation length; as we want to consider {\it
any} $ T >T_c$ we have to take into account the  divergence of the correlation length when approaching $T_c$ (from above). 
The above statement: ``the system is weakly  coupled on a scale larger
than the correlation length" seems a  tautology; in fact it is
not since we need a suitable mathematical setup in order to be able to
implement the above simple observation. The basic idea is to obtain a
perturbative  expansion on the basis of very strong
mixing conditions satisfied by the Gibbs measures; the small parameter
ceases to be the inverse temperature but it will, rather, be related
to the ratio between the correlation length and the scale on which we
are analyzing our system. The geometrical objects (polymers) in terms
of which we perform our perturbative expansions will not live any more
on scale one (like in [C]) but on a scale sufficiently larger than the
correlation length.

A possible notion of strong mixing is the exponential decay of
truncated correlations for any finite volume Gibbs measure with decay
constants uniform in the volume and in the boundary conditions. This
is a stronger notion w.r.t uniqueness of the Gibbs state or
exponential clustering of {\it infinite volume} truncated correlations.
In [DS2], [DS3] Dobrushin and Shlosman introduced and studied the so
called completely analytical interactions showing, in those cases, the
above strong mixing behavior {\it for any} finite or infinite domain of 
arbitrary shape. This complete analyticity turns out to be a
too strong notion in the context of renormalization group
theory. Indeed Dobrushin--Shlosman's complete analyticity implies that
exponential clustering takes place even inside volumes with very
anomalous shapes (for instance with anomalous ratio between boundary
and bulk) so that one is forced to take into account the influence of
boundary conditions up to a scale of order one. 
There are cases of systems perfectly well behaved on regular domains,
say cubes, which, however, do not satisfy D-S complete analyticity
because of their behavior for anomalous shape domains (see example in
[MO2]). 
Another point of view, introduced in [O], [OP], [MO1], [MO2], [MO3] leading to what
can be called restricted complete analyticity, takes into account only
regular domains. In this approach there is a minimal basic length $L$
and one never goes  below $L$ in the sense that one only considers
domains obtained as disjoint unions of cubes of side $L$ (for instance
cubes of side $nL$). 

The algorithm used in the present paper to compute the renormalized
potential is the following.
We start, as basic hypothesis, from restricted complete analyticity
for the constrained, multi--canonical systems, with a minimal length
 $L$ proportional to the scale $\ell$ of
our BAT transformation and with decay constants uniform in the
constraint. We then construct a convergent cluster expansion 
which allows us to compute the renormalized potential.
Since studying directly the mixing properties of a canonical or
multi--canonical measure is a very difficult task we instead deduce it by
using a sharp form of equivalence, or better comparison, between
canonical and grancanonical ensembles. Indeed,
the main key novel technical point of this paper is to get a very
precise notion of equivalence of ensembles, implying the validity of a
finite size condition, which, in turn, will imply a strong mixing
condition for the constrained multi--canonical systems. See also
[DT], [CM], [Y] for a further discussion on the equivalence of ensembles.

Certainly assuming strong mixing for the object system with a given
value $\l$ of the chemical potential is not sufficient to imply the
strong mixing property of the constrained models even at the level of
regular domains. It is, rather, necessary to assume for the object
grancanonical system a strong mixing condition {\it uniform} in
$\l$. Quite surprisingly, this condition is not sufficient in general. 
Indeed  it turns out that what we really need is a strong mixing
condition for a {\it multi--grancanonical} object system; by
multi--grancanonical  we mean a grancanonical measure which is not
translationally invariant because in each cube $Q_\ell(i)$ we put a
different chemical potential $\l_i$ whereas we leave the original,
translationally invariant, mutual interaction.
It happens, as it is shown by an example in Appendix A.2 that uniform
(in $\l$) strong mixing for a grancanonical measure {\it does not
imply } uniform in $\underline \l = \{\l_i\}$ strong mixing for the
multi--grancanonical measure; this pathology is due to the possibility
of a sort of layering phase transition, with long range order, taking
place along the interface between two contiguous large cubes with
different chemical potentials $\l_1, \l_2$ even though, introducing
the same chemical potential $\l_1=\l_2 = \l$ in both cubes the
resulting system, is, $\forall\,\l$, very well behaved. 
On the other side we show that since the interface between two
regular two--dimensional domains is one--dimensional, this layering
phase transition cannot occur when the object system lives in two--dimensions.
Then the result of Theorem \thm[Results] ultimately follows from
strong mixing, uniform in $\l$, exploiting the two--dimensionality of
the Ising system. The latter follows from the general result of
[MOS] saying that in two dimensions the so--called weak mixing implies
strong mixing, provided one is   able to prove weak mixing for our
particular model. 
This is a weaker notion of mixing of a finite volume Gibbs measure
saying, roughly speaking, that the influence of a change in a
conditioning spin on a site $x$ outside a domain $\L$ decays, inside
$\L$, with the distance from the boundary $\de \L$ and not, like it
would be the case assuming strong mixing, with the distance from $x$.
Weak mixing, uniform in the chemical potential $\l$, for Ising model
above $T_c$ has been proved by Higuchi in [H] exploiting general
results by Aizenman {\it et al.} [ABF] about boundedness of susceptibility
above $T_c$. 
We thus use the two--dimensionality in two crucial points: i) in
deducing uniform strong mixing for multi--grancanonical measure from the
same property for simple grancanonical measure and ii) in deducing
strong mixing from weak mixing. 
On the other side, given the strong mixing condition for the
multi--grancanonical measure, the results on the RGT of this paper apply
to any dimension.

The general results about Gibbsianess of renormalized measures that
have been obtained by Haller and Kennedy in [HK], use a strategy very
similar to ours. Indeed their computations, also based on the methods
developed in [O], [OP] are much simpler and more transparent then ours
but apply only to  the case when the  image system is Ising-like;
namely the $\s'$ variables are dichotomic as in Majority rule or
Kadanoff (see [EFS]) transformations. Haller and Kennedy for a given
$\ell$ use the  hypothesis of D--S complete analyticity of the
constrained models to deduce Gibbsianess of the measure resulting
from the application of one transformation. 

We conclude by a brief outline of the various steps needed in the proof of 
Theorem \thm[Results]. 
Higuchi [H]  proves weak mixing, uniform
in $\l$, for Ising model above $T_c$. 
[MOS] proves, in general, that in
two dimensions, for regular domains, weak mixing implies strong mixing. 
In Appendix A.1 we prove that in two dimension strong
mixing uniform in $\l$ for grancanonical measure implies  strong
mixing uniform in $\underline \l = \{\l_i\}$ for the corresponding
multi--grancanonical measure.  
In Section 4 we prove results about  comparison, in a finite volume
$\L$, between multi--grancanonical and multi--canonical measures with
precise estimates of the behavior in $\L$.  From 
this and previous points we deduce that, on a sufficiently large
scale, an {\it effective} (propagating to arbitrarily large, regular domains)
finite  size condition is satisfied for multi--canonical constrained systems.
Then, from this finite size conditions, using the theory developed in [O],
[OP] we are able to perturbatively compute the renormalized Hamiltonian;
and  to extract the potentials. The long range terms of the
interaction potential are computed starting from  a cluster expansion
whose convergence is directly related to the validity of the above
finite size condition. 
Finally the short range terms are handled via a local central limit
theorem for the multi--grancanonical measure.

\newsection Notation and results 

We introduce the general setup: the one of the finite state space,
lattice spin systems. Contrary to the usual treatments
we drop the hypothesis of translation invariance; indeed it will 
be replaced by spatial uniformity of some basic estimates.
We start by giving a list of basic definitions.

\bigskip
\noi{\it 2.1. The lattice.} 

\noi
For $x=(x_1,\cdots,x_d)\in\bR^d$ we let $|x|:=\sup_{k=1,\cdots,d} |x_k|$.
The spatial structure is modeled by the 
$d$--dimensional lattice $\cL:=\Z^d$ in which we let $e_i, i=1, \dots,d$ 
be the coordinate  unit vectors. We shall denote by $x,y,\cdots$ the sites
in $\cL$ and by $\L,\D,\cdots$ subsets of $\cL$.
We consider $\cL$ endowed with the distance 
$d(x,y)=|x-y|$. We use $\L^c :=
\cL\setminus \L$ to denote the complement of $\L$.
For $\L$ a finite subset of 
$\cL$ (we use $\L \subset\subset \cL$ to indicate that $\L$ is finite), 
$|\L|$ denotes the cardinality of $\L$. 
For $x\in\cL$ and $\ell$ an odd integer we
let $Q_\ell (x):=\{ y\in \cL : d(y,x)\le (\ell-1)/2\}$ 
be the cube of side $\ell$ centered at $x$; 
for $\ell$ an even integer we let instead 
$Q_\ell(x):=\left\{y\in \cL : \left| y - (x+\hat e) \right| 
\le \ell/2\right\}$, where $\hat e :=\(1/2,\cdots,1/2\)$, 
be the cube of side $\ell$ centered in $x+\hat e$ 
(which belongs to the dual lattice).
We shall denote $Q_\ell(0)$ simply by $Q_\ell$.
Given $r>0$ and $\L\su\cL$ we introduce the outer 
boundary of $\L$ by $\p_r \L :=\{x\not\in \L : d(x,\L) \le r \}$.
We let also $\overline{\L}^r:= \L\cup\p_r\L$.

Given an  integer $\ell$, we also introduce the rescaled lattice
$\cL_\ell := (\ell\Z)^d$ which is naturally embedded in $\cL$; we shall
therefore regard points in $\cL_\ell$ also as points in $\cL$ without
further mention, more precisely we will make the following identification:
${\cal L}_{\ell}\ni(i_1,\dots,i_d)\equiv (\ell i_1,\dots,\ell i_d)\in{\cal L}$. 
We use $i,j,\cdots$ to denote points in $\cL_\ell$ and $I,\cdots$
to denote subsets of $\cL_\ell$.
Analogously the distance in $\cL_\ell$ is denoted by $d_\ell(i,j)$,
therefore for $i,j\in\cL_\ell$ we have $d(i,j)=\ell d_\ell(i,j)$.

\bigskip
\noi{\it 2.2. The configuration space.} 

\noi
We suppose given a positive integer $\cN\in {\Bbb N}^+$ and, for every 
$x \in \cL $,   a positive integer $\cN_x\leq \cN$.
We then introduce the following:
\item{-}{{\sl Configuration space of a single spin}. For any 
$x \in \cL$ we have a finite set $\O_x$,  $|\O_x| =\cN_x + 1$. 
We identify $\O_x$  with $\{0,1,\dots, \cN_x\}$ which we 
consider endowed with the discrete topology;}
\item{-}{{\sl Configuration space in a subset $\Lambda \subset\cL$}.
We set $\Omega_\Lambda^{(\cN)} := \otimes_{x\in \L} \O_x$;}
\item{-}{{\sl Configuration space in the whole $ \cL$}. We set
$\Omega^{(\cN)} := \otimes_{x\in \cL}\O_x$ and equip it with the
product topology.} 

\noi
We can therefore look at a configuration $\sigma \in \Omega^{({\cal N})}$ 
as a function
$\sigma : \cL \mapsto  \{0,1,\dots,\cN\}$. The integer
$\sigma_x \equiv\sigma(x)$ is called value of the
spin at the site $x\in\Lambda $ in the configuration 
$\sigma$. For $\L\subset\cL$, we use $\s_\L:=\{\s_x\in \O_x, ~x\in\L\}$ 
to denote the collection of spins in $\L$. 
For $x\in\cL$ we define the shift $\th_x$ (acting on $\O^{(\cN)}$) by  
$(\th_x\s)_y := \s_{y-x}$.

We also introduce $C\(\O^{(\cN)}\)$ the space of continuous functions on $\O$
which becomes a Banach space under the norm 
$\|f\| :=\sup_\s |f(\s)|$ and note that the local functions (i.e. 
the functions depending only on a finite number of spins) 
are dense in  $C\(\O^{(\cN)}\)$. For $f$ a local function depending on 
the spins in $\L\subset\subset{\cal L}$, 
i.e. $f(\s)=f(\s_\L)$, we let $S(f)\equiv\supp(f) := \L$ 
be the support of $f$.

In the case $\cN=1$ the spin $\s_x$ takes values in $\{0,1\}$,
i.e. we have a lattice gas. In such a case we use the notation
$\O:=\{0,1\}^\cL$ and denote by $\h,\z,\cdots$ typical elements of  
$\O$; the value $\h_x\in\{0,1\}$ is interpreted as the occupation 
number in $x$.
Given $\h\in \O$ we define a new configuration $\h^x$ which is
obtained from $\h$ by flipping the occupation number in $x$, i.e.
$$
(\h^x)_y := \cases
{
\h_y & if $y\not = x$\cr
1-\h_x & if $y=x$\cr
}
$$

\bigskip
\noi{\it 2.3. The potential.} 

\noi
A potential $\Phi=\{\Phi_\L, \L\subset\subset\cL\}$ is a family of
functions labeled by finite subsets of $\cL$ and $\Phi_\L :
\O_\L^{(\cN)} \mapsto \bR$.   
We introduce the following possible conditions on $\Phi$: \hfill\break
\item{--}{{\it Finite range}. There exists $r>0$ such that 
$\Phi_\L=0$ if ${\rm diam } (\L) > r$;}
\par
\item{--}{{\it Translation invariance}. For each $x\in\cL$, 
$\Phi_{\L} (\sigma) = \Phi_{\L+x} (\th_x \sigma)$.

We note that the potentials  (which do not need to satisfy 
the conditions above) form a linear space. Given $\a\ge 0$, we introduce
in it the norm $\|\cdot\|_\a$ defined by
$$
\| \Phi\|_\a := \sup_{x\in\cL} \sum_{\L \ni x} e^{\a |\L|}
\sup_{\s_\L\in\O_\L^{(\cN)} } \left| \Phi_\L (\s_\L) \right|
$$
We also note that in the translation invariant case we can omit the first
supremum above. 

Given $\L\su\su\cL$ and a potential $\Phi$ with bounded $\|\cdot\|_0$ norm,
the {\it energy} associated to a configuration $\s$  when
the boundary condition outside $\Lambda $ is (the restriction to $\L^c$
of) $\tau\in\Omega$, is given by:
$$
E_\L(\s |\t ) := \sum_{\G \cap \L\neq \emptyset} \Phi_\G( \s \circ_\L \t )
\Eq(1.2)
$$
where
$$
(\s \circ_\L \t)_x := 
\cases{
\s_x  &  if $x \in \L$ 
\cr 
\t_x  & if $x \not\in \L$ \cr
}
$$
Note that the sum on the r.h.s. of \equ(1.2) is absolutely convergent
(uniformly in $\sigma$ and $\tau$) 
by the boundedness of $\|\Phi\|_0$. We also remark that
for a finite range potential the map $\t\mapsto E_\L(\s |\t )$ 
depends only on $\t_{\p_r\L}$.

\bigskip
\noi{\it 2.4. The Gibbs measures.}

\noi
Given a potential $\Phi$ of bounded $\|\cdot\|_0$ norm, for each 
$\L\subset\subset \cL$ we define the (finite volume) Gibbs measure 
in $\L$ with boundary condition $\t$ as
$$
\mu^\t_{\L} (\s) := {1\over Z^\t_\L} 
\exp\left\{ -  E_\L\(\s | \t \) \right\}
$$  
where $ Z^\t_\L$, called partition function, is the normalization constant, i.e.
$$
Z^\t_\L =  Z^\t_\L (\Phi) := \sum_{\s\in \O_\L^{(\cN)}} e^{- 
E_\L\(\s | \t \)}
$$
Note that we have included the inverse temperature in the definition
of energy; in fact it will be kept fixed in our analysis.  

We regard $\mu_\L^\t$ also as a measure on the whole $\O^{(\cN)}$ by giving
zero mass to the configurations $\s$ which do not agree with $\t$ on 
$\L^c$. The (infinite volume) Gibbs states associated to the potential
$\Phi$ are then the measures $\mu$ on $\O^{(\cN)}$ which satisfy the 
DLR equations
$$
\int\! \mu(d\t)\,\, \mu_\L^\t(f) = \mu (f), \quad  {\rm for~ any~~ } \L\ssu
\cL, \,\, f\in C \( \O^{(\cN)} \)
$$

For a translationally invariant lattice gas we observe that
we have $\Phi_{\{x\}}(\h)= - \l\h_x +a$ for some 
constants $\l,a\in\bR$. We neglect
the constant $a$ (which do not affect the definition of the Gibbs
measure) and note that $\l$ is interpreted as the {\it
chemical potential}.  We also introduce the  {\it activity} $z\in\bR^+$
by $z:=e^\l$ which we use to parametrize lattice gases with different
chemical potentials. In such a case we write $\Phi = (z,U)$ where
$U = \{ \Phi_\L, \L\subset\subset \cL ,|\L|>1\}$ and call $U$ the
interaction.  
We shall also write (sometimes) $\mu_{\L,z}^\t$ 
(resp. $Z_\L^\t(z)$) in order to indicate explicitly the dependence 
on the activity $z$.

\bigskip
\noi{\it 2.5. Strong Mixing Conditions.}

\noi
In what follows we recall  notions concerning some mixing properties of
Gibbs measures.
Most of the theory has been, up to now, developed
in the finite range, translationally invariant case.
Extension to not translationally invariant cases, when suitable uniform
conditions hold, is, in most of the cases, straightforward.
In particular we will be concerned with the so--called strong mixing
condition which can be formulated in terms of exponential clustering of
truncated expectation with respect to the Gibbs measures in certain domains
$\L$ with $\t$ boundary conditions when this exponential clustering takes
place uniformly in $\L$ and $\t$.
This strong mixing condition implies uniqueness of infinite volume Gibbs
measure and its exponential clustering. It can be shown that finite volume
strong mixing condition, with constants uniform in the volume and in the
boundary conditions, is strictly stronger than the equivalent infinite volume
notion (see [Sh], [Ba], [DM], [CM]).
As it has been shown by Dobrushin and Shlosman (see [DS2],
[DS3]) this strong mixing condition, supposed to hold   {\it for
any} (finite or infinite) volume $\L$, is equivalent to many other
conditions like analyticity properties of thermodynamic and correlation
functions or tree--decay of semi--invariants.
Interactions giving rise to this kind of nice behavior have been called by
Dobrushin and Shlosman {\it completely analytical}.
Among their equivalent complete analyticity conditions, Dobrushin and
Shlosman have introduced suitable finite size conditions that they call
``constructive conditions". 
They show that, supposing that there exists a finite
domain $\L$ such that  strong mixing condition  is 
satisfied with suitable (depending on
$\L$) decay constants {\it for all} subsets of  $\L$,
then a strong mixing condition holds {\it for all} (finite or
infinite) volumes.

We refer to  [MO2] for a discussion on the applicability of this point of
view.
Indeed often the request of exponential clustering for {\it arbitrary} shape
does not fit with many reasonable applications. There are examples (see [MO1])
where nice exponential mixing properties hold for regular domains (like, for
instance, cubes) and in infinite volume, whereas they fail to hold for
domains with anomalous ratio between boundary and bulk, implying violation of
Dobrushin--Shlosman complete analyticity.
In [O], [OP], [MO1], [MO2], [MO3]
another scenario has been introduced, more suited to the
renormalization group problematic. It can be called ``restricted complete
analyticity" or ``complete analyticity for regular domains"
This point of view refers to exponential mixing in finite volumes {\it
multiples} of a given cube of size $\ell_0$.
In the framework of this theory one can develop a constructive condition of
the following kind: if $\exists \, \ell_0$ such that a suitable (depending on
$\ell_0$) mixing condition holds in the cube             
$Q_{\ell_0}$, then the
same condition (possibly with worse constants) holds for any {\it multiple} of 
$Q_{\ell_0}$.
This possibility of propagation from finite to arbitrarily large (and even
infinite) volumes is called ``effectiveness" in [MO2].
Subsequently many results have been obtained in the framework of restricted
complete analyticity that could have been problematic and even false in the
context of Dobrushin--Shlosman complete analyticity (see, for instance [MO2],
[MO3], [MOS], [SS]).

Given a measure $\mu$ and two square integrable random variables 
$f,g$ we denote by $\mu(f;g):= \mu(fg)-\mu(f)\mu(g)$ the covariance
between $f$ and $g$.
For $\D\ssu\cL$ we introduce $\mu^\t_{\L;\D}$ as the
relativization (projection) of the Gibbs measure $\mu^\t_{\L}$ to
$\O_\D^{(\cN)}$, i.e. 
$$
\mu^\t_{\L;\D} (\s_\D) := \int\! \mu^\t_{\L}(d\z) \; \id_{\z_\D=\s_\D}
$$
We finally recall that the total variation  distance between two
measures $\mu$, $\nu$ on a {\sl finite} set $S$ is given by
$$
\Var(\mu,\nu) := \12 \sum_{\o\in S} | \mu(\o) - \nu(\o) |
\equiv \sup_{X\su S} \left| \mu(X)-\nu(X) \right|
$$
If $a,b\in \bR$, we let $a\mi b:= \min\{a,b\}$. For a finite range
potential we introduce the following strong mixing condition.

\smallskip 
\noindent{\bf Condition SM($\ell_0$)} ({\it Strong Mixing}).
\hfill\break
{\sl Given an integer $\ell_0$ we
say that the potential $\Phi$ satisfies SM($\ell_0$) if there exist
two constants  $A,\g>0$ such that for 
any volume 
$$
\L=\bigcup_{i\in I} Q_{\ell_0}(i),\quad I\su\su \cL_{\ell_0}
\Eq(multiplo)
$$ 
the following bound holds.
For any $x\in\p_r\L$ and any $\D\su\L$ we have
$$
\sup_{\t\in \O^{(\cN)}} \sup_{a\in \O_x} 
\Var\( \mu_{\L;\D}^{\t \circ_x a} , \mu_{\L;\D}^{\t} \) 
\le A e^{-\g d(x,\D)}
\Eq(e:sm)
$$
}

\smallskip
We next discuss {\it finite size} conditions which imply SM($\ell_0$).
Let $m$ be an integer, $m>r$, and consider the cube $Q_{3m} (j)$,
$j\in\cL_m$.  
Given a particular lattice direction $e_i$ we can partition $Q_{3m}
(j)$ into three parallelepipeds having $d-1$ sides equal to  $3m$ and
the last one equal to $m$ (slices) with the minimal side parallel to
the $e_i$ direction  (slice orthogonal to $e_i$).  
We write
$$
Q_{3m} (j)=Q^{(i,-)}_{3m} (j)\cup Q^{(i,0)}_{3m} (j)\cup Q^{(i,+)}_{3m} (j)
\Eq (partizione)
$$
here $Q^{(i,0)}_{3m} (j)$ denotes the central slice.

Let $P^{(i)}_m(j)$ be the set of all subsets of $Q^{(i,0)}_{3m} (j)$ which are
unions of cubes $Q_m(j)$.
For $V \in  P^{(i)}_m (j)$ let $\de ^{(i,+)} V$,
$\de ^{(i,-)} V$ denote the part of  $\de_r V$ 
in the direction of $e_i, -e_i$ respectively (opposite $r$--faces of $V$). 
Given $\s,\z,\t \in \O$, $y \in \de ^{(i,+)} V$, 
$y' \in \de ^{(i,-)} V$, we denote by $\s^{(i,+)},\z^{(i,-)},\t$  
the configuration obtained from $\t$ 
by substituting  in $\de ^{(i,+)} V$,$\de ^{(i,-)}  V$  the restrictions
of  $\s,\z$, respectively:
$$
\( \s^{(i,+)},\z^{(i,-)},\t \)_x
:= \cases
{
\s_x & if $x\in \de^{(i,+)}V$ \cr
\z_x & if $x\in \de^{(i,-)}V$\cr
\t_x & otherwise
}
$$
analogously we denote by  $\s_y,\z^{(i,-)},\t $  the configuration
obtained from $\t$  by substituting  to $\t$ in $y$, $\de ^{(i,-)}  V$
the restrictions of  $\s,\z$, respectively:
$$
\( \s_y,\z^{(i,-)},\t \)_x
:= \cases
{
\s_x & if $x=y$ \cr
\z_x & if $x\in \de^{(i,-)}V$\cr
\t_x & otherwise
}
$$
finally we denote by $
\s_y,\z_{y'},\t 
$
the configuration obtained by substituting 
to $\t$ in $y,y'$ the restrictions of $\s, \z$:
$$
\(\s_y,\z_{y'},\t\)_x
:= \cases
{
\s_x & if $x=y$ \cr
\z_x & if $x=y'$\cr
\t_x & otherwise
}
$$
of course 
$$ \t^{(i,+)},\t^{(i,-)},\t \, \equiv \, \t_y,\t^{(i,-)},\t\equiv
\t_y,\t_{y'},\t \equiv \t.$$
We introduce the notation: $Z_V(\tau):=Z_V^{\tau}$.

\smallskip
\noindent{\bf Condition C1($m, \e_1$)} {\it (see [OP], Eq (1.8)]).}
\hfill\break
{\sl
$$
\sup _{j\in \cL_m} 
\; \sup_{i\in \{ 1,\cdots,d\} }
\; \sup_{V\in P^{(i)}_m (j)}
\; \sup_{\s,\t\in \O^{({\cal N})}}
\left |{  Z_V \( \s^{(i,+)},\s^{(i,-)},\t \)
Z_V\(\t^{(i,+)},\t^{(i,-)},\t\) \over
Z_V\( \s^{(i,+)},\t^{(i,-)},\t \) Z_V \( \t^{(i,+)},\s^{(i,-)},\t \) }
-1 \right|
<\e_1
\Eq (C1)
$$
}

\smallskip
\noindent
{\bf Condition C2($m, \e_2$)}
{\sl
$$
\sup _{j\in \cL_m} 
\; \sup_{i\in \{ 1,\cdots,d\} }
\; \sup_{V\in P^{(i)}_m (j)}
\; \sup _{ {y\in \de ^{(i,+)} V }}
\; \sup_{\s,\t\in \O^{({\cal N})}}
\left |{  Z_V(\s_y,\s^{(i,-)},\t ) Z_V(\t_{y},\t^{(i,-)},\t)\over
Z_V(\s_y,\t^{(i,-)},\t )Z_V(\t_{y},\s^{(i,-)},\t )} -1\right|
<{\e_2\over m^{d-1}}
\Eq (C3/2)
$$
}

\smallskip
\noindent{\bf Condition C3($m, \e_3$)} {\it (see [O]).} \hfill\break
{\sl
$$
\sup _{j\in \cL_m} 
\; \sup_{i\in \{ 1,\cdots,d\} }
\; \sup_{V\in P^{(i)}_m (j)}
\; \sup _{ {y\in \de ^{(i,+)} V \atop  y' \in \de ^{(i,-)} V}}
\; \sup_{\s,\t\in \O^{({\cal N})}}
\left |{  Z_V(\s_y,\s_{y'},\t ) Z_V(\t_{y},\t_{y'},\t)\over
Z_V(\s_y,\t_{y'},\t )Z_V(\t_{y},\s_{y'},\t )} -1\right|
<{\e_3\over m^{2(d-1)}}
\Eq (C2)
$$
}

\smallskip
It is easy to show, using a telescopic argument, that  there exists a
constants $\k$ such that C2($m,\e_2$) implies C1($m,\k \e_2$) and 
C3($m,\e_3$) implies C2($m,\k \e_3$) (see [O]). 
It is also immediate to see that the results proven for
the translationally invariant case in 
[O], [OP], [MO2] extend to the general case when the space uniform 
condition holds. We have indeed the following result. 
Let  
$$
 \e(d) := [3(2^{d+1} +1)]^{-d} \; 2^{-2d} \; e^{-4},
\Eq (CCC)
$$
then condition C1($m,\e(d)$) implies the existence of a 
convergent cluster expansion which, in turn, implies SM($m$). 

We remark that once we have proven the crucial point which is the  {\it
effectiveness}, namely that C1($m,\e(d)$) implies SM($m$),
then, considering the rescaled system
whose new single spin variables, labeled by
$j\in\cL_m$, are the old spin configurations in the blocks $Q_m(j)$, we
can apply Dobrushin--Shlosman's results [DS1], [DS2], [DS3] 
to get all their equivalent
mixing and analyticity properties of the Gibbs state for every
``multiple'' of the $Q_m$'s namely for all volumes $\L$ of the form
\equ(multiplo). 
This is the {\it restricted complete
analyticity} namely the validity of the D--S equivalent properties (see
[DS2], [DS3]) for every volume of the form \equ(multiplo).
In particular SM($\ell_0)$ is equivalent to:

\smallskip
\noindent{\bf Condition SM2($\ell_0$)} 
\hfill\break
{\sl
Given an integer $\ell_0$ 
we say that the potential $\Phi$ satisfies SM2($\ell_0$) if
there exist two constants $A,\g > 0$ such that for
every pair of local functions 
$f,g$ and every volume of the form \equ(multiplo)
$$
\sup _{\t\in \O^{(\cN)}} \; |\mu^{\t}_{\L}(f;g)|
\leq A \; (|S_f|\wedge|S_g|) \; \|f\|\  \; \|g \| \; e^{-\g  d (S_f,S_g)} 
\Eq(sm2)
$$
where we recall $S_f$, $S_g$ are the supports of $f,g$.
}

\smallskip
Indeed the implication C1($\ell_0,\e(d)$) 
$\Rightarrow$ SM2($\ell_0$) can be obtained
directly via cluster expansion by using the methods of references [O], [OP].
We do not reproduce here the results of [O], [OP] but, looking at the
application to the renormalization group problem that will be
developed in next section, the reader could easily understand these results.

\bigskip
\noi{\it 2.6. Lattice gases: Uniform Strong Mixing Conditions.} 

\noi
We here consider just a finite range lattice gas (i.e. 
$\O = \{0,1\}^\cL$)  and 
introduce  some uniform strong mixing conditions which are needed  
to study the RG map. These conditions say -- rougly speaking -- that
SM($\ell_0$) holds {\it uniformly} in the external field (one body
interaction). Unfortunately, as discussed in the Introduction, we need
such a condition also for some non homogeneous external field. Such a
condition plays also a crucial role in the ergodic properties of the
Kawasaki (conservative) dynamics, [Y]. 

Given a finite range lattice gas with translationally invariant 
interaction we introduce the following Condition.
We recall $z=\exp\{\lambda\}$ is the activity. 

\smallskip
\noindent{\bf Condition USM($\cA$)} ({\it Uniform Strong Mixing}). 
\hfill\break
{\sl
Given an open set ${\cal A}\subseteq [0,\infty)$, we say that
the interaction $U$ satisfies USM($\cA$) if
for each $z\in\cA$ there
exists $\ell_0=\ell_0(z)$ such that  
condition SM($\ell_0$) holds for $(z,U)$. 
Furthermore the following is to be satisfied: 
\item{({\it i})}{for any closed set  $\cC\subseteq \cA$ 
we can take the constants 
$\ell_0,A,\g$ uniform for $z\in \cC$;} 
\item{({\it ii})}{we can take $A= A_0 z\mi z^{-1}$ for some other constant 
$A_0$ independent of $z$.} 
}
\smallskip

\noi{\it Remark.} 
We note that for ${\cal A}=[0,\e]\cup[\e^{-1},\infty)$ with $\e$
small enough (depending on $d$, $r$ and $\|\Phi\|_0$) the above
conditions holds. Indeed for $z\wedge z^{-1}$ small, SM(1)
follows from standard perturbative theory (for instance by using
Dobrushin single site criterion [D1]). We can therefore safely replace
the set $[0,\infty)$ in Condition USM(${\cal A}$) by the compact set 
$[\e,\e^{-1}]$. To avoid  delicate continuity questions we introduced
($i$) above  as an independent hypothesis. 
The same argument shows $(ii)$ is automatically satisfied; we have included it
only for convenience.

\smallskip
\noi{\bf Condition GUSM}({\it Global Uniform Strong Mixing})
\hfill\break
{\sl 
We  say that Condition GUSM is satisfied if 
Condition USM($\cA$) holds for $\cA=[0,\infty)$. 
}

\bigskip
\noi{\it 2.7. The multi--grancanonical state.}

\noi
Let  $\ell$ be a positive integer and  $\L\su\cL$ a disjoint
union of cubes of side $\ell$, i.e. 
$$
\L = \bigcup_{i\in I} Q_\ell(i), \quad I\su\cL_\ell
\Eq(La=)
$$
Given a lattice gas with a finite range translationally invariant
interaction $U$, we next define a Gibbs measure in $\O_\L$ which 
has a fixed chemical potential in each cube $Q_\ell$. We call
such a measure a {\it multi--grancanonical} state. 
Let $\bz:= \{z_i\in [0,\infty),\,\, i\in I\}$ the measure
$\mu^\t_{\L,\bz}$ is then 
defined as a Gibbs measure in $\O_\L$ whose potential 
$\Phi^\bz=(\bz,U)$ is given by 
$$
\Phi^\bz_\G(\eta) :=\cases
{
- \h_x \log z_i & if $\G =\{x\}$ and $x\in Q_\ell (i)$ \cr
\Phi_\G(\eta) & if $|\G|>1$ \cr
}
$$
If $I\ssu\cL_\ell$ the finite volume multi--grancanonical measure is thus
defined by 
$$
\mu^\t_{\L,\bz}(\h) = {1\over Z^\t_\L(\bz)}
\prod_{i\in I} z_i^{N_i} \cdot\exp\left\{ 
\sum_{{\G\cap\L\neq\emptyset\atop |\G|>1}}\Phi_\G(\h\circ_\L\t)\right\}
\Eq(dmgcvf)
$$
where $Z^\t_\L(\bz)$ is the normalization constant and 
$$
N_i := \sum_{x\in Q_\ell(i)} \h_x 
\Eq(dNi)
$$
is the total number of particles in $Q_\ell(i)$.
We stress that the multi--grancanonical state $\mu_{\L,\bz}^\t$ does
depend on $\ell$.

We shall need a stronger version of Condition USM which
is formulated in terms of the multi--grancanonical
state.

\smallskip
\noindent{\bf Condition MUSM($\cA$)} 
({\it Uniform Strong Mixing for Multigrancanonical States}).
\hfill\break
{\sl
Given an open set ${\cal A}\subseteq [0,\infty)$, we 
say that the interaction $U$ satisfies MUSM($\cA$) 
if the following condition holds.
For each closed set $\cC\subseteq\cA$ there are constants 
$\ell_0 \in {\Bbb N}, A,\g > 0$ 
such that for any 
$\ell$ integer multiple of $\ell_0$, any $I\su\su\cL_\ell$ 
and any $\bz\in{\cal C}^I$ we have that for any $\L$ of 
the form given in \equ(La=)
the multi--grancanonical  measure $\mu^{\t}_{\L,\bz}$ 
satisfies the bound \equ(e:sm).
}

\smallskip
\noindent{\bf Condition GMUSM} 
({\it Global Uniform Strong Mixing for Multigrancanonical States}).
\hfill\break
{\sl If Condition MUSM($\cA$) holds for $\cA=[0,\infty)$
we finally say that Condition GMUSM is satisfied.}

\smallskip
We also give an effective finite size condition of type C1 which implies
MUSM($\cA$). We note that if $V\in P^{(i)}_m (j)$ we have 
$V= \cup_{k\in\hat V} Q_m(k)$ for some $\hat V \su \cL_m$ uniquely
determined by $V$. We denote below by $Z_{V,\bz} (\t)$ the 
multi--grancanonical
partition
function   as defined in \equ(dmgcvf) 
with $\ell=m$.

\smallskip
\noindent{\bf Condition MUC1($\cA$)} {\it}  
\hfill\break
{\sl
Given an open set ${\cal A}\subseteq [0,\infty]$ we say that MUC1(${\cal A}$)
holds for the interaction $U$ if for each closed set $\cC\subseteq\cA$
there exists an integer $m$ such that
$$
\sup_{i\in \{ 1,\cdots,d\} } \; \sup_{V\in P^{(i)}_m (0)} 
\; \sup_{\bz\in \cC^{\hat V}} \; \sup_{\s,\t\in \O}
\; \left |{  Z_{V,\bz} \( \s^{(i,+)},\s^{(i,-)},\t \)
Z_{V,\bz} \(\t^{(i,+)},\t^{(i,-)},\t\) \over
Z_{V,\bz} \( \s^{(i,+)},\t^{(i,-)},\t \) Z_{V,\bz} 
\( \t^{(i,+)},\s^{(i,-)},\t \) } -1 \right|
\le \e(d)
\Eq (MUC1)
$$
}

\smallskip
Indeed, by exploiting the translationally
invariance of $U$ and following the same argument as the one used in
[O], [OP] it is easy to verify that if MUC1 holds we have that also
MUSM($\cA$) holds.  

\smallskip
\noi{\it Remark 1.}\/ In the high temperature regime, $\|U\|_0\le \e$ with
$\e$ small enough, it is not difficult to show (by using, for instance
Dobrushin's single site condition [D1]) that Condition MUC1($[0,\infty)$) 
holds. 

\noi{\it Remark 2.}\/ We recall that  
By [DS2], [DS3] if 
SM($\ell_0$) holds for the potential $(z,U)$ we can find a neighborhood
$\cO_\e(z)$ of $z$ such that MUSM($\cO_\e(z)$) holds for the
interaction $U$.

\smallskip
We stress that the above Remark 2 gives only a local implication.
On the global side the relationship between MUSM($\cA$)
and USM($\cA$) is not trivial. It is in fact possible to have a
sort of layering phase transition which prevents MUSM($\cA$)
to hold even though  USM($\cA$) does hold.
On the positive side, following an argument in the same spirit as
[MOS] (i.e. that no phase transition may happen in a one--dimensional
boundary of a regular two--dimensional domain), 
we rule out such a possibility in $d=2$.
We have in fact the following Theorem.

\nproclaim Theorem [moslike].
Let $d=2$. Then
USM($\cA$) $\Longrightarrow$ MUSM($\cA$). 

On the negative side we show that the aforementioned pathology may
indeed happen. In Appendix A.2 we give in fact an {\it ad hoc} example of an
interaction $U$ (in $d=3$) such that:
\item{--}{GUSM holds;}
\item{--}{there exist  $\bz$ and $\L$ of the form \equ(La=) 
such that the multi--grancanonical measure associated to $(\bz,U)$ 
exhibits long range order. In particular there exist $\t,\t'$ such that 
$$
\liminf _{\ell \to \infty}
\Var\( \mu_{\L,\bz;\{0\}}^{\t'} , \mu_{\L,\bz;\{0\}}^{\t} \) > 0
$$

We finally mention that, in the context of the two--dimensional Ising
model, the above Theorem implies the following. Consider a standard
Ising model with a non--homogeneous external field which is however
constant in cubes of side $\ell$; then for each $\beta<\beta_c$, there
exist $\ell$ and $L\gg \ell$ such that SM($L$) holds uniformly in the
external field.

\bigskip
\noi{\it 2.8. Block Averaging Transformation (BAT)}

\noi
Let $\mu_z$ be the (unique) infinite volume Gibbs measure of a finite
range translationally invariant lattice gas satisfying Condition
SM($\ell_0$) and $\ell$ an integer. In this case we can define the
block averaging transformation directly in infinite volume.
We partition the lattice $\cL\equiv \bZ^d$ into cubes of side $\ell$, i.e.
$\cL = \bigcup_{i\in \cL_\ell} Q_\ell(i)$. 
We recall that the random variable $N_i$ has been defined in \equ(dNi);
it takes values in the set 
$$
\Omega_i^{(\ell)}:=\{0,1,\dots,\ell^d\}
\Eq(omega-rinor)
$$
We then define the centered and renormalized random variables $M_i$
as in \equ(Centr); it takes values in the set $\bar \O^{(\ell)}_i $
defined in \equ(Ombi). We finally let $\bM :=\{M_i,\; i\in{\cal L}_{\ell}\}$.
The BAT renormalized measure, that we denote by $\mu_z^{(\ell)}$,
is then the (joint) probability distribution of $\bM$
under $\mu_z$. 
Denoting the renormalized configuration by 
$\bm=\{m_i\in \bar \O^{(\ell)}_i \;, i\in{\cal L}_{\ell}\}$, 
the measure $\mu_z^{(\ell)}$ is formally given by 
$\mu_z^{(\ell)}(\bm) =\mu_z (\bM=\bm)$.
We avoid the troublesome issue of describing Gibbs measures on non compact 
single spin space (see [EFS] for a discussion) and consider
$\mu^{(\ell)}_z$ only for finite $\ell$. Therefore the setup previously 
described applies to the finite single spin space $\bar \O^{(\ell)}_i$. 

It is also possible to use a finite volume setup.
Given the integer $p$ we will denote by $\L_p\su\su \cL$
a cube with side $2 d \ell p$. We have
$\L_p= \bigcup_{i\in I_p} Q_\ell(i)$ where 
$I_p\subset\subset \cL_\ell$ is a cube of side $2 d p$. 
Let  $\m_{\L_p,z}^\t$ be the finite volume  Gibbs measure for our lattice 
gas with activity $z$  enclosed in $\L_p$ with $\t$ boundary condition.
We denote by  $\m^{(\ell,\t)}_{I_p,z}$ the finite
volume renormalized measure arising from the application to $\m_{z,p}$ of 
the Block Averaging Transformation on scale $\ell$; it is defined as: 
$$
\m^{(\ell,\t)}_{I_p,z} \( \{ m_i, \; i\in I_p \} \)
:=\m_{\L_p,z}^\t  \( \left\{ M_{i}=m_i,\; i\in I_p \right\} \), 
\quad \quad m_i \in \bar \O^{(\ell)}_i 
\Eq (2.9)
$$ 

\bigskip
\noi{\it 2.9. Main results.}

\noi
We first discuss the case when the global Condition GMUSM
holds. The most relevant example is the standard two--dimensional
Ising model for $T>T_c$. In such a case we are able to prove that the
renormalized measure $\mu^{(\ell)}_z$ is, for each (finite)
$\ell$ large enough, Gibbsian w.r.t. a potential $\Phi^{(\ell)}$ 
of  bounded $\|\cdot\|_\a$ norm (for suitable $\a >0$). 
We can furthermore control the $\ell$ dependence of the norm
$\|\Phi^{(\ell)}\|_\a$. We note (see [IS], [N]) that $\mu^{(\ell)}_z$
converges weakly to $\otimes_{i\in \cL_{\ell}} \varphi_i$  where
$\varphi_i$ denotes a standard Gaussian measure. 
Accordingly, $\Phi^{(\ell)}$ should converge to
the interaction of independent harmonic oscillators. 
Unfortunately, as the limiting interaction has not finite norm (since
the limiting single spin space is unbounded), this convergence cannot
be described in the $\|\cdot\|_\a$ norm. 
However this lack of convergence affects only the (somehow trivial) short 
range part of the interaction; we will decompose the potential into a
short and a long range part $\F^{(\ell)}=\F^{(\ell),sr}+\F^{(\ell),lr}$
($\exists\kappa\in{\Bbb N}:$ $\Phi_{X}^{(\ell),sr}\equiv 0$ if
$\diam(X)\ge\kappa$).
We then introduce a large field cutoff 
(diverging as $\ell\ten\infty$) to control the short range part:
it will converge to the potential of independent harmonic
oscillator for values of the image variables within the cutoff.
We note that this result would be false for large image
variables.
The precise statement is given in the following Theorem. 

\nproclaim Theorem [teog].
Let $U$ satisfy GMUSM. Then there exists $\a>0$ such that
for any $z\in (0,\infty)$ and $\ell$ large enough $\mu^{(\ell)}_z$ is
a translationally invariant Gibbs measure w.r.t. a potential
$\Phi^{(\ell)}$ for which
$$
\left\| \Phi^{(\ell)} \right\|_{\a} \le K(\ell) 
$$
for some constant $K(\ell) < \infty $. 
\hfill\break
Furthermore it is possible to decompose the potential 
into a short and a long range part, 
$\F^{(\ell)}=\F^{(\ell),sr}+\F^{(\ell),lr}$,
such that $\exists\kappa\in{\Bbb N}:$ $\Phi_{I}^{(\ell),sr}\equiv 0$ if
$\diam(I)\ge\kappa$ and the following holds:  
\item{{\it (i)}}{for the same $\a$ as before
$$
\lim_{\ell\ten\infty} \; \left\| \Phi^{(\ell),lr} \right\|_\a =0
$$
}
\item{{\it (ii)}}{there exists a constant $a>0$ such that
$$
\eqalign{
&\lim_{\ell\ten\infty} 
\sup_{m_I \in \bar \O^{(\ell)}_I \atop |m_I| \le \ell^a}
\left| \Phi^{(\ell),sr}_I (m_I) \right| = 0,
\quad\quad {\rm for ~any}~~ I\su\su \cL_\ell, ~|I|\ge 2
\cr
&\lim_{\ell\ten\infty} 
\sup_{m_i \in \bar \O^{(\ell)}_i \atop |m_i|\le \ell^a}
\left| \Phi^{(\ell),sr}_{\{i\}} (m_i)  + \12 m_i^2 \right| = 0
\quad\quad {\rm for ~any}~~ i\in \cL_\ell
}
$$
}

When we assume only the local Condition MUSM($\cA$) our results
are much weaker. 
Before discussing them, let us first note that for the
standard two--dimensional Ising model this Condition holds for $T\le
T_c$ away from the phase coexistence line $z=z^*$ ($z^*$ corresponds
to zero magnetic field in the spin variables), i.e. for each $T\le
T_c$, MUSM($\cA$) holds for $\cA=[0,\infty)\setminus \{z^*\}$.
We are not able to deal directly with the BAT defined in infinite
volume, but we have to start from the finite volume transformation and
take the thermodynamic limit. Moreover, we need to introduce a large
field cutoff also in the long range part of the interaction. These
difficulties are of course related to the limiting single spin space
for which the usual (i.e. uniform in all possible b.c.) Gibbsian
formalism do not apply. We refer to [EFS] for a discussion on the
problems connected with the  introduction of a norm for interactions defined 
on a non compact spaces. Our results are formulated as follows.

\nproclaim Theorem [teol].
Let $U$ satisfies Condition MUSM($\cA$) and $z>0$, $z\in\cA$.
Let also $\Phi^{(\ell,\t)}$ be the (finite volume) potential associated to
the (finite volume) renormalized measure $\mu^{(\ell,\t)}_{I_p,z}$. Then
it is possible to decompose the potential 
into a short and a long range part, 
$\F^{(\ell,\t)}=\F^{(\ell,\t),sr}+\F^{(\ell,\t),lr}$,
such that $\exists\kappa\in{\Bbb N}:$ $\Phi_{I}^{(\ell,\t),sr}\equiv 0$ if
$\diam(I)\ge\kappa$ and the following holds. 
There is a constant $\e=\e(z)>0$  such that for any $I\su\su\cL_\ell$
and any $\ell$ large enough
$$
\eqalign{
& \exists\lim_{p\ten\infty} \Phi^{(\ell,\t),lr}_I (m_I) 
=:\Phi^{(\ell),lr}_I (m_I)
,\quad\quad {\rm uniformly ~for}~~ 
m_I \in \bar \O^{(\ell)}_I,~ |m_i|\le \e \sqrt{\chi |Q_\ell|}, ~\t\in\O
\cr
& \exists\lim_{p\ten\infty} \Phi^{(\ell,\t),sr}_I (m_I) 
=: \Phi^{(\ell),sr}_I (m_I)
\quad,\quad {\rm uniformly ~for}~~m_I \in \bar \O^{(\ell)}_I,~\t\in\O
}
\Eq(AB)
$$
\hfill\break
Furthermore, there are $\a=\a(z)>0, a=a(z)>0$ 
such that for the same $\e=\e(z)$ as before 
$$
\eqalign{
&\lim_{\ell\ten\infty} 
\sum_{I\ni 0} e^{\a|I|} 
\sup_{m_I \in \bar \O^{(\ell)}_I \atop |m_I|\le \e \sqrt{\chi |Q_\ell|}}
\left| \Phi^{(\ell),lr}_I (m_I) \right| =0
\cr
&\lim_{\ell\ten\infty} 
\sup_{m_I \in \bar \O^{(\ell)}_I \atop |m_I| \le \ell^a}
\left| \Phi^{(\ell),sr}_I (m_I) \right| =0 \quad  \quad
\quad\quad {\rm for ~any}~~ I\su\su \cL_\ell, ~|I|\ge 2
\cr
&\lim_{\ell\ten\infty} 
\sup_{m_i \in \bar \O^{(\ell)}_i \atop |m_i|\le \ell^a}
\left| \Phi^{(\ell),sr}_{\{i\}} (m_i) + \12 m_i^2  \right| =0
\quad\quad {\rm for ~any}~~ i\in \cL_\ell
\cr
}
$$

\bigskip
\noi{\it Warnings.} 

\item{--}{Taking advantage of the symmetry of our Conditions w.r.t. the map 
$z\mapsto z^{-1}$, we shall assume, without loss of
generality, that all the activities are bounded by 1. This will be used
extensively without further mention.}
\item{--}{We denote by $C$ a generic positive constant whose 
numerical value can change from line to line. From the statements it
will appear clear from which parameters it depends on.}

\newsection Computing  the renormalized potential via cluster expansion

In this section we discuss the BAT transformation in finite volume. 
We will compute the renormalized interaction via a cluster expansion:
the convergence of the expansion will be ensured by the 
validity of condition C1($m,\e(d)$) for the constrained 
(multi--canonical) systems. This condition C1, in turn, will be deduced from 
the MUSM property of the original system in Section 5.

To simplify notation we write the Boltzmann's factor (with $\t$ boundary
condition) for a configuration $\h$ in the volume 
$\L$, $\h \in \{ 0,1\}^{\L}$, as $\exp \left(+H_{\L}(\h|\t) \right)$
where
$$
H_{\L}(\h|\t):= - E_{\L}(\h|\t)
\Eq (2.10'')
$$
Let us set $L:= d\ell$; given the odd integer $p$, let $\L_p$ be the cube 
with side $2d\ell p$ given by
$$
\L_p:=
\cases{
\{ x =(x_1,\dots, x_d) \in \cL : 
-d\ell\left(p+{1\over 2}\right)+d\ell+1
\leq x_j\leq
+d\ell\left(p+{1\over 2}\right),
j=1,\dots,d
\}
&$d\ell$ even\cr
&\cr
\{ x =(x_1,\dots, x_d) \in \cL : 
-\left(d\ell p+{d\ell-1\over 2}\right)+d\ell
\leq x_j\leq
d\ell p+{d\ell-1\over 2},
j=1,\dots,d
\}
&$d\ell$ odd\cr
}
$$
We can write $\L_p = \cup_{i \in I_p} Q_{\ell}(i) $
where $I_p\su\subset\cL_\ell$ is the cube of side $2dp$ given by
$$
I_p= \cases{
\{ i\in\cL_{\ell}: 
-d\left(p+{1\over 2}\right)+d+1
\leq x_j\leq
+d\left(p+{1\over 2}\right),
j=1,\dots,d
\}
&if $d$ is even\cr
&\cr
\{ i\in\cL_{\ell}: 
-\left(dp+{d-1\over 2}\right)+d
\leq x_j\leq
dp+{d-1\over 2},
j=1,\dots,d
\}
&if $d$ is odd\cr
}
$$
Let us introduce the quantity:
$$
Z^{(\ell,\t)}_{\L_p,\underline {n}}  
:= \sum _{\h \in \O^{(\underline {n})}_{\L_p}} e^{H_{\L_p}(\h|\t)}
\Eq (2.10)
$$
where
$\underline {n} = \{n_i, ~i\in I_p\} \in 
\O_{I_p}^{(\ell)} := \otimes_{i\in I_p} \O^{(\ell)}_i \equiv
\{0,1,\dots,\ell^d\}^{I_p}$ and
$$
\O^{(n_i)}_i:= \left\{ \h \in \{0,1\}^{Q_\ell(i)} ~:~ 
 N_i(\h) = n_i \right\}
,\quad\quad
\O^{(\underline {n})}_{\L_p}:= \bigotimes_{i\in I_p} \O^{(n_i)}_i
\Eq(dOni)
$$
It is convenient to look at the renormalized measure 
$\m^{(\ell,\t)}_{I_p,z}$ in \equ(2.9) in terms of the variables $\bn$;
such measure is Gibbs w.r.t. to the renormalized Hamiltonian given by
$$
H^{(\ell,\t)}_{\L_p} \( \underline {n} \) = 
\log Z^{(\ell,\t)}_{\L_p,\underline {n}} 
\Eq (2.14)
$$
Given $\underline{n} \in \O_{I_p}^{(\ell)}$; we can look at  the quantity
$Z^{(\ell,\t)}_{\L_p,\underline {n}} $ defined in
\equ(2.10) as the partition function of a (generally not
translationally invariant) system which is  the original lattice 
gas {\it constrained} to have fixed values of the total number of
particles in each block $Q_\ell(i)$, $i\in I_p$.
Its elementary configurational
variables are the original spin configurations in each block  $Q_\ell(i)$
compatible with the assigned value $n_i$ of $N_{i}$ namely the 
set $\O^{(\underline {n})}_{\Lambda_p}$ defined in \equ(dOni).
The elements of $\O^{(n_i)}_i$ will be called 
{\it block variables} not to be confused with the renormalized variables
$n_i$. We also call {\it multi--canonical} these constrained  systems.

We will consider blocks of these block variables of size $d$; these 
corresponds to the blocks $Q_{L}(i)$ with $L=d \ell$ in the 
original variables. The reason for this choice
will appear clear in the following sections: it 
corresponds to the minimal size for which we are able to prove, for the
constrained model, the validity of our Condition C1($m,\e(d)$).
In other words, to meet Condition C1($m,\e(d)$) we have 
to choose $m=d$ and $\ell$ sufficiently large.
With respect to the general setting of Section 2 we have
$\O_i=\O^{(n_i)}_i$ whereas the potential is the one inherited by the
original model. 
In particular, if we choose   $\ell $ larger that the range $r$ of the
original interaction, then only contiguous blocks will interact. 
We repeat that the size of the blocks that in Section 2 was  
generically called $m$ now equals $d$. 
The main result of this section is stated as follows, where, for 
$V\in P^{(k)}_{L}(i)$, we let $\hat V \subset\subset \cL_\ell$ be 
such that $V=\cup_{j\in\hat V}Q_{\ell}(j)$. 

\nproclaim Theorem [summa].
Consider a $d$--dimensional lattice gas with finite range, translationally 
invariant interaction.
Let $\ell\in{\Bbb N}$ and suppose there exists a closed
${\cal D}\subseteq [0,1]$ such that 
$$
\sup_{i\in I_p} ~~
\sup_{k=1,\cdots,d} ~~
\sup_{V\in P^{(k)}_{L}(i)}~~
\sup_{\underline n \in \cD^{(\ell)}_{\hat V}} ~~
\sup_{\s,\z,\t} 
\left| { Z_{V,\bn}\(\s^{(k,+)},\s^{(k,-)},\t\) 
         Z_{V,\bn}\(\z^{(k,+)},\z^{(k,-)},\t\) 
\over    Z_{V,\bn}\(\s^{(k,+)},\z^{(k,-)},\t\)  
         Z_{V,\bn}\(\z^{(k,+)},\s^{(k,-)},\t\)
}
-1\right|
\le \d ( \ell )
\Eq(CC)
$$
where $ \cD^{(\ell)}_{\hat V} := \( |Q_\ell| \cD \)^{\hat V} 
\cap \O_{\hat V}^{(\ell)}$ and $\d ( \ell ) \to 0$ as $\ell\to \infty$.
\hfill\break
Then, the measure $\mu_{I_p,z}^{(\ell,\t)}$ defined in 
\equ(2.9) is Gibbsian w.r.t. a potential 
$\Phi^{(\ell,\t)} = \{ \Phi_{X}^{(\ell,\t)},\; X\subset I_p\}$.
Let
$$
\cM^{(\ell)}_X := \( { |Q_{\ell}| \cD - \rho (z) \over 
\sqrt{\chi |Q_{\ell}|}} \)^X \bigcap \bar \O^{(\ell)}_X
$$ 
We have the following:
\item{{\it (i)}}
For each $X\subset I_p$ with $d_\ell (X,I_p^c)> 3d$ and 
$m_X\in \cM^{(\ell)}_X$, $\Phi_{X}^{(\ell,\t)}$ does not depend on
$\t$ (and $I_p$). 
In particular for each $X\subset\subset\cL_{\ell}$
$$
\exists\lim_{p\to\infty} 
\Phi_{X}^{(\ell,\t)}(m_X)=:\Phi_{X}^{(\ell)}(m_X),
\quad \quad \hbox{uniformly for $m_X\in\cM^{(\ell)}_X$} , \t\in \O
\Eq (ltpp)  
$$
\item{{\it (ii)}}
Let $\Phi^{(\ell)}=\{\Phi_X^{(\ell)},\; X\subset\subset\cL_{\ell}\}$, 
we have a decomposition 
$\Phi^{(\ell)}=\Phi^{(\ell),sr}+\Phi^{(\ell),lr}$
where $\Phi_X^{(\ell),sr}\equiv 0$ if $\diam_\ell(X) >3d$ and  
there are constants $\alpha>0$, $C$ such that:
$$ 
\sum_{X\ni 0} e^{\alpha |X|} \sup_{ m_X \in \cM^{(\ell)}_X }
\left| \Phi_X^{(\ell),lr} (m_X) \right| \le C \delta(\ell)
\Eq(e:summa)
$$

\noi{\it Remark.}\/
The potential $\Phi_X^{(\ell,\t)}$ will be explicitly constructed 
(see \eqf(2.78) and \eqf(2.73/7)) below. In Section 5 we will show that
we can take $\Phi_{\{i\}}^{(\ell),sr}(m_i)=-m_i^2/2$ and there exists a
constant $a>0$ such that for each $X$,  $|X|\ge 2$
$$
\sup_{m_X \in\cM^{(\ell)}_X \atop |m_X|<\ell^a}
\left| \Phi_X^{(\ell),sr}(m_X)\right|< \gamma(\ell)
$$
with $\gamma(\ell)\to 0$ if $\ell\to\infty$.   

\bigskip
Similarly to what has been done in [HK], 
in order to compute the
renormalized potential and prove Theorem \thm[summa],   
we are going to apply to the
constrained systems the method developed in [O], [OP].
To simplify the exposition we will treat in detail
only the  two--dimensional case. An analogous treatment can be 
developed for the $d$--dimensional case along the lines of [OP]. 
For the same reason, we discuss only the case of periodic boundary 
condition in $\L_p$; the case of general boundary condition can be
treated along the same lines with minor changes giving rise to
estimates uniform in $\tau$.

In the rest of this section we will express the coordinates of points and
components of vectors in
$\cL_L$ in $L$ units.
Let us denote by $e_1,e_2$, respectively, the 
coordinate unit vectors in $\cL_L$: $e_1=(1,0)$ horizontal,
$e_2=(0,1)$ vertical. 
Recall that since now $d=2$, we have $L=2\ell$.
We further partition $\cL_L$ into four sub--lattices of spacing $2L$:
$$
\cL_L =\cL_{2L} ^A \cup \cL_{2L} ^B \cup \cL_{2L} ^C \cup\cL_{2L} ^D
$$
where:
$$
\eqalign{
\cL_{2L} ^A &:= \{ i= (i_1,i_2) \in \cL_L: i_1 = 2 j_1,i_2 = 2 j_2,\hbox 
{for some integers}\;\; j_1,j_2\}
\cr
\cL_{2L}^B &:= \cL_{2L} ^A + e_2
\cr
\cL_{2L} ^C &:= \cL_{2L} ^A + e_1 + e_2 = \cL_{2L} ^B + e_1 
\cr
\cL_{2L} ^D &:= \cL_{2L} ^A + e_1 = \cL_{2L} ^C + e_2 = \cL_{2L} ^B + e_1 + e_2
\cr
}
\Eq (2.7')
$$
We also set, for $i \in \cL_L$:
$$
A_i := Q_{L}(2i),\;\;\; 
B_i := Q_{L}(2i +e_2)\;\;\;
C_i := Q_{L}(2i+e_1 +e_2)\;\;\;
D_i := Q_{L}(2i+e_1).
\Eq (2.8')
$$
(See Fig. 1).

\midinsert
\vskip 8 truecm\noindent
\includegraphics{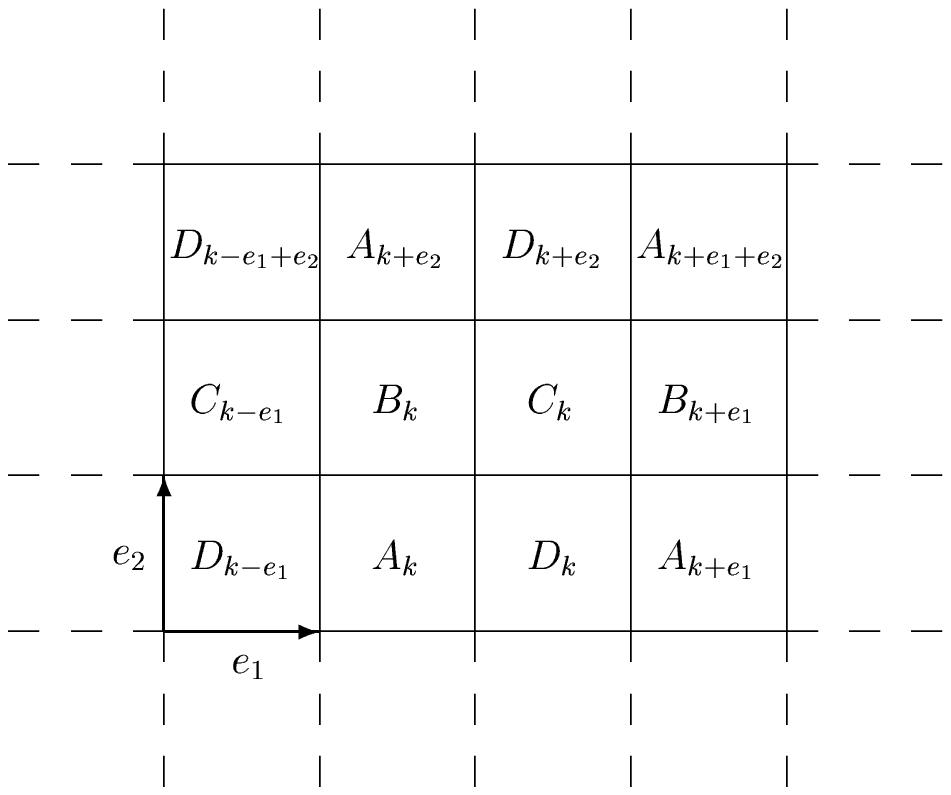} 
\par\noindent 
{\centerline {\bf Fig. 1}} 
\endinsert
\par

Then we can partition the torus $\L_p$ into the union of the $L$--blocks of the
four types: $A, B, C, D$:
$$
\L_p = \cA_p \cup \cB_p \cup \cC_p \cup \cD_p
$$
where
$$
\cA_p := 
\{A_i: \; |i_j| \,\leq \,{p-1 \over 2},\; j=1,2\}
$$
and similarly for $\cB_p,\cC_p,\cD_p$.\par
\par
Given a renormalized configuration of our multi--canonical  model and a 
block $A_i$ we denote
by $\a_i$ a generic original lattice gas  configuration compatible with the 
four renormalized configurations 
$\{n_j,\; j\in\cL_\ell:\; Q_{\ell} (j) \subset A_i \}$; 
in other words: 
$\a_i \in \otimes_{j:Q_{\ell}(j) \subset A_i } \O^{(n_j)}_j$ 
(recall that $\ell={L\over 2}$).
Similarly for $\b_i,\g_i, \d_i$. 
We simply denote by $\a,\b,\g,\d$ the configurations in
$\cA_p,\cB_p,\cC_p,\cD_p$, respectively.\par 
\par
Let us now quickly describe our strategy.
We want to transform the constrained system into a  polymer  system (see, for
instance, [GrK], [KP], [D3]) which, by
condition \equ(CC) will turn out to be in the
small activity region.  
To be more precise we shall  prove the following formula:
$$
Z^{(\ell)}_{\L_p,\underline {n}} =\bar Z^{(\ell)}_{\L_p,\underline {n}} 
\X ^{(\ell)}_{\L_p,\underline {n}} 
\Eq (2.15)
$$
where $\bar Z^{(\ell)}_{\L_p,\underline {n}} $ is factorized 
in the sense that it  has the form of a product  of
partition functions in suitable volumes not depending on $p$; 
the dependence
on $\underline {n}$ of the single factors is local. The partition function
$\bar Z^{(\ell)}_{\L_p,\underline {n}} $ describes the 
{\sl reference system} around which we perform a perturbative expansion.
On the other hand, $\X ^{(\ell)}_{\L_p,\underline {n}} $ is the partition
function of a gas of polymers; it has the form
$$
\X ^{(\ell)}_{\L_p,\underline {n}} = 1 
+ \sum _{k \geq 1}\sum _{R_1, \cdots, R_k} \prod _{j=1}^k
\z_{R_j} (\underline n)
\Eq (2.16)
$$
where the polymers $R_j$, that will be defined  below, are geometrical
local objects living on scale $L=2\ell$; the sum in \equ (2.16) is 
restricted to ``non--intersecting" polymers so that the unique
interaction between polymers is a pairwise hard core exclusion. 
Finally the activity $\z_{R_j} (\underline n)$ depends only on the
$n_i$'s with $i$ localized on the polymer. 
It is already clear from this preliminary
discussion that expression \equ (2.16) is well suited to compute
renormalized potential: in order to get good estimates of the
norm of the renormalized potential we shall need that the polymer
system described by $\X ^{(\ell)}_{\L_p,\underline {n}} $ is in the small
activity region.\par 
\par
To get expression \equ(2.15) we will perform a sequence of block decimations
like in [O], [OP].
We start by 
integrating over the $\d$--variables, then 
the $\g$--variables, the $\b$--variables and, finally,
the $\a$--variables. 
Using Condition \equ(CC) we will be able to prove that at each step of
decimation  
the system described by the surviving variables  is weakly coupled.\par 
\smallskip
We use the following notation for the interaction (which is defined
independently of the multi--canonical constraints) between two sets
$\L_1$ and $\L_2$: 
$$
W_{\L_1,\L_2}(\h_{\L_1}|\h_{\L_2}) := W(\h_{\L_1}|\h_{\L_2})= 
H_{\L_1\cup\L_2}(\h_{\L_1},\h_{\L_2}) \,-\,
H_{\L_1}(\h_{\L_1})\, -\, H_{\L_2}(\h_{\L_2})
\Eq (2.17)
$$
where 
$\h_{\L_1},\h_{\L_2}\in \{0,1\}^{\L_1}, \{0,1\}^{\L_2}$, respectively.
Recalling that $L$ is larger than the range of the interaction, we can write:
$$
\eqalign{
H_{\L_p}(\s_{\L_p}) 
=& \sum _{k_1:A_{k_1} \in \cA_p}H_{A_{k_1}} (\a_{k_1})
+
\sum_ {k_2:B_{k_2} \in \cB_p} H_{B_{k_2}} (\b_{k_2}) + W_{B_{k_2},
\cA_p} (\b_{k_2} |
\a) 
\cr
&+
\sum_ {k_3:C_{k_3} \in \cC_p} H_{C_{k_3}} (\g_{k_3}) + 
W_{C_{k_3},\cA_p\cup \cB_p
 } (\g_{k_3} | \b
,\a) 
\cr
&+ \sum_ {k_4:D_{k_4} \in \cD_p} H_{D_{k_4}} (\d_{k_4}) +
W_{D_{k_4},\cA_p\cup \cB_p \cup \cC_p}
(\d_{k_4} |\g, \b,\a) 
\cr
}
\Eq (2.18)
$$
Again the above decomposition of $H$ holds independently of the
constraints on the number of particles in the blocks; in \equ (2.18) we have only used
that
$L>r$ so that there is no direct interaction between blocks belonging to the same
sub--lattice.\par  

To simplify notation we  shall often omit from
$H, W$ the subscripts referring to the various domains; the symbols used for the
arguments of the functions $H$, $W$ should be sufficiently clarifying;
moreover we will also omit the explicit extensions of the sums (or
products) over $k_1,k_2,k_3,k_4$ as well as the one over $\a \in
\otimes_{i:Q_\ell(i) \subset \cA_p} \O^{(n_i)}$, and similarly for
$\b,\g,\d$.  We have:  
$$
\eqalign{
Z^{(\ell)}_{\L_p,\underline {n}}  
=&
\sum_{\a} \prod _{k_1}
\exp\left(H(\a_{k_1}) \right)
\sum_{\b} \prod _{k_2}
\exp\left(H(\b_{k_2}) + W(\b_{k_2}| \a) \right)
\cr
&\times
\sum_{\g} \prod _{k_3}
\exp\left(H(\g_{k_3}) + W(\g_{k_3}|\b,\a) \right)
\sum_{\d} \prod _{k_4}
\exp\left(H(\d_{k_4}) + W(\d_{k_4}|\g,\b,\a) \right)
\cr
}
\Eq (2.19)
$$
We first perform the sum over $\d$ variables;
using that the sums over different $\d_{k_4}$ are decoupled since the
size $L$ of the blocks is larger than the range of the interaction, we
get:  
$$
Z^{(\ell)}_{\L_p} =\sum_{\a} \dots\sum_{\b} \dots\sum_{\g} \dots\prod_{k_4} 
Z_{D_{k_4}}\left((\b,\g)^u,(\b,\g)^d,\a)\right)
\Eq (2.20)
$$
where by $Z_{D_{k_4}}\left((\b,\g)^u,(\b,\g)^d,\a)\right)$ we denote
the partition function in $D_{k_4}$ with boundary conditions
$(\b,\g)^u $ on the top (up) and $(\b,\g)^d$ on the bottom (down) of
$D_{k_4}$ (see Fig. 2).  More explicitly $(\b,\g)^u$ is given by the
restriction of $\b,\g$ to (simply called configuration in): 
$Q_{L}(2k_4 +e_2)\cup Q_{L}(2k_4 +e_2+e_1) \cup  Q_{L}(2k_4 +e_2 + 2 e_1)  
\equiv B_{k_4}\cup C_{k_4}  \cup  B_{k_4+e_1}$
whereas  $(\b,\g)^d$ is the configuration in 
$Q_{L}(2k_4 - e_2)\cup Q_{L}(2k_4 - e_2 + e_1) \cup  Q_{L}(2k_4 - e_2
+ 2 e_1)\equiv B_{k_4-e_2}\cup C_{k_4 -e_2} \cup  B_{k_4 + e_1 -e_2}$. 
Finally $\a$ in \equ (2.20) denotes the configuration in 
$Q_{L}(2k_4 )\cup Q_{L}(2k_4 + 2 e_1)\equiv A_{k_4}\cup A_{k_4+e_1}$.\par
\midinsert
\vskip 7 truecm\noindent
\includegraphics{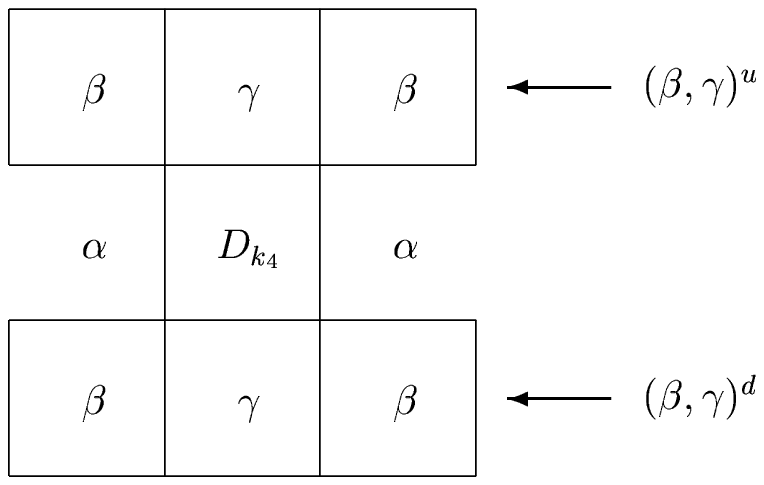} 
\vskip -1 truecm 
\par\noindent 
{\centerline {\bf Fig. 2}} 
\endinsert
\par
Notice that we are also presently omitting the explicit dependence on
$\underline{n}$ and $L$.  
Let $0$ denote a given reference configuration in  $\O^{(\underline
{n})}_{\L_p}$. We write 
$$
\eqalign{ 
Z_{D_{k_4}}  
\left( (\b,\g)^u, (\b,\g)^d, \a) \right)  
= & \left ( { Z_{D_{k_4}} \left((\b,\g)^u,(\b,\g)^d,\a) \right) 
Z_{D_{k_4}}\left((0)^u,(0)^d,\a)\right)   \over
Z_{D_{k_4}}\left((\b,\g)^u,(0)^d,\a)\right) 
Z_{D_{k_4}}\left((0)^u,(\b,\g)^d,\a)\right)} -1 +1 \right) 
\cr
& \times { Z_{D_{k_4}}\left((\b,\g)^u,(0)^d,\a)\right)
Z_{D_{k_4}}\left( (0)^u,(\b,\g)^d,\a) \right) \over
Z_{D_{k_4}}\left((0)^u,(0)^d,\a)\right) } 
\cr
}
\Eq (2.21)
$$
Where by $(0)^u,(0)^d,\a$ we mean the boundary condition on $D_{k_4}$
obtained from $(\b,\g)^u,(\b,\g)^d,\a$ by substituting $(\b,\g)$ with
$(0)$ both in the ``up" and ``down" blocks; similarly
$(\b,\g)^u,(0)^d,\a\;$; $(0)^u,(\b,\g)^d,\a$ denote the boundary
conditions on $D_{k_4}$ obtained from $(\b,\g)^u,(\b,\g)^d,\a$ by
substituting $(\b,\g)$ with $(0)$ only in the ``down", ``up"  blocks,
respectively.
We call the above operation ``splitting" of the partition function 
$Z_{D_{k_4}}\left((\b,\g)^u,(\b,\g)^d,\a)\right)$ in the
vertical $e_2$ direction. 
We set 
$$
\F_D^{(4)}(\a,\b,\g) :=
{ Z_{D_{k_4}}\left((\b,\g)^u,(\b,\g)^d,\a)\right)
Z_{D_{k_4}}\left((0)^u,(0)^d,\a)\right)   \over
Z_{D_{k_4}}\left((\b,\g)^u,(0)^d,\a)\right) 
Z_{D_{k_4}}\left((0)^u,(\b,\g)^d,\a)\right)} -1
\Eq (2.22)
$$
The quantity
$\F_D^{(4)}(\a,\b,\g) $ can be considered as an effective interaction
potential between $\a,\b,\g$ variables coming from decimation of the
$\d$ variables.  
In what follows we will exploit condition \equ(CC) above to deduce that
$\F_D^{(4)}(\a,\b,\g) $ and other similar quantities are uniformly small.\par

We can write:
$$
\eqalign{
Z^{(\ell)}_{\L_p} =
& \sum_{\a} \dots\sum_{\b} \dots\sum_{\g} 
\prod_{k_3} \exp\left(H(\g_{k_3}) + W(\g_{k_3}|\b,\a) \right)
\cr
&\times
Z_{D_{k_3 +e_2}}\left((0)^u,(\b, \g_{k_3})^d,\a \right)
Z_{D_{k_3 }}\left( (\b, \g_{k_3})^u, (0)^d,\a \right)
\cr
&\times
\prod_{k_4} 
\left[ Z_{D_{k_4}} \left( (0)^u,(0)^d,\a) \right) \right]^{-1}
\prod_{k_4} \left( 1 + \F^{(4)}_{D_{k_4}}(\a,\b,\g) \right)
\cr
}
\Eq (2.23)
$$
In \equ(2.23) above we 
associated to every $C_{k_3}$ block in
$\cC_p$ the two terms 
$Z_{D_{k_3 +e_2}}\left((0)^u,(\b, \g_{k_3})^d,\a \right)$, 
 $Z_{D_{k_3 }}\left( (\b, \g_{k_3})^u, (0)^d,\a \right)$
coming from the splitting of the original partition functions over the
volumes $D_{k_3 +e_2},D_{k_3}$, respectively.  
Notice that 
$$
\eqalign{
& \sum_{\g_{k_3}} 
 \exp\left(H(\g_{k_3}) + W(\g_{k_3}|\b,\a) \right)
Z_{D_{k_3 +e_2}}\left((0)^u, (\b, \g_{k_3})^d,,\a \right)
Z_{D_{k_3}}\left((\b, \g_{k_3})^u, (0)^d,\a \right)
\cr
&~~~
= Z_{\widetilde C_{k_3}} \left((0)^u,(0)^d,\a,\b  \right)
\equiv Z_{\widetilde C_{k_3}} \left((0)^u,(0)^d,(\a,\b)^l, (\a,\b)^r\right)
\cr
}
\Eq (2.24)
$$
Where: 
$$
\widetilde C_{k_3}:= C_{k_3}\cup D_{k_3}\cup D_{k_3 + e_2}
\Eq (2.25)
$$
is a $3L \times L$ rectangle $DCD$ centered at $C_{k_3}$ (see Fig 3)
and 
$Z_{\widetilde C_{k_3}} \left((0)^u,(0)^d,(\a,\b)^l, (\a,\b)^r\right)$
is the partition function in $\widetilde C_{k_3}$ with $(0)$ boundary
condition on the top and on the bottom; $(\a,\b)^l$ on the left and
$(\a,\b)^r$ on the right of $C_{k_3}$. 
Here by ``on the left" of $\widetilde C_{k_3}$ we mean ``in
$A_{k_3+e_2}\cup B_{k_3}\cup A_{k_3}$" and by ``on the right of
$\widetilde C_{k_3}$ we mean ``in $A_{k_3+e_2+e_1}\cup B_{k_3+e_1}\cup
A_{k_3+e_1}$"; see Fig. 3. 
In what follows we will continue to use ``on the top", ``on the
bottom", ``on the left" and ``on the right" for the boundary
conditions to a volume in a  similar sense.\par
\midinsert
\vskip 10 truecm\noindent
\includegraphics{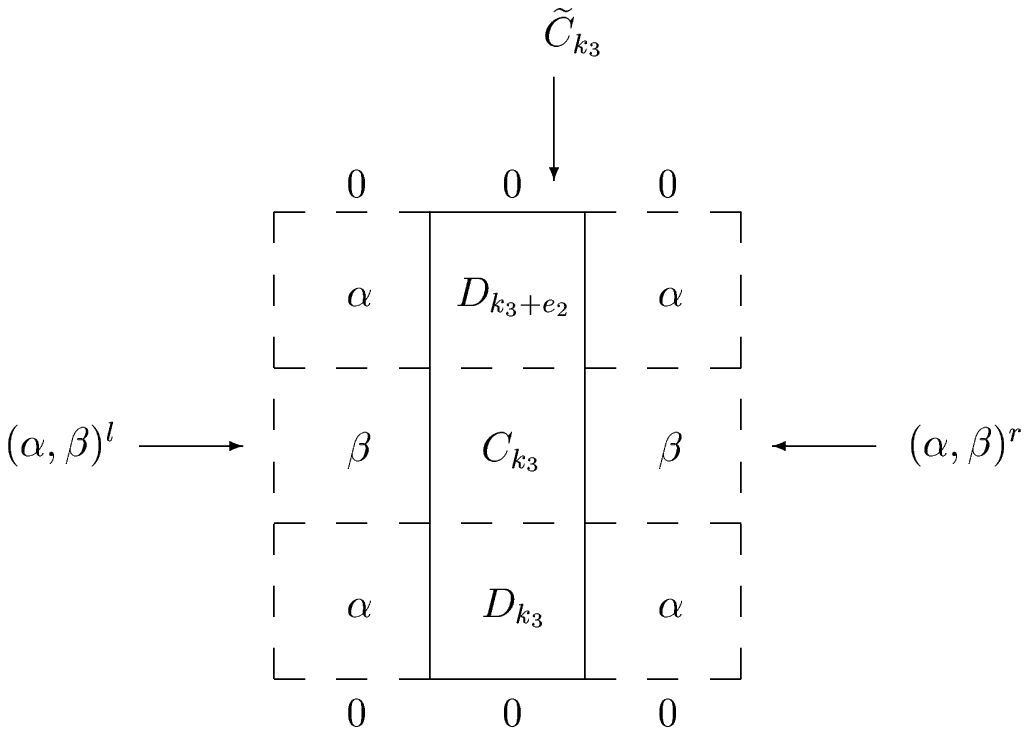} 
\vskip -1 truecm 
\par\noindent 
{\centerline {\bf Fig. 3}} 
\endinsert
\par
The operation described by equation \equ(2.24) above is called ``gluing" of
the partition functions  
$Z_{D_{k_3 +e_2}}\left((0)^u, (\b, \g_{k_3})^d,\a \right)$, 
$Z_{D_{k_3 }}\left((\b, \g_{k_3})^u, (0)^d,\a \right)$ on $C_{k_3}$
in the vertical $e_2$ direction.\par 

Now if in \equ (2.23) we multiply and divide
by 
$$
\prod _{k_3} 
Z_{\widetilde C_{k_3}} \left((0)^u,(0)^d,(\a,\b)^l, (\a,\b)^r  \right),
$$
we get:
$$
\eqalign{
Z^{(\ell)}_{\L_p,\underline {n}}  
= & \sum_{\a} \prod _{k1} \exp\left(H(\a_{k_1}) \right)
\sum_{\b} \prod _{k2}
\exp\left(H(\b_{k_2}) + W(\b_{k_2}| \a) \right)
 \prod _{k_3}
Z_{\widetilde C_{k_3}} \left((0)^u,(0)^d,(\a,\b)^l, (\a,\b)^r  \right)
\cr
&\sum_{\g} \m^{\a,\b}_3 (\g) 
\prod _{k_4}
\left ( 1 + \F^{(4)}_{D_{k_4}}(\a,\b,\g)\right)
\prod _{k4}\left [ Z_{D_{k_4}}\left ( (0)^u,(0)^d,(\a) \right )\right]^{-1}
\cr
}
\Eq (2.26)
$$
where $\m^{\a,\b}_3 (\g)$ is the product (Bernoulli) probability
measure on $\g$ parametrically depending on $\a,\b$ given by:
$$
\m^{\a,\b}_3 (\g) := \prod _{k_3} \m^{\a,\b}_{C_{k_3}} (\g_{k_3})
\Eq (2.27)
$$
where
$$
\eqalign{
\m^{\a,\b}_{ C_{k_3}} (\g_{k_3})
:= & {1\over Z_{\widetilde C_{k_3}}
\left((0)^u,(0)^d,(\a,\b)^l, (\a,\b)^r  \right)}
\exp \left ( H(\g_{k_3}) + W(\g_{k_3}|\b,\a) \right)
\cr
& \times 
 Z_{D_{k_3 +e_2}}\left((0)^u, (\b, \g_{k_3})^d,\a \right)
Z_{D_{k_3 }}\left((\b, \g_{k_3})^u, (0)^d,\a \right)
\cr
}
\Eq (2.28)
$$
At this moment we  operate again a ``splitting" but now we act on 
the partition function 
$Z_{\widetilde C_{k_3}}\left((0)^u,(0)^d,(\a,\b)^l, (\a,\b)^r
\right)$ in the horizontal $e_1$ direction; namely  we write: 
$$
\eqalign{
&
Z_{\widetilde C_{k_3}} \left((0)^u,(0)^d,(\a,\b)^l,(\a,\b)^r  \right)
\cr
&~~=
\left ( {Z_{\widetilde C_{k_3}} \left((0)^u,(0)^d,(\a,\b)^l,(\a,\b)^r  \right)
Z_{\widetilde C_{k_3}} \left((0)^u,(0)^d,(0)^l,(0)^r  \right)
\over
Z_{\widetilde C_{k_3}} \left((0)^u,(0)^d,(\a,\b)^l,(0)^r  \right)
Z_{\widetilde C_{k_3}} \left((0)^u,(0)^d,(0)^l,(\a,\b)^r  \right) 
}
-1+1
\right) 
\cr
&~~~\times
{Z_{\widetilde C_{k_3}} \left((0)^u,(0)^d,(\a,\b)^l,(0)^r  \right)
Z_{\widetilde C_{k_3}} \left((0)^u,(0)^d,(0)^l,(\a,\b)^r  \right)
\over 
Z_{\widetilde C_{k_3}} \left((0)^u,(0)^d,(0)^l,(0)^r  \right)}
\cr}
\Eq (2.29)
$$
We set 
$$
\left ( {Z_{\widetilde C_{k_3}} \left((0)^u,(0)^d,(\a,\b)^l,(\a,\b)^r  \right)
Z_{\widetilde C_{k_3}} \left((0)^u,(0)^d,(0)^l,(0)^r  \right)
\over
Z_{\widetilde C_{k_3}} \left((0)^u,(0)^d,(\a,\b)^l,(0)^r  \right)
Z_{\widetilde C_{k_3}} \left((0)^u,(0)^d,(0)^l,(\a,\b)^r  \right) 
}
-1
\right)=:\F_{C_{k_3}}^{(3)}(\a,\b)
\Eq (2.30)
$$

We remark that:
$$
\eqalign{
& \sum_{\b_{k_2}} 
 \exp\left(H(\b_{k_2}) + W(\b_{k_2}|\a) \right) \times
Z_{\widetilde C_{k_2 -e_1}}\left((0)^u,(0)^d,(0)^l,(\a,\b_{k_2})^r  \right) 
\cr
&~~Z_{\widetilde C_{k_2 }}\left((0)^u,(0)^d,(\a,\b_{k_2})^l,(0)^r  \right)
= Z_{\widetilde B_{k_2}} \left((0),\a  \right)
\cr
}
\Eq (2.31)
$$
where 
$\widetilde B_{k_2}$ is the set, centered at $B_{k_3}$, having the
shape of a capital H given by:
$$
\widetilde B_{k_2}:=  B_{k_2}\cup C_{k_2}\cup C_{k_2-e_1}
\cup D_{k_2}\cup D_{k_2-e_1}\cup D_{k_2-e_1+e_2}\cup D_{k_2+e_2}
\Eq (2.32)
$$
see Fig 4.
The above operation, described in \equ (2.31) above, is a ``gluing"
of the partition functions 
$Z_{\widetilde C_{k_2 -e_1}}\left((0)^u,(0)^d,(0)^l,(\a,\b_{k_2})^r
\right)$,   
$Z_{\widetilde C_{k_2 +e_1}}\left((0)^u,(0)^d,(\a,\b_{k_2})^l,(0)^r  \right)$
on $B_{k_2}$ in the $e_1$ direction.
\midinsert
\vskip 9 truecm\noindent
\includegraphics{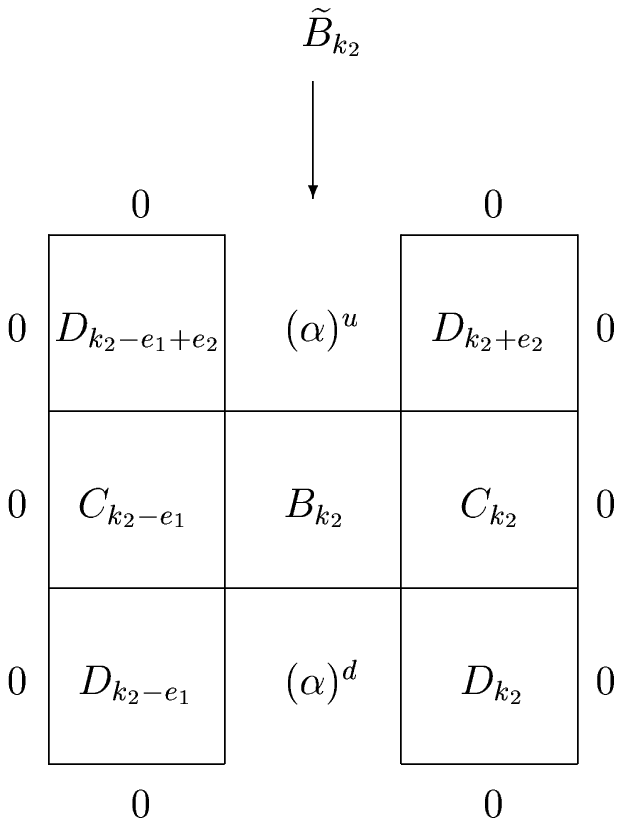} 
\par\noindent 
{\centerline {\bf Fig. 4}} 
\endinsert
\par
The boundary condition on $\widetilde B_{k_2}$ in the partition
function $Z_{\widetilde B_{k_3}} \left((0),\a  \right)$ are $0$
everywhere except for the $A$--blocks $A_{k_2+e_2}$, $A_{k_2}$
touching on the top and on the bottom, respectively, the block $
B_{k_3}$. 
We 
write:
$$ 
Z_{\widetilde B_{k_2}} \left((0),\a  \right)=:
Z_{\widetilde B_{k_2}} \left((0),(\a )^u, (\a )^d\right)
\Eq (2.33)
$$
with $(\a )^u$, $(\a )^d$,  given, respectively, by the restriction
of $\a$ to $A_{k_2+e_2}$, $A_{k_2}$.\par

Similarly for the term $Z_{D_{k_4}}\left((0)^u,(0)^d,\a)\right)$
appearing (at the power $-1$) in \equ (2.23) we can write
$$
Z_{D_{k_4}}\left((0)^u,(0)^d,\a)\right) =
Z_{D_{k_4}}\left((0)^u,(0)^d,(\a)^l, (\a)^r)\right) \equiv 
Z_{D_{k_4}}\left((0),(\a)^l, (\a)^r)\right)
\Eq (2.34)
$$
where by $(0),(\a)^l, (\a)^r$ we mean the boundary conditions, outside
$D_{k_4}$ given by $0$ everywhere except for the two blocks $A_{k_4}$,
$A_{k_4+e_1}$, contiguous to $D_{k_4}$; $(\a)^l, (\a)^r$ are the
restrictions of $\a$ to $A_{k_4}$, $A_{k_4+e_1}$, respectively.\par

Now we perform a ``splitting" in the $e_1$ direction of the quantity
$\left [Z_{D_{k_4}}\left((0),(\a)^l,
(\a)^r)\right)\right]^{-1}$; namely we write: 
$$
\eqalign{
\left [Z_{D_{k_4}}\left((0),(\a)^l, (\a)^r)\right)\right]^{-1} 
=&
\left ( {Z_{D_{k_4}}\left((0),(\a)^l, (0)^r)\right)Z_{D_{k_4}}\left((0),(0)^l,
(\a)^r)\right)
 \over
 Z_{D_{k_4}}\left((0),(\a)^l, (\a)^r)\right)Z_{D_{k_4}}\left((0),(0)^l,
(0)^r)\right)} -1 +1
\right)
\cr
&\times
{Z_{D_{k_4}}\left((0),(0)^l, (0)^r)\right)
\over Z_{D_{k_4}}\left((0),(\a)^l, (0)^r)\right)Z_{D_{k_4}}\left((0),(0)^l,
(\a)^r)\right)}
\cr
}
\Eq (2.35)
$$
We set:
$$
\left ( {Z_{D_{k_4}}\left((0),(\a)^l, (0)^r)\right)Z_{D_{k_4}}\left((0),(0)^l,
(\a)^r)\right)
 \over
 Z_{D_{k_4}}\left((0),(\a)^l, (\a)^r)\right)Z_{D_{k_4}}\left((0),(0)^l,
(0)^r)\right)} -1 
\right)=: \Psi^{(4)} _{D_{k_4}} (\a)
\Eq (2.36)
$$

We introduce the product probability measure $\m^{\a}_2(\b)$ on $\b$,
parametrically dependent on $\a$, given by:
$$
\m^{\a}_2(\b) := \prod _{k_2} \m^{\a}_{ B_{k_2}}(\b_{k_2})
\Eq (2.37)
$$
where
$$
\eqalign{
\m^{\a}_{ B_{k_2}} (\b_{k_2})
:= & {1\over 
Z_{\widetilde B_{k_2}} \left((0),(\a)^u, (\a )^d\right)}
\exp 
\left ( H(\b_{k_2}) + W(\b_{k_2}|\a) \right) 
\cr
&\times
Z_{\widetilde C_{k_2-e_1 }} \left((0)^u,(0)^d,(0)^l (\a,\b_{k_2})^r, \right)
Z_{\widetilde C_{k_2}} \left((0)^u,(0)^d,(\a,\b_{k_2})^l,(0)^r  \right)
\cr
}
\Eq (2.38)
$$

Now we proceed similarly to the step leading to \equ(2.26).
We multiply and divide the expression on the r.h.s. of \equ (2.26) by 
$$
\prod _{k_2} Z_{\widetilde B_{k_2}} \left((0),(\a )^u, (\a )^d\right);
$$
by inserting in the r.h.s. of \equ (2.26) the expression  given by
\equ(2.38) and after operating the splitting described in \equ(2.29),
the gluing described in \equ(2.31) and the splitting described in
\equ(2.35), we get:  
$$
\eqalign{
Z^{(\ell)}_{\L_p,\underline {n}}  = 
& \sum_{\a} \prod _{k_1}
\exp\left(H(\a_{k_1}) \right)
\left [Z_{D_{k_1+e_1}}\left((0),(\a_{k_1})^l, (0)^r)\right)Z_{D_{k_1}}\left((0),(0)^l,
(\a_{k_1})^r)\right)\right]^{-1}
\cr
&\times
\prod _{k_2}
Z_{\widetilde B_{k_2}} \left((0),(\a )^u, (\a )^d  \right)
\sum_{\b} \m^{\a}_2(\b)
\prod _{k_3}
\left [ Z_{\widetilde C_{k_3}} \left((0)\right) \right]^{-1}
\prod _{k_3}
\left ( 1 +\F^{(3)}_{C_{k_3}}\right )
\sum_{\g} \m^{\a,\b}_3 (\g) 
\cr
&\times
\prod _{k_4}
\left ( 1 + \F^{(4)}_{D_{k_4}}(\a,\b,\g)\right)
\prod _{k_4}
\left ( 1 + \Psi^{(4)}_{D_{k_4}}(\a)\right)
\prod _{k_4}
\left [ Z_{D_{k_4}}\left ( (0)\right)\right ]^{-1}
\cr
}
\Eq (2.39)
$$
where we used the shorthand notation 
$Z_{\widetilde C_{k_3}} \left((0)\right)$ for 
$Z_{\widetilde C_{k_3}} \left((0)^u,(0)^d,(0)^l, (0)^r  \right)$ and  
$Z_{D_{k_4}}\left ( (0) )\right)$ for 
$Z_{D_{k_4}}\left ( (0)^u,(0)^d,(0)^l, (0)^r \right )$.\par

\bigskip
Now we perform,  on the partition function 
$Z_{\widetilde B_{k_2}} \left((0),(\a )^u, (\a )^d\right)$,  
a splitting a bit different with
respect to the previous ones.
Let $F_{k_2}$ be the horizontal $L\times 3L$ rectangle $CBC$ contained in 
$\widetilde B_{k_2}$:
$$
F_{k_2}=  B_{k_2}\cup C_{k_2}\cup C_{k_2 - e_1}
\Eq (2.40)
$$
We can write:
$$
\eqalign{
Z_{\widetilde B_{k_2}} \left((0),(\a )^u, (\a )^d\right)
= &
\sum_{(\d)^u,(\d)^d} \exp \left[H((\d)^u) +  (H((\d)^d) 
+ W((\a)^u|(\d)^u)) +  W((\a)^d|(\d)^d))    \right] 
\cr
&\times
\left ( {Z_{ F_{k_2}} \left( (0),(\a,\d)^u,(\a,\d)^d \right)
Z_{ F_{k_2}} \left( (0),(0)^u,(0)^d \right)
\over
Z_{ F_{k_2}} \left( (0),(\a,\d)^u,(0)^d \right)
Z_{ F_{k_2}} \left( (0),(0)^u,(\a,\d)^d \right) 
}
-1+1
\right) 
\cr
&\times
{Z_{ F_{k_2}} \left( (0),(\a,\d)^u,(0)^d \right)
Z_{ F_{k_2}} \left( (0),(0)^u,(\a,\d)^d \right)
\over 
Z_{ F_{k_2}} \left( (0),(0)^u,(0)^d \right)}
\cr
}
\Eq (2.41)
$$
Where, for a generic
$\d \in \otimes_{j:Q_{\ell}(j) \subset \cD_p }\O^{(n_j)} $
we denote  by $(\d)^u$  the restriction of $\d$ to $D_{k_2-e_1 +e_2}
\cup D_{k_2 +e_2}$ whereas we denote  by $(\d)^d$ the restriction of
$\d$ to $D_{k_2-e_1 } \cup D_{k_2}$; by $(0),(\a,\d)^u,(\a,\d)^d$ we
mean boundary conditions on $F_{k_2}$ given by $(\a,\d)^u$ on the top,
$(\a,\d)^d$ on the bottom and $0$ elsewhere (see Fig. 5).
\midinsert
\vskip 8 truecm\noindent
\includegraphics{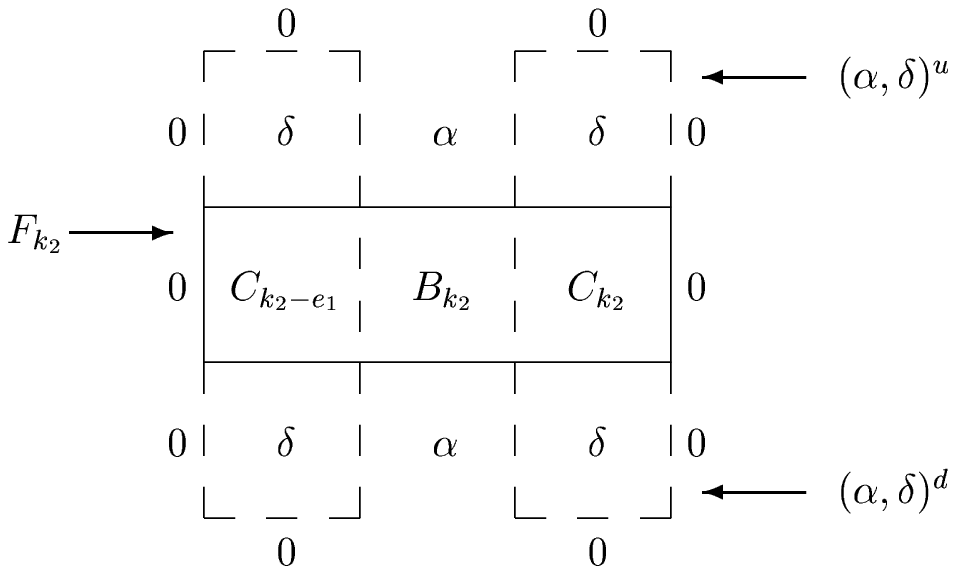} 
\vskip -0.5 truecm 
\par\noindent 
{\centerline {\bf Fig. 5}} 
\endinsert
\par
Let 
$F^{(u)}_{k_2}$, $F^{(d)}_{k_2}$ be the ``horseshoe" shaped domains given by:
$$
\eqalign{
&
F^{(u)}_{k_2} :=
B_{k_2}\cup C_{k_2}\cup C_{k_2-e_1}
\cup D_{k_2-e_1+e_2}\cup D_{k_2+e_2},
\cr
& F^{(d)}_{k_2}:=B_{k_2}\cup C_{k_2}\cup C_{k_2-e_1}
\cup D_{k_2}\cup D_{k_2-e_1}
}
\Eq(2.43)
$$
(see Fig (6)). From \equ (2.41) we easily get:
$$
Z_{\widetilde B_{k_2}} \left((0),(\a )^u, (\a )^d\right)
={Z_{F^{(u)}_{k_2}} \left((0),(\a )^u\right)
Z_{F^{(d)}_{k_2}} \left((0),(\a )^d\right)
 \over
Z_{F_{k_2}} \left((0),(0)^u, (0)^d\right) }
\left (  1 +  \F^{(2)} _{B_{k_2}}(\a)\right)
\Eq (2.44)
$$
\midinsert
\vskip 9 truecm\noindent
\includegraphics{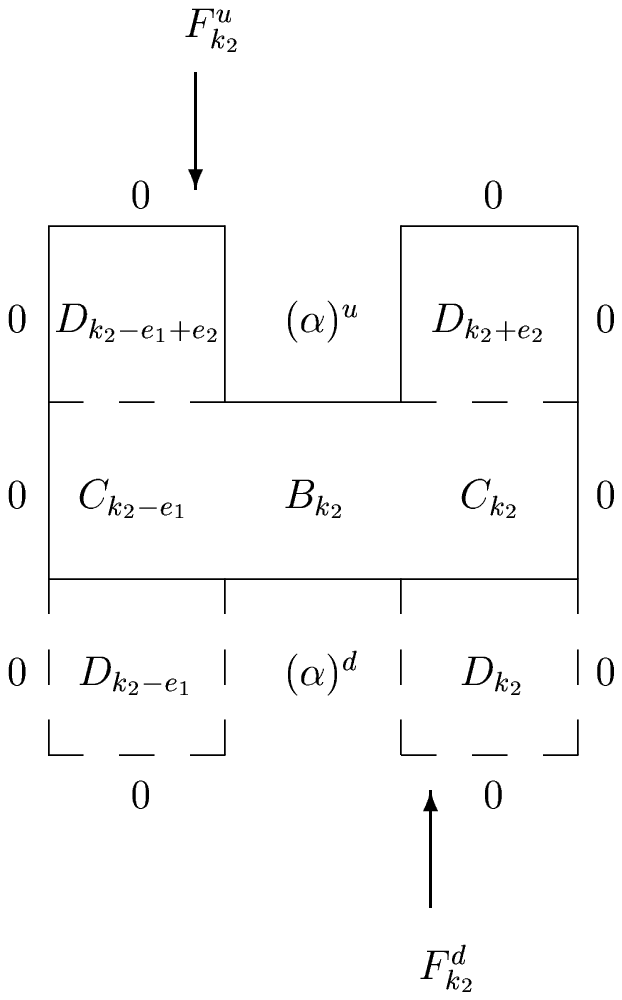} 
\vskip 2 truecm 
\par\noindent 
{\centerline {\bf Fig. 6}} 
\endinsert
\par
where  $Z_{F^{(u)}_{k_2}} \left((0),(\a )^u\right)$ is the partition function on the domain $F^{(u)}_{k_2}$ with boundary conditions $0$ everywhere except for
$A_{k_2+e_2}$ where they take the value $(\a )^u ( \;\equiv $ the
restriction of $\a$ to  $A_{k_2+e_2}$); similarly $Z_{F^{(d)}_{k_2}}
\left((0),(\a )^d\right)$ is the partition function on the domain
$F^{(d)}_{k_2}$ with boundary conditions $0$ everywhere except for
$A_{k_2}$ where they take the value $(\a )^d ( \;\equiv $ the
restriction of $\a$ to  $A_{k_2}$); finally $ \F^{(2)} _{B_{k_2}}(\a)$
is defined as: 
$$
\F^{(2)} _{B_{k_2}}(\a) :=
 \sum_{(\d)^u,(\d)^d} \tilde \m^{\a}_{k_2}((\d)^u,(\d)^d)
\left ( {Z_{ F_{k_2}} \left( (0),(\a,\d)^u,(\a,\d)^d \right)
Z_{ F_{k_2}} \left( (0),(0)^u,(0)^d \right)
\over
Z_{ F_{k_2}} \left( (0),(\a,\d)^u,(0)^d \right)
Z_{ F_{k_2}} \left( (0),(0)^u,(\a,\d)^d \right) 
}
-1
\right)
\Eq (2.45)
$$
where $\tilde \m^{\a}_{k_2}((\d)^u,(\d)^d)$ is a probability measure on 
$\otimes_{i:Q_l(i) \subset \cD_p \cap \widetilde B_{k_2}} \O^{(n_i)}$
parametrically dependent on 
$\a_{k_2 +e_2}, \a_{k_2 }$
given by:
$$
\eqalign{
\tilde \m^{\a}_{k_2}((\d)^u,(\d)^d) = &
\exp \left(H((\d)^u) +  (H((\d)^d) 
+ W((\a)^u|(\d)^u)) +  W((\a)^d|(\d)^d))    \right)
\cr
&\times{
Z_{ F_{k_2}} \left( (0),(\a,\d)^u,(0)^d \right)
Z_{ F_{k_2}} \left( (0),(0)^u,(\a,\d)^d \right) 
\over
Z_{F^{(u)}_{k_2}} \left((0),(\a )^u\right)
Z_{F^{(d)}_{k_2}} \left((0),(\a )^d\right) }
\cr
}
\Eq (2.46)
$$
Indeed $\tilde \m^{\a}_{k_2}((\d)^u,(\d)^d)$ has the form of a product
measure over the ``up" and ``down" variables but in \equ (2.45) we are
averaging, with respect to $\tilde \m^{\a}_{k_2}$, a function which
couples these variables so that the result is a 
$\F^{(2)}_{B_{k_2}}(\a)$ which is a  non--factorized function of
$(\a)^u,(\a)^d$. 

By inserting \equ (2.44) into \equ (2.39) we get
$$
\eqalign{
Z^{(\ell)}_{\L_p,\underline {n}}  = &
\sum_{\a} \prod _{k_1}
\exp
\left( H(\a_{k_1})  \right)
\left [
Z_{D_{k_1}}
\left(
(0),(\a_{k_1})^l, (0)^r)
\right)
Z_{D_{k_1}-e_1}\left((0),(0)^l,
(\a_{k_1})^r)\right)
\right ]^{-1}
\cr
&\times
Z_{F^{(u)}_{k_1-e_1}} \left((0),(\a_{k_1})^u\right)
Z_{F^{(d)}_{k_1}} \left((0),(\a_{k_1} )^d\right)
\prod _{k_2}\left ( 1 +\F^{(2)} _{B_{k_2}}(\a) \right)
\left[ Z_{ F_{k_2}} (0)\right ]^{-1}
\cr
&\times
\sum_{\b} \m^{\a}_2(\b)
\prod _{k_3}
\left [ Z_{\widetilde C_{k_3}} \left((0)\right) \right]^{-1}
\prod _{k_3}
\left ( 1 +\F^{(3)}_{C_{k_3}}(\a,\b)\right )
\cr
&\times
\sum_{\g} \m^{\a,\b}_3 (\g) 
\prod _{k_4}
\left ( 1 + \F^{(4)}_{D_{k_4}}(\a,\b,\g)\right)
\prod _{k_4}
\left ( 1 + \Psi^{(4)}_{D_{k_4}}(\a)\right)
\prod _{k_4}
 Z_{D_{k_4}}\left ( (0)\right)
\cr
}
\Eq (2.47)
$$
where we have used the shorthand forms 
$Z_{D_{k_1}}\left((0),(\a_{k_1})^l)\right)$, respectively 
$Z_{D_{k_1}-e_1}\left((0), (\a_{k_1})^r)\right)$, for 
$Z_{D_{k_1}}\left((0),(\a_{k_1})^l, (0)^r)\right)$,
$Z_{D_{k_1}-e_1}\left((0),(0)^l,(\a_{k_1})^r)\right)$ and  
$Z_{ F_{k_2}} (0)$ for $Z_{ F_{k_2}} \left( (0),(0)^u,(0)^d \right)$.\par

We notice  that if in \equ (2.47) above we neglect all the ``small
quantities" $\F$ and $\Psi$ and we use that  $\m^{\a}_2(\b)$ and
$\m^{\a,\b}_3 (\g)$ are normalized measures, then,  by performing the
sum over the $\g, \b$ variables, we get a factorized partition
function describing a system of independent $\a$ variables. So we
substantially have already reached our goal; we want now to manipulate
a little bit these factorized terms (the product over $k_1$) in order
to get a simpler expression with a more transparent physical meaning.\par 

We use the notation $\widetilde A_{k_1}$ to denote the $3L\times 3L$
cube centered at the block $A_{k_1}$:
$$
\widetilde A_{k_1} := Q_{3L}(2k_1 ), \;\;\;\;\;\;\; k_1 \, \in \, \cL_L
\Eq (2.48')
$$

Let $G_{k_1}$ denote the annulus obtained from $\widetilde A_{k_1} $
by removing the block $A_{k_1}$ itself:
$$
G_{k_1} :=  \widetilde A_{k_1} \setminus A_{k_1} \equiv
 B_{k_1}\cup B_{k_1-e_2}\cup C_{k_1}\cup C_{k_1-e_1}
\cup C_{k_1-e_1-e_2}\cup C_{k_1-e_2}
\cup D_{k_1}\cup D_{k_1-e_1}
\Eq (2.48)
$$
 
We denote by $Z_{G_{k1}} ((0),\a_{k_1})$ the partition function on
$G_{k_1}$ with boundary conditions $\a_{k_1}$ on the ``hole" $A_{k_1}$
and $0$ elsewhere. 
Moreover let $Z_{D_{k_1 - e_1}\cup D _{k_1}}\left(
(0),\a_{k_1},(\b\g)^u,(\b\g)^d\right)$ denote the partition function
on the (non--connected) set $D_{k_1 - e_1}\cup D _{k_1}$ with boundary
conditions $\a_{k_1}$ on $A_{k_1}$, $(\b\g)^u$  on the up part of
$G_{k_1}\setminus (D_{k_1 - e_1}\cup D _{k_1})$ (namely in $C_{k_1
-e_1}\cup B_{k_1}\cup C_{k_1 }$), $(\b\g)^d$ in the down part  $C_{k_1
-e_1 -e_2}\cup B_{k_1-e_2}\cup C_{k_1-e_2 }$ and $0$ elsewhere. (see
Fig (7)). 
\midinsert
\vskip 8 truecm\noindent
\includegraphics{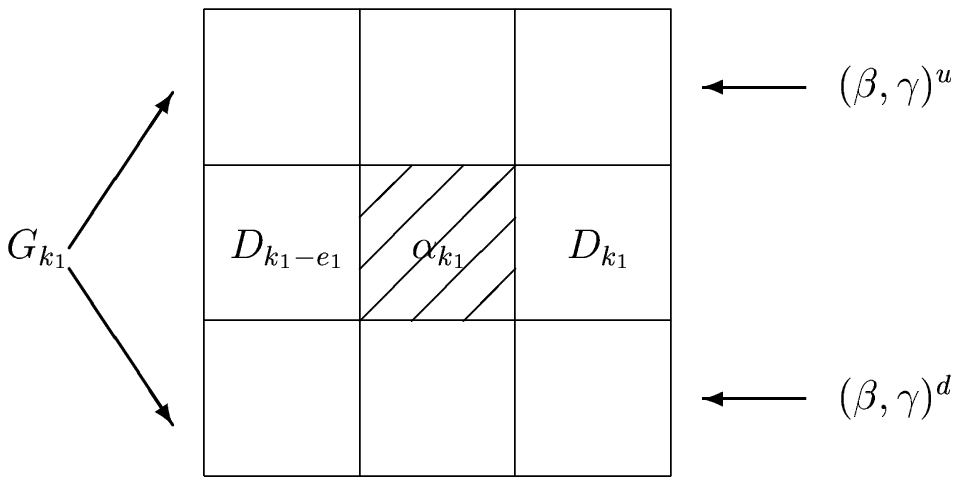} 
\vskip -1 truecm
\par\noindent 
{\centerline {\bf Fig. 7}} 
\endinsert
\par
Indeed we have  the following factorization:
$$
Z_{D_{k_1 - e_1}\cup D _{k_1}} \! \left( (0),
\a_{k_1},(\b\g)^u,(\b\g)^d\right)
=
Z_{D_{k_1 - e_1}} \! \left( (0), \a_{k_1},(\b\g)^u,(\b\g)^d\right)
Z_{ D _{k_1}}\! \left( (0), \a_{k_1},(\b\g)^u,(\b\g)^d\right)
\Eq (2.49)
$$
We have:
$$
Z_{G_{k1}} ((0),\a_{k_1})=
\sum _{(\b\g)^u,(\b\g)^d} \exp \left (
H((\b\g)^u) +H((\b\g)^d)
 \right)
Z_{D_{k_1 - e_1}\cup D _{k_1}}\left( (0), \a_{k_1},(\b\g)^u,(\b\g)^d\right)
\Eq (2.50)
$$
We can write:
$$
\eqalign{
Z_{G_{k1}} ((0),\a_{k_1})
& =
\sum _{(\b\g)^u,(\b\g)^d} \exp \left (
H((\b\g)^u) +H((\b\g)^d) \right)
\cr
&~\times
\left [
{Z_{D_{k_1 - e_1}\cup D _{k_1}}\left( (0), \a_{k_1},(\b\g)^u,(\b\g)^d\right)
Z_{D_{k_1 - e_1}\cup D _{k_1}}\left( (0), \a_{k_1},(0)^u,(0)^d\right)
\over
Z_{D_{k_1 - e_1}\cup D _{k_1}}\left( (0), \a_{k_1},(\b\g)^u,(0)^d\right)
Z_{D_{k_1 - e_1}\cup D _{k_1}}\left( (0), \a_{k_1},(0)^u,(\b\g)^d\right)
} -1+1
 \right]
\cr
&~\times
{Z_{D_{k_1 - e_1}\cup D _{k_1}}\left( (0), \a_{k_1},(\b\g)^u,(0)^d\right)
Z_{D_{k_1 - e_1}\cup D _{k_1}}\left( (0), \a_{k_1},(0)^u,\b\g)^d\right)
\over
Z_{D_{k_1 - e_1}\cup D _{k_1}}\left( (0), \a_{k_1},(o)^u,(0)^d\right)
} 
\cr
&=
{ Z_{F^{(u)}_{k_2}} \left((0),( \a_{k_1} )^u\right)
Z_{F^{(d)}_{k_2}} \left((0),( \a_{k_1})^d\right)
\over
Z_{D_{k_1 - e_1}\cup D_{k_1}}\left( (0), \a_{k_1},(0)^u,(0)^d\right)
} \left( 1 + \F^{(1)}_{A_{k_1}}(\a_{k_1}) \right)
\cr
}
\Eq (2.51)
$$
where
$$
\eqalign{
\F^{(1)}_{A_{k_1}}(\a_{k_1}) := &
\sum _{(\b\g)^u,(\b\g)^d}
\tilde \m_{k_1}^{(\a_{k_1})}
((\b\g)^u,(\b\g)^d) 
\cr
&\times
\left [
{Z_{D_{k_1 - e_1}\cup D _{k_1}}\left( (0), \a_{k_1},(\b\g)^u,(\b\g)^d\right)
Z_{D_{k_1 - e_1}\cup D _{k_1}}\left( (0), \a_{k_1},(0)^u,(0)^d\right)
\over
Z_{D_{k_1 - e_1}\cup D _{k_1}}\left( (0), \a_{k_1},(\b\g)^u,(0)^d\right)
Z_{D_{k_1 - e_1}\cup D _{k_1}}\left( (0), \a_{k_1},(0)^u,(\b\g)^d\right)
} -1
\right]
\cr
}
\Eq (2.52)
$$
and 
$$
\eqalign{
& \tilde \m_{k_1}^{(\a_{k_1})}
((\b\g)^u,(\b\g)^d) := 
\exp \left ( H((\b\g)^u) +H((\b\g)^d)\right)
\cr
&~~~~~\times
 {Z_{D_{k_1 - e_1}\cup D _{k_1}}\left( (0), \a_{k_1},(\b\g)^u,(0)^d\right)
Z_{D_{k_1 - e_1}\cup D _{k_1}}\left( (0), \a_{k_1},(0)^u,(\b\g)^d\right)
\over
Z_{F^{(u)}_{k_2}} \left((0),( \a_{k_1} )^u\right)
Z_{F^{(d)}_{k_2}} \left((0),( \a_{k_1})^d \right)
}
\cr
}
\Eq (2.53)
$$

We write:
$$
\Psi^{(1)}_{A_{k_1}}(\a_{k_1}):= (1+\F^{(1)}_{A_{k_1}}(\a_{k_1}))^{-1} -1
\Eq (2.54)
$$
 From \equ (2.51),\equ (2.52),\equ (2.53),\equ (2.54) we get 
$$
{Z_{F^{(u)}_{k_2}} \left((0),( \a_{k_1} )^u\right)
Z_{F^{(d)}_{k_2}} \left((0),( \a_{k_1})^d\right)
\over
Z_{D_{k_1 - e_1}\cup D _{k_1}}\left( (0), \a_{k_1},(0)^u,(0)^d\right)
}
=
Z_{G_{k_1}} ((0),\a_{k_1})\left (
1 + \Psi^{(1)}_{A_{k_1}}(\a_{k_1})
\right)
\Eq (2.54')
$$

We define the Bernoulli probability measure $\m_1 (\a)$ as
$$
\m_1 (\a) :=\prod _{k_1} \m_{A_{k_1}} (\a_{k_1})
\Eq (2.55)
$$
where
$$
\m_{A_{k_1}} (\a_{k_1}):=
{1\over
Z_{\widetilde A_{k_1}}((0))}
\exp ( H_{A_{k_1}} ( \a_{k_1})
Z_{G_{k_1}} ((0),\a_{k_1})
\Eq (2.56)
$$
in which by $Z_{\widetilde A_{k_1}}((0))$ we denote the partition function in 
$\widetilde A_{k_1}$ with $0$ boundary conditions.\par

In conclusion, from \equ (2.47), \equ (2.48'), \equ (2.54'), \equ
(2.55) and \equ (2.56) we get : 
$$
\eqalign{
Z^{(\ell)}_{\L_p,\underline {n}}  = &
\prod _{k_1} 
Z_{\widetilde A_{k_1}}((0)) 
\prod _{k_2}
\left[ Z_{ F_{k_2}} (0)\right ]^{-1}
\prod _{k_3}
\left [ Z_{\widetilde C_{k_3}} \left((0)\right) \right]^{-1}
\prod _{k_4}
 Z_{D_{k_4}}\left ( (0)\right)
\cr
&\times
\sum_{\a} \m_1(\a)
\prod _{k_1}
\left (
1 + \Psi^{(1)}_{A_{k_1}}(\a_{k_1})
\right)
\prod _{k_2}
\left ( 1 +\F^{(2)} _{B_{k_2}}(\a) \right)
\prod _{k_4}
\left ( 1 + \Psi^{(4)}_{D_{k_4}}(\a)\right)
\cr
&\times
\sum_{\b} \m^{\a}_{B_{k_2}}(\b_{k_2})
\prod _{k_3}
\left ( 1 +\F^{(3)}_{C_{k_3}}(\a,\b)\right )
\sum_{\g} \m^{\a,\b}_3 (\g)
\prod _{k_4}
\left ( 1 + \F^{(4)}_{D_{k_4}}(\a,\b,\g)\right) 
\cr
}
\Eq (2.57)
$$
We  write 
$$
Z^{(\ell)}_{\L_p,\underline {n}}  =
\bar
Z^{(\ell)}_{\L_p,\underline {n}}\X^{(\ell)}_{\L_p,\underline {n}}
\Eq (2.58)
$$
with
$$
\bar Z^{(\ell)}_{\L_p,\underline {n}}
:= \prod _{k_1} 
Z_{\widetilde A_{k_1}}((0)) 
\prod _{k_2}
\left[ Z_{ F_{k_2}} (0)\right ]^{-1}
\prod _{k_3}
\left [ Z_{\widetilde C_{k_3}} \left((0)\right) \right]^{-1}
\prod _{k_4}
 Z_{D_{k_4}}\left ( (0)\right)
\Eq (2.59)
$$
and
$$
\eqalign{
\X^{(\ell)}_{\L_p,\underline {n}} 
= & \sum_{\a} \m_1(\a)
\prod _{k_1}
\left (
1 + \Psi^{(1)}_{A_{k_1}}(\a_{k_1})
\right)
\prod _{k_2}
\left ( 1 +\F^{(2)} _{B_{k_2}}(\a) \right)
\prod _{k_4}
\left ( 1 + \Psi^{(4)}_{D_{k_4}}(\a)\right)
\cr
&\times
\sum_{\b} \m^{\a}_{B_{k_2}}(\b_{k_2})
\prod _{k_3}
\left ( 1 +\F^{(3)}_{C_{k_3}}(\a,\b)\right )
\sum_{\g} \m^{\a,\b}_3 (\g)
\prod _{k_4}
\left ( 1 + \F^{(4)}_{D_{k_4}}(\a,\b,\g)\right) 
\cr
}
\Eq (2.60)
$$

We are now ready to express $\X^{(\ell)}_{\L_p,\underline {n}}$ as the
partition function of a gas of polymers whose only interaction is a
hard core exclusion.\par 

We have to analyze the various interaction terms (the $\F$'s and
$\Psi$'s ) appearing in \equ (2.60). 
We see from \equ (2.22) that the term $\F^{(4)}_{D_{k_4}}(\a,\b,\g)$,
involving the $\a,\b,\g$ variables in the annulus $Q_{3L}(2k_4 + e_1)\setminus
D_{k_4}$, corresponds to an ``eight body" interaction among the $A,B,C$ blocks
adjacent to $D_{k_4}$; we see from \equ(2.30) that $\F^{(3)}_{C_{k_3}}(\a,\b)$ is a
six body interaction involving the $A$ and $B$ blocks adjacent to  $C_{k_3}$;
$\F^{(2)}_{B_{k_2}}(\a)$,
$\Psi^{(4)}_{D_{k_4}}(\a)$ are two body terms involving the pair of
$A$ blocks contiguous to $B_{k_2}$ , $D_{k_4}$, respectively. 
Finally $\Psi^{(1)}_{A_{k_1}}(\a_{k_1})$ is  just a one body term.\par

Looking at \equ (2.27), \equ (2.28) we can say  that
$\F^{(4)}_{D_{k_4}}(\a,\b,\g)$ extends its action to all $A$ and $B$
blocks adjacent to the $C$ blocks in $Q_{3L}(2k_4 +e_1 )$ (see
Fig. 8), becoming a ``twelve body" interaction.  
Indeed we have to average $\F^{(4)}_{D_{k_4}}(\a,\b,\g)$ with respect
to the product of the measures $\m^{\a,\b}_{ C_{k_3}} (\g_{k_3})$
which are parametrically dependent on the $\a, \b$ variables adjacent
to $C_{k_3}$ . On the other hand, looking at \equ(2.38), it is easily
seen that we do not have to extend any more the region of influence of  
$\F^{(4)}_{D_{k_4}}(\a,\b,\g)$ because of the parametric dependence on
$\a$ of $\m^{\a}_{ B_{k_2}} (\b_{k_2})$. 
Moreover, still looking at \equ (2.38), we easily see that also the term 
$\F^{(3)}_{C_{k_3}}(\a,\b)$ does not extend at all its influence.
Of course $\F^{(2)}_{B_{k_2}}(\a)$, $\Psi^{(4)}_{D_{k_4}}(\a)$
$\Psi^{(1)}_{A_{k_1}}(\a_{k_1})$ do not extend, as well, their action.
\midinsert
\vskip 10 truecm\noindent
\includegraphics{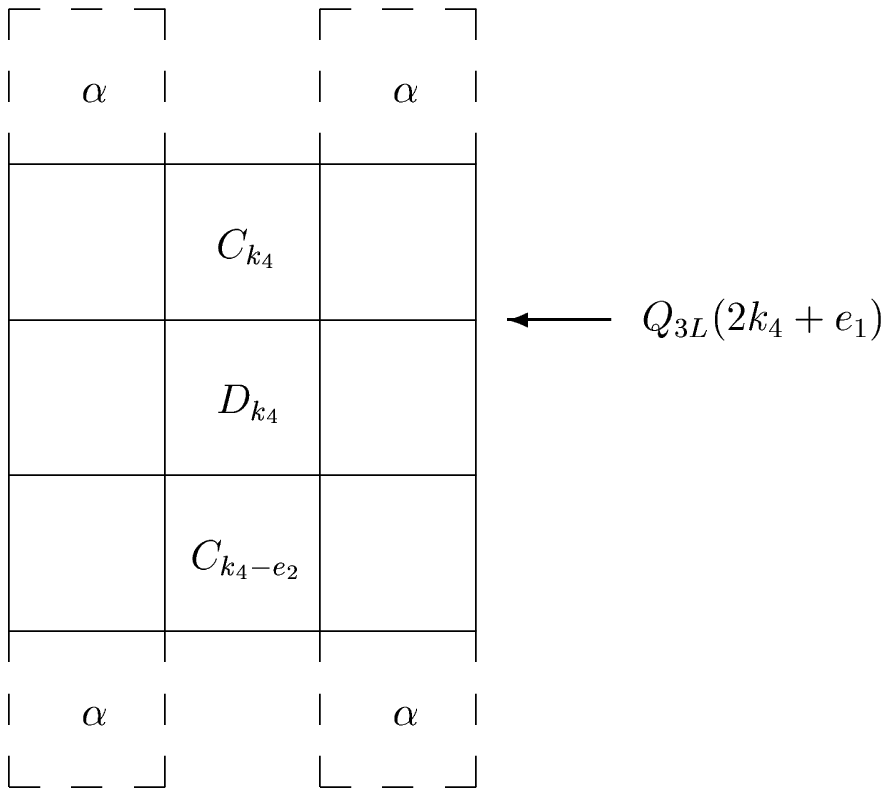} 
\par\noindent 
{\centerline {\bf Fig. 8}} 
\endinsert
\par
So it is natural to define different kind of (many body) bonds
corresponding to the above interaction terms. 
As a consequence of the above discussion we have the following kind of bonds;
the bond $D^{(\F)}_{k_4}$, to which corresponds the weight 
$\F^{(4)}_{D_{k_4}}(\a,\b,\g)$ which is given
by the set of $A$, $B$ and $C$ blocks contiguous to $D_{k_4}$ united with the 
other $A$  blocks adjacent from the exterior of
$Q_{3L}(2k_4 +e_1 )$ to the already considered $B$ blocks. So a
$D^{(\F)}_{k_4}$--bond contains twelve blocks.
We similarly define (now without any extension) the 
bond $C^{(\F)}_{k_3}$ with weight $\F^{(3)}_{C_{k_3}}(\b,\a)$; 
the bond $B^{(\F)}_{k_2}$ with weight $\F^{(2)}_{B_{k_2}}(\a)$; 
the bond $D^{(\Psi)}_{k_4}$ with weight $\Psi^{(4)}_{D_{k_4}}(\a)$ and
the bond $A^{(\Psi)}_{k_1}$ with weight $\Psi^{(1)}_{A_{k_1}}(\a_{k_1})$.

Given a bond $b$ of one of the above kinds we define its support
$\tilde b$ as the subset of $\cL$ obtained as the union of the $Q_L$
blocks making part of $b$.
For any bond $b$  we denote by $\xi_b$ the corresponding weight.
Notice that  $\xi_b$  will be, in general, a function of the $\a,\b,\g$
variables associated to the blocks in  $\tilde b$.
For instance a bond $b = D^{(\F)}_{k_4}$ can be seen as  an element of
$(\cL_L)^{12}$ whereas $\tilde b$ is a subset of the original lattice
$\cL$ given by the union of the twelve interacting  blocks.\par 

We say that two bonds $b_1$,$b_2$ are {\it connected} if $\tilde b_1
\cap \tilde b_2 \neq \emptyset$. 
A {\it polymer} $R$ is a set of bonds $b_1,\dots,b_k$ which is
connected in the sense that $\forall\;i,\; j$: $1\leq i < j \leq k$
there exists a chain of connected bonds in $R$ joining $b_i$ to $b_j$
namely $\exists \;b_{i_1},\dots b_{i_h}; b_{i_m }\; \in R, \;m=1,\dots
h$, $b_{i_1}=b_i,b_{i_h}= b_j$: $\tilde b_{i_m }\cap \tilde b_{i_{m+1}
} \neq \emptyset, \;m=1,\dots h-1$. 

The support $\widetilde R$ of a polymers $R = b_1,\dots,b_k$ is simply
$\widetilde R = \cup_{i=1}^k \tilde b_i$. We call $\cR _{\L_p}$ the
set of all possible polymer with support in $\L_p$ and $\cR $ the set
of all possible polymers with arbitrary support in $\cL$. Two polymers
$R_1,R_2$ are said to be compatible if  $\widetilde R_1\cap \widetilde
R_2 = \emptyset$; otherwise they are called incompatible.\par 

Given a polymer $R = b_1,\dots, b_k$ we define its {\it activity} $\z_R$ as:
$$
\z_R\;:=\;
\sum _{\a}\m_1 (\a)\sum _{\b}\m^{\a}_2 (\b)\sum _{\g}\m^{\a,\b}_3 (\g)
\prod_{i=1}^k \xi_{b_i}(\a,\b,\g)
\Eq (2.61)
$$
Notice that, due to the Bernoulli character of the above probability
measures, we can, as well, write:
$$
\sum _{\a_{\tilde R}}\m_{1,\tilde R} (\a_{\tilde R})\sum _{\b_{\tilde
R}}\m^{\a_{\tilde R}}_{2,
\tilde R} (\b_{\tilde R})\sum _{\g_{\tilde R}}\m^{\a_{\tilde R},\b_{\tilde
R}}_{3,\tilde R} (\g_{\tilde R})
\prod_{i=1}^k \xi_{b_i}(\a_{\tilde R},\b_{\tilde R},\g_{\tilde R})
\Eq (2.62)
$$
where $\a_{\tilde R},\b_{\tilde R},\g_{\tilde R}$ denote the $\a,\b,\g$
variables in $\tilde R$;
$$
\m^{\a_{\tilde R},\b_{\tilde R}}_{3,\tilde R} (\g) = \prod _{k_3:C_{k_3} \subset
\tilde R}
\m^{\a,\b}_{C_{k_3}} (\g_{k_3})
\Eq (2.63)
$$
and so on.\par
Going back to the specific structure of our multi--canonical  model it is
immediately seen that the activity of a polymer $R$ is a function of the renormalized
variables $n_i$ ($\equiv $ number of particles   fixing the constraint in the block
$Q_{L}(i)$)  {\it only} for $Q_{L}(i) \in \; \widetilde R$. To make explicit this dependence
we write
$$
\z_R = \z_R(n_{\tilde R})
\Eq (2.64)
$$
where $n_{\tilde R}= \{n_i\}_{Q_{L}(i)\subset \tilde R}$.\par
 From \equ (2.60), \equ (2.62) 
we get the desired expression:
$$
\X^{(\ell)}_{\L_p,\underline {n}}\; = \;
1 + \sum _{k\geq 1}\sum _{R_1, \dots
, R_k:\tilde R_i \subset \L_p,\atop \tilde R_i\cap \tilde R_j
=\emptyset, i<j=1,\dots,n}  
\prod_{i=1}^k \z_{R_i} (n_{\tilde R_i})
\Eq (2.65)
$$

Now we state a Proposition referring to a general class of polymer
systems. Its proof, which is based on the standard methods of the 
theory of the cluster expansion, can be found in [O] 
(see also [GMM], [KP], [D3], [NOZ]).

\nproclaim Proposition [prop2.1].
Consider a general polymer system (see [GrK], [KP], [D])  where the only
in\-te\-rac\-tion is a hard core exclusion forbidding overlap of the
supports $\widetilde R$ of the polymers $R$. Its partition function is:
$$
\X_{\L}\; = \;
1 + \sum _{k\geq 1}\sum _{R_1, \dots
, R_k:\tilde R_i \subset \L,\atop \tilde R_i\cap \tilde R_j
=\emptyset, i<j=1,\dots,n}  
\prod_{i=1}^k \z_{R_i} 
\Eq (2.66)
$$
Suppose that:
\item{i)}{$\exists \;\k>0$ such that the number of different polymers
$R$ with $m$ bonds (we write $|R| = m$) and support $\widetilde R$
containing a fixed point (say the origin) is bounded by $\k^m$;}    
\item{ii)}{$\exists \;\e > 0$ such that $|\z_R| < \e ^{|R|}$.} \hfill\break
Let
$$
\f_T(R_1,\dots,R_n) = {1\over n!} \sum_{g\in G(R_1,\dots,R_n)}(-1)^{ \# \;
{\rm edges \;in} \;g}
\Eq (2.67)
$$
where $G(R_1,\dots,R_n)$ is the set of connected graphs with $n$
vertices $(1,\dots,n)$ and edges 
$i,j$ corresponding to pairs $R_i, R_j$ such that $\widetilde
R_i\cap\widetilde R_j\neq \emptyset$ 
(we set the sum equal to zero if $G$ is empty and one if $n=1$).
If
$$
\e < \left. {1\over \k}\; 
 {x\over 1+x} e^{-x} \right|_{ x= ( 5^{{1\over 2}} - {1\over 2}) }
\Eq (2.68)
$$
then there exists a positive constant $C(\e)$ such that
$$
\sum _{R_1, \dots, R_n : \tilde R_i \subset \L \atop \exists R_i =R} 
| \f_T(R_1,\dots,R_n) | \prod_{i=1}^n  |\z_{R_i}|
\leq
C(\e) \left(
\e \, \exp \left\{  { \sqrt{5} - 1  \over 2 } \right\} \right)^{|R|}
\Eq (2.69)
$$
$$
\X_{\L} = \exp \left\{
\sum _{n\geq 1}
\sum _{R_1, \dots, R_n:\tilde R_i \subset \L }
\f_T(R_1,\dots,R_n) \prod_{i=1}^n \z_{R_i}
\right\}
\Eq (2.70)
$$

In our context, it is  clear that we can find a constant $\k$ so that 
the hypothesis {\it (i)}\/ of Proposition \thm[prop2.1] holds. 
It is also clear from
\equ (2.22), \equ(2.30), \equ (2.36), \equ (2.45), \equ (2.52),
\equ(2.54) that there exists a universal constant $C$ such that hypothesis 
{\it (ii)}\/ holds with $\e = C \d(\ell)$ 
(recall that $\d(\ell) \to 0$ as $\ell\to \infty$) so that 
\equ (2.68) holds for any $\ell$ sufficiently large.
In fact in the two--dimensional case we use a weaker condition: 
we do not need, in the left hand side of \equ(CC) to take the 
supremum over $V\in P^{(i)}_{L} (j)$, but only the analogous
condition only for the squares $Q_{L}$ and for the 
$L\times 3L$ rectangles.

Then, using the results of Proposition \thm[prop2.1],
we can compute the renormalized potential and 
perform the thermodynamic limit.
Suppose, instead of considering periodic boundary conditions, 
we had a generic b.c. $\t$ outside our cube $\L_p$.
It is clear that we can apply the same procedure (block decimation 
and cluster expansion) that we have used above in the case of periodic 
boundary conditions and get very similar results. 
Let us briefly sketch the differences.

Recall that our square $\L_p$ has a side  being an integer multiple of
the elementary square $Q_L$ 
with side $L =2\ell$; then certainly we will have a horizontal edge of
$\de \L_p$ adjacent to a row 
(of thickness $L$) made by $C$ and $B$  blocks (a $CB$ row) and a
horizontal edge adjacent to a $DA$ row. 
Similarly we will have a vertical edge adjacent to an  $AB$ column and
one adjacent to a $CD$ column.

It is easy to convince ourselves that even with generic $\t$ b.c. we
can repeat the same sequence of 
splitting and gluing, following the same ``path" joining the 4
sub--lattices of $\cL_L$ namely $D\to C\to B\to A$.
In the bulk, namely where the sets $D, \tilde C, \tilde B, \tilde A$
do not touch the boundary, we get the same results as in the case of
periodic b.c. For the blocks close to $\de \L_p$ we get the following
modifications:  

\item {{\it i)}}
The various sets $\tilde C, \tilde B, \tilde A$ of the bulk are substituted by their ``truncations
in $\L_p$" namely by $\tilde C\cap \L_p, \tilde B\cap \L_p, \tilde A\cap \L_p$ with the proper 
$\t$ b.c. on their part touching $\de \L_p$ and $0$, like in the bulk,
otherwise. 
\item {{\it ii)}}
The various probability measures 
$\m_{C_{k_3}} ^{\a,\b}, \m_{B_{k_2}} ^{\a}$ are defined similarly
to what is done in the bulk with the difference that, in their
definitions, the  terms corresponding to partition functions on regions 
lying totally (resp. partially) 
outside $\L_p$ are absent (resp. truncated); moreover 
the configuration on which they depend
parametrically:
$\a,\b$ in
$\m_{C_{k_3}} ^{\a,\b}$; $\a$ in $\m_{B_{k_2}} ^{\a}$ may contain $\t$; 
notice that 
$\m_{A_{k_1}}$   stays unchanged.
\item {{\it iii)}}
Some of the bonds, close to $\de\L_p$, are consequently modified
and their weights
can depend on $\t$. By an abuse of notation, we still denote them by
$D^{(\F)}_{k_4}$, $C^{(\F)}_{k_3}$, $B^{(\F)}_{k_2}$,
$D^{(\Psi)}_{k_4}$, $A^{(\Psi)}_{k_1}$. 
 
Indeed the splitting operation is very similar in the bulk and close
to the boundary; the true difference is the following.
When we have some term produced by a splitting
that, following the ``bulk rule", 
we would like to glue with some other term outside $\L_p$ or coming
from $\L_p$, simply we omit the 
gluing and in this way we 
construct some new domains just consisting in the parts of the
corresponding bulk domains 
($\tilde C, \tilde B, \tilde A$), lying inside $\L_p$. 

Let us describe an example.
Suppose that the upper horizontal side of $\de \L_p$ 
is adjacent from the exterior to a $CB$ row
(which, indeed, is the case with our choice of the location of $\L_p$).
After integrating over $\d$ variables and splitting like in \equ(2.21)
we do not glue on the
blocks $C_{k_3}$ sitting on the top row like in \equ(2.24) 
but we make an analogous operation combining
the term $Z_{D_{k_3}}$ (coming from the splitting on the $D_{k_3}$
block in $\L_p$ ) with the 
self--interaction in $C_{k_3}$ and its interaction with the exterior
configuration $\t$. 
In other words we use a formula analogous to \equ(2.24) but without
the term $Z_{D_{k_3 + e_2}}$ which, now, is absent. 
In this way the set corresponding to $\tilde C_{k_3}$ in the bulk, just
consists, now, of $C_{k_3}\cup D_{k_3}$.
Accordingly we define $\m_{C_{k_3}} ^{\a,\b}$ by omitting the factor
$Z_{D_{k_3 + e_2}} (\g _{k_3})$ in its definition.
When we continue with the splitting on the horizontal direction and
the gluing, say, on  
$B_{k_3 +e_1}$ we end up with the construction of a set, playing the
role of $\tilde B_{k_3 +e_1}$, 
obtained by removing from $\tilde B_{k_3 +e_1}$ the two $D$ blocks
exterior to $\L_p$ where the ``top" b.c. are given by $\t$ whereas the
other b.c are still given by the reference configuration $0$ like in the bulk. 
Of course also the error terms (of $\F$ or $\Psi$ type) are,
accordingly,  modified. 

In this way we can repeat the transformation of our system into a
polymer gas. 
We just have to introduce the obvious modifications 
in the terms appearing in the expression of the partition function of
the reference system  
$\bar Z^{(\ell)}_{\L_p,\underline n}$  
(see \equ(2.59)) as well as in the bonds 
$D^{(\F)}_{k_4}$, $C^{(\F)}_{k_3}$, $B^{(\F)}_{k_2}$,
$D^{(\Psi)}_{k_4}$, $A^{(\Psi)}_{k_1}$ close to the
boundary and in the   measures $\m_{C_{k_3}} ^{\a,\b}$,  $\m_{B_{k_2}}
^{\a}$ when  $C_{k_3}, B_{k_2}$ happen to be adjacent to the boundary
$\de \L_p$; then, accordingly,  we modify the definition of the
polymers and of their activity, $\z^\t_R= \z^\t_R (\bn_{\tilde R})$ 
(see  \equ(2.61)) which, now, will in general depend on
the location of the polymer and on the b.c. $\t$.  
Anyway if $d_\ell\( \tilde R, I_p^c \) > d$ the activity  $\z^\t_R$ of $R$ 
is the same as in the bulk and does not depend on $\t$.

\medskip
\noi{\it Proof of Theorem \thm[summa].}\/
Let us take the logarithm of \equ (2.58). By using \equ (2.59)
\equ(2.65) and \equ(2.70) we get the following expression for the
renormalized Hamiltonian.  
$$
\eqalign{
H^{(\ell,\t)}_{I_p} (\underline {n}) := &
\log \left [Z^{(\ell,\t)}_{\L_p,\underline {n}}\right] =
\sum _{k_1} \log \left [ 
Z_{\widetilde A_{k_1}}((0))\right ]-  
\sum _{k_2} \log
\left[ Z_{ F_{k_2}} (0)\right ] -
\sum _{k_3}\log 
\left [ Z_{\widetilde C_{k_3}} \left((0)\right) \right] 
\cr
&+ \sum _{k_4}\log
 \left [ Z_{D_{k_4}}\left ( (0)\right)\right]
+ \sum _{k\geq 1}
\sum _{R_1, \dots, R_k:\tilde R_i \subset \L_p }
\f_T(R_1,\dots,R_n) \prod_{i=1}^k \z_{R_i}^\t (n_{\tilde R_i})
\cr
}
\Eq (2.71)
$$
We have
$$
H^{(\ell,\t)}_{I_p} (\underline {n}) = {\rm const} + 
\sum_{X\su I_p} \F^{(\ell,\t),sr}_{X} (m_X)
+ \sum_{X\su I_p} \F^{(\ell,\t),lr}_{I_p,X} (m_X)
\Eq (2.72)
$$
where, with $A_{k_1}\in \cA_p,$ $B_{k_2} \in \cB_p,$ 
$C_{k_3} \in \cC_p$, $D_{k_4} \in \cD_p$ and $d_\ell(X,I_p^c)>d$
(see the above discussion for $d_\ell(X,I_p^c)\le d$), 
we set
$$
\F^{(\ell,\t),sr}_{ X} (m_X):=\cases{
{\phantom -}\log \left [\m^0 _{\widetilde A_{k_1}, z} (M _i =m_i,
Q_{\ell}(i) \su \widetilde A_{k_1})\right ]
& if  $X:\, \cup_{i\in X} Q_\ell(i) = \widetilde A_{k_1}$\cr
&\cr
-\log \left [\m^0 _{F_{k_2}, z} (M_i =m_i, Q_{\ell}(i)
\su F_{k_2})\right ]
& if  $X:\, \cup_{i\in X} Q_\ell(i) = F_{k_2}$\cr
&\cr
-\log \left [\m^0 _{\widetilde C_{k_3}, z} (M_i =m_i,
Q_{\ell}(i)
\su \widetilde C_{k_3})\right] 
& if  $X:\, \cup_{i\in X} Q_\ell(i) = \widetilde C_{k_3}$\cr
&\cr
{\phantom -}\log \left [\m^0 _{D_{k_4}, z} (M_i =m_i, Q_{\ell}(i) \su
D_{k_4}) \right] 
& if  $X:\, \cup_{i\in X} Q_\ell(i) = D_{k_4}$\cr
&\cr
{\phantom -}0 & otherwise\cr
}
\Eq(2.73/7)
$$
and 
$$
\F^{(\ell,\t),lr}_{I_p,X} (m_X):=
\sum _{R_1, \dots, R_k:\cup_i \tilde R_i = X }
\f_T(R_1,\dots,R_n) \prod_{i=1}^k \z_{R_i}^\t (n_{\tilde R_i})
\Eq (2.78)
$$

By the above discussion on the dependence of the activity 
on the boundary condition, for each $X \subset \subset \cL_\ell$ such that 
$d_\ell (X,I_p^c) > d$, $\F^{(\ell,\t)}_{ X}$ is independent of 
$\t$. 
Therefore the limit in \equ(ltpp) exists and is actually reached for
a finite $p$. 
Finally the estimate \equ(e:summa) is a direct consequence of 
\equ(2.78) and Proposition \thm[prop2.1].
\qed

\newsection The multi--canonical measure.

Given a positive integer $\ell$ and a volume $\L\su\su\cL$ 
of the form \equ(La=) 
we want to study the {\sl multi--canonical} state which is obtained from the
multi--grancanonical one by fixing the total number of particles in each
cube $Q_\ell(i)$, $i\in I$. Let thus $\bN =\{N_i, i\in I\}$ be the random
variables defined in \equ(dNi)
and, given $\bn =\{n_i=0,\cdots, |Q_\ell|, \,\, i\in I\}$, 
 the multi--canonical
state $\nu_{\L,\bn}^\t$ is given by
$$
\nu_{\L,\bn}^\t(\cdot) :=
\mu_{\L,\bz}^\t \( \cdot \left| \bN=\bn \right.\)
$$
which, in the RG context, represents the {\it constrained} model.
Note that $\nu_{\L,\bn}^\t$ is independent on $\bz$.

\bigskip
\noi{\it 4.1. Thermodynamic relationships.}

\noi
We need to compare the multi--canonical and  multi--grancanonical
state. We start here by discussing some thermodynamic relationships
between them. With respect to the usual treatment we work in finite
volume and take advantage of the strong mixing condition  to obtain explicit
bounds.

Let the volume $\L$ be of the form \equ(La=) for some $I\ssu\cL_\ell$ and 
$\mu_{\L,\bz}^\t$ be a multi--grancanonical state satisfying
Condition MUSM($\cA)$. Introduce the map
$\cA^I\ni\bz\mapsto \br^\t(\bz)\in [0,1]^I$ defined by  
$$
\r_i^\t(\bz)= \r_i^{\t,(\ell)}(\bz) 
:= {1\over |Q_\ell|} \mu_{\L,\bz}^\t \( N_i \), \quad i\in I
\Eq(zmtr)
$$

\nproclaim Proposition [thtr].
For each $I\ssu \cL_\ell$ and each closed $\cC \su\cA$ 
there is a constant $C>0$ such that for any boundary
condition $\t$, any $\bz\in \cC^I$
and all $\ell$ multiple of $\ell_0$
$$
{1\over C} \le {\p \over \p z_i} \r_i^\t(\bz) \le C 
\Eq(t1) 
$$
$$
\left| {\p \over \p z_j} \r_i^\t(\bz) \right| 
\le C  z_i {1+ \left|{\overline Q}^r_\ell(i) \cap 
{\overline Q}^r_\ell(j)\right| \over | Q_\ell|} 
e^{- d\( Q_\ell(i), Q_\ell(j)\)/C}, \quad
i\neq j 
\Eq(t2)
$$
$$ 
\left|  \r_i^{\t^x}(\bz)- \r_i^{\t}(\bz)\right| 
\le C {z_i \over |Q_\ell|} e^{-d\(x,Q_\ell(i)\) / C }
\Eq(t3)
$$

The proof of the lower bound in \equ(t1) is based on the following
Gaussian bound on the characteristic function (see
[DS4,\S 2.3] and  [Y,\S 9]) which will be
extensively used in the sequel. For $\bt\in\bR^{|I|}$, we use the
notation 
$\<\bt,\bN\> :=\sum_{i\in I} t_i N_i$.

\nproclaim Lemma [gaussbound].
For each $I\ssu\cL_\ell$ there is a constant $C>0$ such that 
for any $\ell$ and $\bt\in [-\pi,\pi]^{|I|}$
$$
\left|
\mu_{\L,\bz}^\t \( \exp \left\{ i \<\bt, \bN \>\right\} \) 
\right| 
\le \exp\left\{ -{1\over C}\12 |Q_\ell| \sum_{i\in I}  z_i |t_i|^2\right\}
\Eq(e:bcf)
$$

\noi{\it Proof.}\/ Before starting we stress that the proof is based
only on the finite range and boundedness of the interaction and does not
use Condition MUSM($\cA$). 

Let $\L'\subset\L$ be a subset of a sub--lattice of $\cL$ with 
spacing parameter larger than the range $r$ of the interaction.
This means that for any $x,y\in \L'$ we have  $d(x,y)> r$ 
but nonetheless  $|\L'| \ge  |\L|/C $ for some constant $C=C(r)\ge 1$. 
If we set  $Q'_\ell(j) :=\L' \cap Q_\ell(j)$, we then have
$$
\eqalign{
\left|
\mu_{\L,\bz}^\t \( e^{ i \<\bt,\bN \>} \)
\right|
&
=
\left|
\int \mu_{\L,\bz}^\t(d\z)  \mu_{\L',\bz}^\z
\( \prod_{j=1}^{|I|} \prod_{x_j\in Q_\ell(j)}
e^{ i  t_j \h_{x_j} }
\)
\right|
\cr &
\le
\sup_\z
\left| \mu_{\L',\bz}^\z\( 
\prod_{j=1}^{|I|} \prod_{x_j\in Q_\ell'(j)}
e^{ i  t_j \h_{x_j} }\)\right|
=
\prod_{j=1}^{|I|} \prod_{x_j \in Q_\ell'(j)}
\sup_\z
\left|\mu_{\L',\bz}^\z\( 
e^{ i  t_j\(  \h_{x_j} \) }
\)
\right| 
\cr
}
$$
since $\mu_{\L',\bz}^\z$ is a product measure.

Let $p_x(\z):= \mu_{\L',\bz}^\z \(\h_x=1\)$. Since the interaction 
is bounded we get, for some constant $C=C(\|U \|)>0$
independent on $x$, $\bz$ and $\z$,
$ z_j/C \le  p_x(\z) \le C z_j $ for $x\in Q_\ell(j)$.
A simple computation on Bernoulli variables shows now
that for $|t|\le \pi $, $x_j\in Q_\ell(j)$
$$
\left|\mu_{\L',\bz}^\z \( 
e^{ i  t_j \h_{x_j}   }
\) \right|
\le \exp\left\{ -  {1\over C} \12  z_j t_j^2 \right\}
$$
the bound \equ(e:bcf) follows.
\qed

\medskip
\noi{\it Proof of Proposition \thm[thtr].}\/ 
We first note that
$$
{\p \over \p z_j} \r_i^\t(\bz) = 
{1\over z_j |Q_\ell|} \mu_{\L,\bz}^\t\(N_i;N_j\)
$$
Let 
$$
v_{i,j} =v_{i,j}^{\t,(\ell)} (\bz) := \mu_{\L,\bz}^\t\(N_i;N_j\)
$$
the lower bound in \equ(t1) follows by noticing that Lemma
\thm[gaussbound] implies the quadratic form estimate
$$
 \sum_{i,j\in I} t_i t_j v_{i,j} 
\ge  {1\over C}  |Q_\ell| \sum_{i\in I} z_i t_i^2  
\Eq(qfe)
$$

To prove the upper bound in \equ(t1) and \equ(t2) we instead use 
Condition MUSM($\cA$) to get
$$
\left|v_{i,j} \right| \le 
\sum_{x\in Q_\ell(i)} \sum_{y\in Q_\ell(j)}
\left| \mu_{\L,\bz}^\t\(\h_x ,\h_y\) \right|
\le \sum_{x\in Q_\ell(i)\cap Q_\ell(j)} 
\mu_{\L,\bz}^\t\(\h_x ,\h_x\) 
+ C z_i z_j \sum_{{x\in Q_\ell(i) \atop y\in Q_\ell(j),y\neq x}}
e^{-d(x,y)/C}
$$
and that for $x\in Q_\ell(i)$, by the same argument as in Lemma
\thm[gaussbound],  $\mu_{\L,\bz}^\t\(\h_x ,\h_x\) \le C z_i$.
The proof of \equ(t3) is analogous and we omit it.
\qed

\medskip
Let $\mu_{z}$ be the infinite volume Gibbs state associated to the
(translation invariant) interaction $(z,U)$ satisfying Condition 
MUSM($\cA$). 
We introduce the (one dimensional) map 
$\cA \ni z \mapsto \r (z) \in [0,1]$ by
$\r(z) = \mu_{z}(\h_x)$ and denote by $\r\mapsto z(\r)$ 
the inverse map which is analytical as a consequence of the strong 
mixing assumption. Let finally $\cB \subseteq [0,1]$ be defined by 
$\cB:= \r (\cA)$ where $\cA$ is as given in Condition
MUSM($\cA$); we note $\cB = [0,1]$ if $\cA= [0,\infty)$.  

Recall that the map $\bz\mapsto \br^{\t,(\ell)}(\bz)$ 
has been defined in \equ(zmtr). We need an inverse map 
$\br\mapsto \bz^{\t,(\ell)} (\br)$ defined 
for all possible boundary condition $\t$.
When $\cB$ is a proper subset of $[0,1]$ we take $\ell$ large 
enough and define it on a subset of $\cB$.
By using strong mixing and Proposition \thm[thtr] 
it is easy to deduce that for each closed $\cC\su\cA$ 
$$
\lim_{\ell\ten\infty} \r_i^{\t,(\ell)} (\bz)= \r(z_i)
,\quad {\rm uniformly~for}~\t\in\O, \bz\in\cC^I
\Eq(limr)
$$
and that for each closed set $\cD\su \cB$ and any $\ell$ large enough
(depending on $\cD$) we have 
$$
\cD^I\su \bigcap_\t \br^{\t,(\ell)}(\cA^I) 
$$
Finally, by \equ(qfe), 
the Jacobian of the map $\bz\mapsto \br^{\t,(\ell)}(\bz)$ is not
degenerate uniformly in $\tau$ and $\ell$. 
Let $\cD\su \cB$ be a closed set and $\ell$ large enough; we can therefore 
define the inverse map on the set $\cD^I$,
i.e. the map $\cD^I\ni\br\mapsto \bz^\t(\r)= \bz^{\t,(\ell)}(\r)$ 
such that 
$$
\br^\t\(\bz^\t(\br) \) = \br
$$
for any $\br \in \cD^I$, $\t\in\O$. 

When $\cB=[0,1]$ we can instead define the inverse map for any $\ell$.
Indeed we have
$$
\lim_{z_i\to 0} \r^{\t,(\ell)}_i(\bz) = 0,
\quad
\lim_{z_i\to +\infty} \r^{\t,(\ell)}_i(\bz) = 1,
\quad
{\rm uniformly~for}~\t\in\O, \{z_j\in [0,\infty), j\neq i\}
$$
which, together with \equ(qfe), implies 
$$
\br^{\t,(\ell)} \([0,\infty)^I \) = [0,1]^I
$$

We prove below some estimate on the Jacobian of the map
$\br\mapsto \bz^{\t,(\ell)} (\br)$; in order to describe 
them we need some more notation. Let $\{\o_h, h=0,\cdots,k\}$ be a path on the
rescaled lattice $\cL_\ell$ such that 
$d_\ell(\o_{h-1},\o_{h})=1$, $h=1,\cdots,k$. We introduce 
$q(\o_{h-1},\o_{h}) := |{\overline Q}^r_\ell(\o_{h-1})\cap  
{\overline Q}^r_\ell(\o_{h})| / |Q_\ell|$.

\nproclaim Proposition [thtr'].
For each $k\in\Z^+$, $I\ssu \cL_\ell$ and each closed $\cD\subseteq \cB$ 
there is a constant $C>0$ such that for any $\t\in\O$, $\br\in\cD^I$,
$x\in\p_r\L$ and all $\ell$ large enough
$$
{1\over C} \le {\p \over \p \r_i} z_i^\t(\br) \le C 
\Eq(t1') 
$$ 
$$
\left| {\p \over \p \r_j} z_i^\t(\br) \right| 
\le C \r_i \left\{ \sup_{1\le k'\le k} 
\sup_{{\o : \atop {\o_0=i, \atop \o_{k'}=j}}} 
\prod_{h=1}^{k'} q(\o_{h-1},\o_h)
+ {1\over \ell^{k+1}}
\right\}, \quad i\neq j 
\Eq(t2')
$$
Moreover
$$
\left|  z_i^{\t^x}(\br)- z_i^{\t}(\br)\right| 
\le {C\r_i \over |Q_\ell|}
\(e^{-d(x,Q_\ell(i))/C} + 
\sup_{j: x\in\p_r Q_\ell(j)} \sup_{1\le k'\le k} 
\sup_{{\o : \atop {\o_0=i, \atop \o_{k'}=j}}} 
\prod_{h=1}^{k'} q(\o_{h-1},\o_h)
+ {1\over \ell^{k+1}}
\)
\Eq(t3')
$$

\noi{\it Proof.}\/ Let 
$$
\bJ_{i,j}= \bJ_{i,j}^{\t,(\ell)}(\bz)  :={\p \over \p z_j} \r_i^\t(\bz) 
$$
be the Jacobian of the map $\bz\mapsto \br^\t(\bz)$. We split in in
its diagonal and off diagonal part; $\bJ =\bD + \bA$ where
$$
\bD_{i,j} := \d_{i,j} {\p \over \p z_i} \r_i^\t(\bz) 
$$
and note that from \equ(t1), \equ(t2) it follows $\bD\ge 1/C$, 
$\|\bA\|\le C \ell^{-1}$. 

In order to prove the bounds \equ(t1'), \equ(t2') we need to invert
the Jacobian $\bJ$. We use the above splitting and Neumann series to get
$$
\bJ^{-1} = \bD^{-1} \( \id + \bA \bD^{-1} \)^{-1}
= \bD^{-1} \( \sum_{h=0}^k (-1)^h (\bA \bD^{-1})^h + \bR_{k+1} \)
$$ 
where $\| \bR_{k+1} \|\le C \ell^{-(k+1)}$. 
Since $\bD$ is bounded from below and $\bA_{i,j}$ is exponentially
small for $d_\ell(i,j)>1$, \equ(t2') follows easily from \equ(t2).

To prove \equ(t3') we note that, by  definition of the map
$\br\mapsto \bz^{\t}(\br)$ we have
$$
\r_i^{\t^x}\( \bz^{\t^x}(\br) \) = \r^{\t}_i\( \bz^{\t}(\br) \) 
,\quad \quad i\in I
\Eq(costante)
$$
By using the invertibilty (uniform in $\t\in\O$ and $\ell$) of 
$\bz\mapsto\br^{\t,(\ell)}(\bz)$ and \equ(limr), it is not diffucult to see 
that \equ(costante) implies that, for $\ell$ large enough, 
$\bz^{\t^x}(\br)$ and $\bz^{\t}(\br)$ are in the same connected 
component of $\bz^{\t^x}\(\cD^I\)\cup \bz^\t\(\cD^I\)$. 

On the other hand, by Lagrange theorem
$$
\r_i^\z(\bz^2) - \r^\z_i(\bz^1)
= \sum_{j\in I} {\p \over \p z_j} \r^\z_i(\bar \bz)
\cdot [z_j^2-z_j^1]
$$
where $\bar \bz\in \cA^I$ if $ \bz^1$, $\bz^2$ are the same connected 
component of $\cA^I$.
Whence, by usig \equ(costante), 
$$
z_i^{\t^x}(\br)- z^{\t}_i(\br)
= \sum_{j\in I}  \( \bJ^\t(\bar \bz) \)^{-1}_{ij}
\cdot \left[
\r_j^\t\(\bz^\t(\br)\) - \r^{\t^x}_j\(\bz^\t(\r)\) \right]
$$
and \equ(t3') follows from \equ(t3) and \equ(t2').
\qed

\vfill\eject
\noi{\it 4.2. Comparison of ensembles in finite volumes.}

\noi
We here discuss the equivalence of multi--grancanonical and
multi--canonical ensembles. We shall work in finite volume with the aim
of obtaining explicit bounds as a consequence of the
strong mixing assumption.

Let $I\ssu\cL_\ell$, and $\L$ as in  \equ(La=). 
We want to compare the measures $\mu_{\L,\bz}^\t$ and 
$\nu_{\L,\bn}^\t$ where the activity $\bz$ is chosen, depending on 
$\bn$, $\L$ {\it and} $\t$, as (recall that the fuction 
$\br\mapsto \bz^\t(\r)$ as been defined above)
$\bz= \bz^\t\( {\bn/ |Q_\ell|}\) $, i.e. so that 
$\mu_{\L,\bz}^\t \( \bN\) =\bn$.
We have the following result. Recall that $\cB = \r (\cA)$.

\nproclaim Theorem [eqen'].
Assume $\mu_{\L,\bz}^\t$ satisfies Condition MUSM($\cA$). Then 
for each closed $\cD\subseteq\cB$, each $I\ssu \cL_\ell$ and each local
function $f$, there is a constant $C$ depending on the
constants in Condition MUSM($\cA$), $\cD$, $|I|$, 
${\rm diam }\(S(f)\)$,
$\|f\|$, such that for any b.c. $\t$, any $\bn\in \cD^I$ and all 
$\ell$ multiple of $\ell_0$ the following bound 
holds
$$
\left| \nu_{\L,\bn}^\t f - \mu_{\L,\bz}^\t f \right| \le C {1\over |Q_\ell|}
\;\;\; .
\Eq(e:eqen')
$$

The proof of this theorem is based on the DS complete analyticity
conditions [DS1], [DS2], [DS3]. Although originally formulated for
arbitrary volumes 
their theory carries over to our strong mixing for regular domains 
as already remarked. 

More precisely we need the following condition [DS3, Condition Ib]
which is equivalent to SM($\ell_0$). 
There is a constant $\e>0$ such that for all complex
interactions $\tilde\Phi$ in an $\e$--neighborhood of $\Phi$, i.e.
$$
\tilde\Phi \in \cO_\e(\Phi):= \{ \| \tilde\Phi - \Phi\|_0 < \e \}
$$
and all finite volumes $\L$ as in \equ(multiplo) the analytic functions 
$Z_\L^\t(\tilde\Phi)$ are non--vanishing. 
Moreover, there is another constant $A' <\infty$ such that  for all
$\tilde\Phi_1,\tilde\Phi_2 \in\cO_\e(\Phi)$ we have the bound 
$$
\sup_{\t\in\O} 
\left| P_\L^\t(\tilde\Phi_1) -  P_\L^\t(\tilde\Phi_2) \right| 
< A' \left|  {\overline \L}^r  \cap  
\supp \(\tilde\Phi_1 - \tilde\Phi_2 \) \right| 
\Eq(ds1b)
$$ 
where the pressure $P$ is defined by 
$$
P_\L^\t(\tilde\Phi):= \log Z_\L^\t(\tilde\Phi)
\Eq(pressure)
$$
and $\supp \(\Phi \) :=\cup_{\D \,:\, \Phi_\D \neq 0} \D$.

\medskip
\noi{\it Proof of Theorem \thm[eqen'].}\/
Since the b.c. $\t$ is kept fixed we drop it from the notation.
We also assume, without loss of generality, that $\|f\|$ is small
enough. 

\smallskip
\noi{\it Step 1.}\/ We express here the difference between 
multi--grancanonical and multi--canonical states by introducing the
Fourier transform of the indicator $\id_{\bN=\bn}$. 

By  definition of the multi--canonical state $\nu_{\L,\bn}$, we have
$$
\nu_{\L,\bn}\(f\) -\mu_{\L,\bz}\(f\)  
={ \mu_{\L,\bz} \(  \( f - \mu_{\L,\bz}\(f\)  \) \id_{\bN=\bn}\)
\over \mu_{\L,\bz} \( \bN=\bn \) }
={ \mu_{\L,\bz} \( (1+u) \id_{\bN=\bn}\) \over \mu_{\L,\bz} 
\( \bN=\bn \)} -1
\Eq(nu-mu)
$$
where we introduced $u:=f - \mu_{\L,\l}\(f\)$ which has the same 
support as $f$ and is mean zero 
w.r.t. $\mu_{\L,\bz}$. 

We next introduce the perturbed probability measure 
$d\mu_{\L,\bz}^u :=(1+u) d\mu_{\L,\bz}$. 
We regard it as the Gibbs measure w.r.t. an interaction $\Phi^u$. 
Since $f$ is a local function, we have that $\Phi^u$ has range bounded
by $\max\{r,{\rm diam}\,(\supp(f))\}$. Moreover, by taking $\|f\|$
small (depending on $\e$) we have that $\Phi^u \in \cO_\e(\Phi)$.

By taking the Fourier transform on the r.h.s. of \equ(nu-mu), we have
(recall that $\mu_{\L,\bz}\( \bN \) = \bn$ by the choice of $\bz$)
$$
\eqalign{
\nu_{\L,\bn}\(f\) -\mu_{\L,\bz}\(f\) & =
{ 
\int_{|\bt|\le \pi} \! d\bt \, 
e^{-i \<\bt,\mu_{\L,\bz} \bN\>} 
\mu_{\L,\bz}^u \( e^{i \<\bt,\bN\>} \) 
\over
\int_{|\bt|\le \pi} \! d\bt \, 
e^{-i \<\bt,\mu_{\L,\bz} \bN\>} 
\mu_{\L,\bz} \( e^{i \<\bt,\bN\>} \) 
} -1
\cr
&={ 
\int_{|\bt|\le \pi} \! d\bt \, 
e^{\psi_\L(\bt,\bz) -i \<\bt,\mu_{\L,\bz} \bN\> } 
\left[ e^{ \psi_\L^u(\bt,\bz) - \psi_\L(\bt,\bz) 
} -1 \right]
\over 
\int_{|\bt|\le \pi} \! d\bt \, 
e^{\psi_\L(\bt,\bz) -i \<\bt,\mu_{\L,\bz} \bN\> } 
}
\cr
}
\Eq(inu)
$$
where, indicating with a superscript the dependence on
the perturbation $u$ and inside the parentheses the dependence on the
complex activity, we introduced
$$
\psi_\L(\bt,\bz) := 
\log \mu_{\L,\bz} \( e^{i \<\bt,\bN\>}\)
= P_\L\( \{ z_j e^{it_j} \}_{j\in I}\) - P_\L\( \bz \)
\Eq(psi=P)
$$
where the second identity holds by expressing the l.h.s. in terms of 
ratio of partition functions. The definition of $\psi_\L^u(\bt,\bz)$ is
analogous, it is enough to consider the pressure of the perturbed interaction.

\smallskip
\noi{\it Step 2.}\/ Here we estimate from below the denominator on
the r.h.s. of \equ(inu). 

Let us introduce the variances
$$
v_i^2=v_i^{\t,(\ell)}(\bz)^2 := \mu_{\L,\bz}^\t \( N_i; N_i\)
$$
and note that from Proposition \thm[thtr] we have 
$C^{-1}z_i |Q_\ell|\le v_i^2 \le Cz_i |Q_\ell|$. This bound will be
used extensively in the sequel.

We shall prove the following bound. There is a constant 
$C$ independent on $\t$, $\ell$ and $\bz$ such that for $\ell$ large enough
$$
\mu_{\L,\bz}\(\bN=\bn\) =
{1\over (2 \pi)^{|I|}}
\int_{|\bt|\le \pi} \! d\bt \, e^{\psi_\L(\bt,\bz)-i \<\bt,\mu_{\L,\bz} \bN\>} 
\ge {1\over C} {1\over \prod_{i\in I} v_i}
\Eq(bfb)
$$ 
where we recall $\bz$ has been chosen so that $\mu_{\L,\bz}\(\bN\) = \bn$.

By a change of variables we get
$$
\mu_{\L,\bz}\(\bN=\bn\) = 
{1\over \prod_{j\in I} {2 \pi} v_j}
\int_{|s_j|\le \pi v_j } 
\! d\bs \, e^{ \psi_\L \( \bs / v ,\bz\) 
-i \< {\bs/v} ,\mu_{\L,\bz} \bN\>} 
$$ 
where we used the notation $\bs/v$ to denote the variables
$\{s_j/v_j, \,\, j\in I\}$ 

Let $K$  be a large constant. 
We take advantage of the Gaussian
bound in Lemma \thm[gaussbound] to get
$$
\eqalign{
\left| 
\int_{\exists j :   K\mi (\pi v_j) \le |s_j| \le \pi v_j } 
\! d\bs \, e^{ \psi_\L\( \bs / v ,\bz\) 
-i \< {\bs/v} ,\mu_{\L,\bz} \bN\>}  
\right|
& \le 
\int_{\exists j : |s_j|\ge K\mi (\pi v_j)}
\! d\bs \,\, 
\exp\left\{ - \12 {1\over C} |Q_\ell| 
\sum_{i\in I} z_i {s_i^2\over v_i^2} \right\}
\cr
&\le C e^{- K^2/C}
\cr
}
\Eq(ugb)
$$

By the above bound we can restrict ourselves to bounded $\bs$. 
We need however to treat separately the Gaussian scaling in which 
$v_i$ diverges with $\ell$ and the very low density case in which
it remains bounded. 
Let $M$ be another large constant ($1\ll K\ll M \ll \ell$)
and introduce 
$I_g := \{ i\in I : v_i^2 \ge M \}$, $I_p:= I\setminus I_g$.
Let also $\bs_g := \{ s_i, i\in I_g\}$ 
(resp. $\bs_p := \{ s_i, i\in I_p\}$); we use an analogous notation
for $\bz$.
We shall prove the following expansion on the logarithm of the
characteristic function.
$$
\eqalign{
& \psi_\L\( \bs /v ,\bz\) -
 i \< {\bs/v} ,\mu_{\L,\bz} \bN\> 
\cr
&~~~= \sum_{j\in I_p} \( e^{i s_j/v_j} -1 - i s_j/v_j \) \mu_{\L,\bz} N_j 
- \12 \sum_{j j' \in I_g} \mu_{\L,\bz} \(N_j; N_{j'}\)  
{ s_j \over v_j} {s_{j'} \over v_{j'}} 
+ R_\L\( \bs,\bz\)
\cr
}
\Eq(expsi)
$$
where
$$
\sup_{|\bs|\le K}  
\left| R_\L\( \bs,\bz\) \right|
\le  C \(  {K^3 \over \sqrt{M}} +  { M^2 \over |Q_\ell|}
+ {K M \over \sqrt{|Q_\ell|}} + { K^2 M  \over {|Q_\ell|}} \)
$$
Note that on the r.h.s. of \equ(expsi) the first term corresponds to a Poisson
limit for $N_j$, $j\in I_p$ and to a (joint) 
Gaussian limit for $N_j$, $j\in I_g$.

Postponing the proof of \equ(expsi), let us first show that, together
with \equ(ugb), it implies the bound \equ(bfb).
It is enough to notice that if $Z$
is a Poisson r.v. with mean $\l \in \bZ^+$ we have
$$
{1\over 2\pi} \int_{|s|\le \pi u} \! ds \,\, 
e^{ \( e^{i s/u } -1 - i s/u \) \l } 
= u \,\,{\rm Prob} \( Z=\l\) = u { e^{-\l} \l^\l \over \l!}
$$
By using the bounds $v_i^2 \ge z_i |Q_\ell|/C $, 
$\mu_{\L,\bz} N_i \le C z_i |Q_\ell|$, Stirling's formula and estimating 
the Gaussian integral (recall \equ(qfe)) we thus get
$$
\int_{|s_i| \le K \mi (\pi v_i) } \!d\bs  \,\,
\exp\left\{ 
\sum_{j\in I_p} \( e^{i s_j/v_j} -1 - i s_j/v_j \) \mu_{\L,\bz} \(N_j\) 
- \12 \sum_{j j' \in I_g} \mu_{\L,\bz} \(N_j; N_{j'}\)  
{ s_j \over v_j} {s_{j'} \over v_{j'}} 
\right\}
\ge {1\over C}
$$
and \equ(bfb) follows since we can make the remainder as small as we want.

In order to prove \equ(expsi) let us first expand $\psi_\L$ in power 
series of $\bs_g$ and get
$$
\eqalign{
\psi_\L\( \bs/v , \bz\) = &
\psi_\L\( 0,\bs_p/v , \bz\) 
+ \sum_{i\in I_g}  {\p \over \p t_i } 
\psi_\L\( 0,\bs_p/v, \bz\)  \,\, {s_i\over v_i}
\cr
&+ \12
\sum_{i,i'\in I_g} {\p^2 \over \p t_i \p t_{i'}} 
\psi_\L\( 0,\bs_p/v,\bz\) \,\, {s_i\over v_i} {s_{i'}\over v_{i'}}
+R_\L^1\(\bs,\bz\)
\cr
}
\Eq(epsx)
$$

We note that by Condition [DS3,Ic], still equivalent to SM($\ell_0$), 
$$
\left|  {\p \over \p t_i}  \psi(\bt,\bz) \right| 
= \left| \mu_{\L,\bz,\bt}^\t \( N_i \) \right| 
\le C z_i |Q_\ell|
\Eq(s1d)
$$
here $\mu_{\L,\bz,\bt}^\t$ denotes the complex measure defined by 
$$
\mu_{\L,\bz,\bt}^\t (f) := 
{ \mu_{\L,\bz}^\t \(e^{i \<\bt,\bN\>} f \) \over
\mu_{\L,\bz}^\t \(e^{i \<\bt,\bN\>} \) }
$$
We remark that in [DS3,Ic] does not include $z_i$ on the r.h.s. of
\equ(s1d). However, by the remark following Condition USM($\ell_0$),
we can easily verify that \equ(s1d) holds.

Recall that the pressure $P_\L(\bz)$ is holomorphic in an
$\e$--neighborhood of $\bz$. Therefore (see \equ(psi=P))
$\psi_\L(\bt,\bz)$ is holomorphic in a neighborhood of $\bt=0$.
By taking $K/\sqrt{M}$  
small enough we can thus
use Cauchy integral formula and bound
the third order derivatives (w.r.t. to $\bt$) of $\psi(\bt,\bz)$ in
terms of the first one. By applying \equ(s1d) we get
$$
\sup_{|\bs| \le K} \left| R_\L^1 \( \bs,\bz\) \right| 
\le C \sup_{|\bs| \le K} \sum_{i,j,k\in I_g} \min\{z_i,z_j,z_k\} |Q_\ell|  
{|s_i s_j s_k| \over v_i v_j v_k}
\le C K^3 {1\over \sqrt{M}}
$$

We next expand the other terms on the r.h.s. of \equ(epsx)  in power
series of $\bz_p$. Note in fact that for $i\in I_p$ 
we have $z_i \le C M / | Q_\ell|$.  
Let us consider the first one. We get
$$
\psi_\L\( 0,\bs_p/ v ,\bz \) 
= \psi_\L\( 0,\bs_p/v, \bz_g,0\)
+ \sum_{j \in I_p} z_j  {\p \over \p z_j}  \psi_\L\( 0,\bs_p/v,\bz_g,0\) 
+ R_\L^2\(\bs,\bz\)
$$
Noticing that $|\psi_\L(\bt,\bz)| \le C |Q_\ell|$ and using again the Cauchy
integral formula, we can bound the remainder as follows
$$
\left| R_\L^2\(\bs,\bz\) \right|
\le C \sum_{j,j'\in I_p}  z_j z_{j'} | Q_\ell|
\le C {M^2 \over |Q_\ell|}
$$
We next observe that $\psi_\L\( 0,\bs_p/v,\bz_g,0\)=0$. On the other hand, by 
\equ(psi=P)
$$
{\p \over \p z_j} \psi_\L\( 0,\bs_p/v,\bz_g,0\)  
= \(e^{i s_j/v_j} -1 \) {\p \over \p z_j}  P_\L \(\bz_g,0 \)
$$
By the analyticity of the pressure (see \equ(s1d)) we also have
$$ 
\left| {\p \over \p z_j}  P_\L\( \bz_g,0 \) - 
{\p \over \p z_j}  P_\L\( \bz_g,\bz_p\)
\right| \le C  |Q_\ell|\sum_{i\in I_p} z_i \le C M
$$
Since $ z_j {\p \over \p z_j}  P_\L\( \bz\) = \mu_{\L,\bz}\(N_j\)$,
we thus get
$$
\psi_\L\( 0,\bs_p/ v ,\bz \) 
= \sum_{j\in I_p} \( e^{i s_j/v_j} -1 \) \mu_{\L,\bz} \(N_j\) 
+ R_\L^3(\bs,\bz)
,\quad\quad
\left| R_\L^3(\bs,\bz) \right| \le C {M^2 \over |Q_\ell|}
$$

We expand similary the other two terms in \equ(epsx). For the second
one we have
$$
{\p \over \p t_i} \psi_\L\( 0,\bs_p/v,\bz\)
= {\p \over \p t_i} \psi_\L\( 0,\bs_p/v,\bz_g,0\) 
+R_{\L,i}^4 \( \bs,\bz\) 
$$
where, by using again \equ(s1d) and the analyticity of 
$\psi_\L\(\bt,\bz\)$, we have
$$
\left| R_{\L,i}^4 \( \bs,\bz\) \right| 
\le C z_i |Q_\ell| \sum_{j\in I_p}  z_j  
\le C M z_i 
$$ 
Furthermore, since by setting $z_i=0$ $\psi_\L$ becomes independent of
$s_i$,  
$$
\left| {\p \over \p t_j} \psi_\L\( 0,\bs_p/v,\bz_g,0\)
- {\p \over \p t_j} \psi_\L\( 0,0,\bz_g,\bz_p\)
\right|
= \left| {\p \over \p t_j} \psi_\L\( 0,\bs_p/v,\bz_g,0\)
- i \mu_{\L,\bz} \(N_j\)  \right|
\le C M z_j
$$
so that
$$
\sum_{j\in I_g}  {\p \over \p t_j } 
\psi_\L\( 0,\bs_p/v, \bz\)  \,\, {s_j\over v_j}
=i \sum_{j\in I_g}    {s_j\over v_j} \mu_{\L,\bz} \(N_j\)  
+ R_{\L}^5 \( \bs,\bz\) 
,\quad\quad
\sup_{|\bs|\le K} \left| R_{\L}^5 \( \bs,\bz\) \right|
\le C {K M \over \sqrt{|Q_\ell|}}
$$

By the same argument we  finally have
$$
{\p^2 \over \p t_i \p t_{i'} } \psi_\L \( 0,\bs_p/v,\bz_g,\bz_p\)
= {\p^2 \over \p t_i  \p t_{i'}} \psi_\L\( 0,\bs_p/v,\bz_g,0\) 
+R_{\L,i,i'}^6 \( \bs,\bz\) 
,\quad
\left| R_{\L,i,i'}^6 \( \bs,\bz\) \right| 
\le C M z_i\mi z_{i'}
\Eq(ed2')
$$ 
Moreover, as before,
$$
\eqalign{
&\left| {\p^2 \over \p t_i \p t_{i'}} \psi_\L\( 0,\bs_p/v,\bz_g,0\) 
-  {\p^2 \over \p t_i \p t_{i'} } \psi_\L\( 0,0,\bz_g,\bz_p\) \right| 
\cr
&~~~~
=\left|  {\p^2 \over \p t_i \p t_{i'}} \psi_\L\( 0,\bs_p/v,\bz_g,0\) 
+ \mu_{\L,\bz}\(N_i;N_{i'}\) \right| \le C M z_i \mi z_{i'}
}
$$
which gives us
$$
\sum_{i,i'\in I_g} {\p^2 \over \p t_i \p t_{i'}} 
\psi_\L\( 0,\bs_p/v,\bz\) \,\, {s_i\over v_i} {s_{i'}\over v_{i'}}
= -\12 \sum_{i,i'\in I_g}  \mu_{\L,\bz}\(N_i;N_{i'}\) 
\,\, {s_i\over v_i} {s_{i'}\over v_{i'}}
+ R_{\L}^7 \( \bs,\bz\) 
$$
where
$$
\sup_{|\bs|\le K} \left| R_{\L}^7 \( \bs,\bz\) \right|
\le C {K^2 M \over \sqrt{|Q_\ell|}}
$$
The proof of \equ(expsi) is now complete.

\smallskip
\noi{\it Step 3.}\/ We finally here estimate from above the numerator on
the r.h.s. of \equ(inu). 

Let $K_\ell :=\log |Q_\ell|$. We make the change of variables
$\bt=\bs/v$ and use Lemma \thm[gaussbound] (which holds also for the
perturbed measure $\mu^u_{\L,\bz}$) to get
$$
\eqalign{
&
\left| 
\int_{\exists j : K_\ell \mi (\pi v_j) \le |s_j| \le \pi v_j } 
\! d\bs \,
e^{\psi_\L(\bs/v,\bz) -i \<\bs/v,\mu_{\L,\bz} \bN\> } 
\left[ e^{ \psi_\L^u(\bs/v,\bz) - \psi_\L(\bs/v,\bz) 
} -1 \right]
\right| 
\cr 
&~~\le 
\int_{\exists j : K_\ell \mi (\pi v_j) \le |s_j| \le \pi v_j } 
\! d\bs \,
\left[
 \left| e^{\psi_\L(\bs/v,\bz)} \right| 
+\left| e^{\psi_\L^u(\bs/v,\bz)} \right| 
\right]
\le C e^{-K_\ell^2 /C} \le C {1\over |Q_\ell|}
}
$$
We can thus consider the case $ |s_j|\le K_\ell \mi (\pi v_j)$. 

Since $ z_i |Q_\ell|/C \le v_i^2 \le  C z_i |Q_\ell|$, 
either $s_i/v_i$  or $z_i$ is small. We can
therefore apply the bound \equ(ds1b).
We get
$$
\left| P_\L^u(\{ e^{i s_j/v_j} z_j \}) -P_\L (\{ e^{i s_j/v_j} z_j \}) 
\right|
\le C 
$$

We next expand the difference $\psi_\L^u(\bs/v,\bz) - \psi_\L(\bs/v,\bz)$
in power series of $\bs$.
Since $\mu_{\L,\bz}^u\( N_k\) - \mu_{\L,\bz}\( N_k\)= 
\mu_{\L,\bz} \( f;N_k \)$, we get
$$
\psi_\L^u(\bs/v,\bz) - \psi_\L(\bs/v,\bz)
= i \sum_{k\in I} \mu_{\L,\bz} \( f;N_k \)  {s_{k} \over v_{k}} 
+ R^1_\L(\bs,\bz)
$$
where
$$
R^1_\L(\bs,\bz) = \12 \sum_{i,j\in I}
\left. {\p^2 \over \p t_i\p t_j} \left[
\psi_\L^u(\bt,\bz) - \psi_\L(\bt,\bz)\right] \right|_{\bt=\bar \bs/v}
{s_i s_j \over v_i v_j }
$$
We note that, by \equ(psi=P),
$$
{\p \over \p t_k} \left[\psi_\L^u(\bt,\bz) - \psi_\L(\bt,\bz)\right]
= iz_k e^{it_k} 
\left. {\p \over \p z'_j} \left[ P_\L^u(\bz') - P_\L(\bz')\right]
\right|_{z'_j = z_j e^{it_j}}
$$
By the analyticity of $P_\L^u(\bz') - P_\L(\bz')$, for $\bt=\bs/v$
we can bound the r.h.s. above by $C z_k$. We thus have
$$
\left| R^1_\L(\bs,\bz) \right|
\le C  \sum_{i,j\in I} z_i \mi z_j {s_i s_j \over v_i v_j }
\le C {|\bs|^2 \over |Q_\ell|}
$$
As $\left| \mu_{\L,\bz} \( f;N_j \)\right| \le C z_j$, 
for $|\bs|\le K_\ell$ we finally have
$$
\exp\left\{ \psi_\L^u(\bs/v,\bz) - \psi_\L(\bs/v,\bz) \right\}
-1=i\; \sum_{k\in I} \mu_{\L,\bz} \( f;N_k \)  {s_{k} \over v_{k}} 
+ R^2_\L(\bs,\bz), \quad\quad  
\left| R^2_\L(\bs,\bz) \right| \le  C {|\bs|^2 \over |Q_\ell|}
\Eq(epu-p)
$$

By Lemma \thm[gaussbound], we have
$$
\left| \int_{|s_j|\le K_\ell\mi (\pi v_j)} \!d \bs\,\,
e^{\psi_\L\(\bs/v,\bz\) - i \< \bs/v, \mu_{\L,\bz} (\bN) \>}
R^2_\L(\bs,\bz)
\right| \le C {1\over |Q_\ell|}
$$

To conclude the proof we consider separately each of the other terms
on the r.h.s of  \equ(epu-p). We want to show that, with a small
error, the function 
$\psi( \bs/v,\bz ) - i \< \bs/v ,  \mu_{\L,\bz} \( \bN \) \> $
is even in $s_k$; hence the integral vanishes by symmetry. 
We thus expand $\psi(\bs/v,\bz)$ as follows
$$
\psi( \bs/v,\bz ) - i \< \bs/v ,  \mu_{\L,\bz} \( \bN \) \>
= \12 \sum_{j,j'} {\p^2 \over \p t_j \p t_{j'}} \psi(\bar \bs/v,\bz)
{s_j s_{j'} \over v_j v_{j'}}
$$
by letting $\bs^{(k)} := \{s_i,\,\, i\in I\setminus\{k\}\}$, we have
$$
 {\p^2 \over \p t_j \p t_{j'}} \psi(\bar \bs/v,\bz) = 
- B_{j,j'} ( \bar \bs^{(k)}) + R^3_{\L,j,j'} (\bs,\bz)
,\quad\quad
B_{j,j'} ( \bar \bs^{(k)}) := - 
{\p^2 \over \p t_j \p t_{j'}} \psi(0,\bar \bs^{(k)}/v,\bz)
$$
and, by \equ(s1d) and the analyticity of $\psi_\L$,
$$
\left| R^3_{\L,j,j'} (\bs,\bz) \right| \le C z_j \mi z_{j'} |Q_\ell|
{s_k\over v_k}
$$
Whence
$$
\psi( \bs/v,\bz ) - i \< \bs/v ,  \mu_{\L,\bz} \( \bN \) \>
= - \12 \sum_{j,j'\in I} 
B_{j,j'} ( \bar \bs^{(k)}){s_j s_{j'} \over v_j v_{j'}}
+R^4_{\L,k} (\bs,\bz), \quad\quad 
\left| R^4_{\L,k} (\bs,\bz) \right| \le C { |s|^3 \over v_k}
\Eq(p=b+r)
$$

We next use the bound
$$
\left| e^{ R^4_{\L,k}(\bs,\bz) } -1 \right|
\le \( 1 + \left| e^{ R^4_{\L,k}(\bs,\bz) } \right| \)  
\left| R^4_{\L,k}(\bs,\bz)  \right|
$$
and \equ(p=b+r) to get
$$
\eqalign{
& \left| \int_{|s_j|\le K_\ell\mi (\pi v_j)} \!d \bs\,\,
e^{\psi_\L\(\bs/v,\bz\) - i \< \bs/v, \mu_{\L,\bz} (\bN) \>}
\mu_{\L,\bz} \( f;N_k \)  {s_{k} \over v_{k}} 
\right| 
\cr
&~~~\le 
\int_{|s_j|\le K_\ell\mi (\pi v_j)} \!d \bs\,\,
\( \left| e^{- \12 \sum_{j,j'\in I} B_{j,j'} (\bar \bs^{(k)}) 
{s_j s_{j'} \over v_j v_{j'}} } \right|
+ \left| e^{\psi_\L\(\bs/v,\bz\)} \right| \)
\left| R^5_{\L,k} (\bs,\bz) \right|
\cr
}
\Eq(<1+2)
$$
where, recalling that $ \left| \mu_{\L,\bz} \( f;N_k \) \right| 
\le C z_k$ and $v_k^2 \ge z_k |Q_\ell| / C$,
$$ 
R^5_{\L,k} (\bs,\bz) := \mu_{\L,\bz} \( f;N_k \)  
{s_{k} \over v_{k}}  R^4_{\L,k} (\bs,\bz)
,\quad\quad
\left| R^5_{\L,k} (\bs,\bz) \right| \le C {|\bs|^4 \over |Q_\ell|}
$$

By applying again Lemma \thm[gaussbound] we have
$$
\int_{|s_j|\le K_\ell\mi (\pi v_j)} \!d \bs\,\,
\left| e^{\psi_\L\(\bs/v,\bz\)} \right|
\left| R^5_{\L,k} (\bs,\bz) \right|
\le  C {1 \over |Q_\ell|}
$$

It now remains only to estimate the other term on the r.h.s. of
\equ(<1+2). Let $M_\ell := \ell^{1/4}$ and introduce 
$I^{(k)}_g := \{ i\in I\setminus\{k\} : v_i^2 \ge M_\ell\}$,
$I^{(k)}_p := I \setminus \( \{k\} \cup I^{(k)}_g \)$. We have
$$
B_{j,j'} ( \bar \bs^{(k)}) 
=- {\p^2 \over \p t_j \p t_{j'}} \psi(0,\bar \bs^{(k)}_p/v,\bz)
+  R^6_{\L,k,j,j'} (\bs,\bz)
$$
where
$$
\sup_{|\bs|\le K_\ell} \left| R^6_{\L,k,j,j'} (\bs,\bz) \right| 
\le C z_j \mi z_{j'} |Q_\ell| \sum_{i\in I^(k)_g} {|s_i|\over v_i}
\le C z_j \mi z_{j'} {K_\ell \over \sqrt{M_\ell} }
$$
so that, by using also \equ(ed2'),
$$
\sum_{j,j'\in I}  B_{j,j'} (\bar \bs^{(k)}) 
{s_j s_{j'} \over v_j v_{j'}}
= \sum_{j,j'\in I}  \mu_{\L,\bz} \( N_j,N_{j'}\)
{s_j s_{j'} \over v_j v_{j'}}
+  R^7_{\L,k} (\bs,\bz)
$$
where
$$
\sup_{|\bs|\le K_\ell} \left| R^7_{\L,k} (\bs,\bz) \right| 
\le C \( 
{K_\ell^3 \over \sqrt{M_\ell}} + {K_\ell^2 M_\ell \over |Q_\ell|}
\)
$$
Hence, recalling \equ(qfe),
$$
\int_{|s_j|\le K_\ell\mi (\pi v_j)} \!d \bs\,\,
\left| e^{- \12 \sum_{j,j'\in I} B_{j,j'} (\bar \bs^{(k)}) 
{s_j s_{j'} \over v_j v_{j'}} } \right|
\left| R^5_{\L,k} (\bs,\bz) \right|
\le C {1 \over |Q_\ell|}
$$
which concludes the proof. 
\qed

\bigskip
\noi{\it 4.3. Local Central Limit Theorem with multiplicative error}

\noi
In order to obtain the convergence of the short range part of the
renormalized potential to the one of independent harmonic oscillators we need
a local central limit theorem which will allow us to compute the
asymptotic behaviour (as $\ell\ten\infty$) of the r.h.s. of \equ(2.73/7).
Since we are interested in the logarithm of the partition function we
do need a local CLT in which the error appears in a multiplicative
way. It can be proven by applying the theory of moderate deviations
as developed in [DS4]; alhough  these results are stated only
for very high temperature, the proof is based only on the analyticity
properties of the thermodynamic functions which hold under 
Condition MUSM($\cA$).

Let us recall that $\mu^\t_{\L,\bz}$ is the
multi--grancanonical state in a volume $\L\su\su\cL$ of the form
\equ(La=). We denote by $v^{(\ell)} = v^{\t,(\ell)}(\bz)$ the
covariance matrix of the total number of particles in each cube
$Q_\ell(i)$, i.e. 
$v^{\t,(\ell)}(\bz)_{i,j} := \mu^\t_{\L,\bz} \(N_i;N_j \)$,
where $N_i$ has been defined in \equ(dNi). We have the following local
central limit theorem.

\nproclaim Theorem [lclt]. 
Let $U$ satisfy MUSM($\cA$) and 
$\br^{(\ell)}=\br^{\t,(\ell)}(\bz) := 
\mu^\t_{\L,\bz} \(\bN\) / |Q_\ell|$. For each $\L$ of
the form \equ(La=) and $\bz\in \cA$, $\e>0$ there are constants 
$\d=\d(\bz,I,\e)>0$, $C=C(\bz,I,\e)<\infty$ such that for any integer 
$\ell$ we have
$$
\eqalign{
\mu^\t_{\L,\bz} \(\bN=\bn \) = 
\left[ (2\pi)^{|I|} \det v^{(\ell)} \right]^{-\12}
& \exp\left\{ -\12 \< \( \bn - \br^{(\ell)} |Q_\ell|\) , 
\( v^{(\ell)} \)^{-1}\( \bn - \br^{(\ell)} |Q_\ell|\)\> \right\}
\cr
&\times \left\{1 + R_\L^\t(\bn) \right\}
\cr}
\Eq(elclt)
$$
where
$$
\sup_{\t\in\O}\; \sup_{\bn \,: \atop 
\left| \bn - \br^{(\ell)} |Q_\ell| \right| \le |Q_\ell|^{2/3 - \e}}
\; \left| R_\L^\t(\bn) \right| 
\le C {1 \over |Q_\ell|^\d}
\Eq(srlclt)
$$

This Theorem is essentially contained in [DS4]; however to
make the paper selfcontained we give below a brief sketch of the
proof.  Given $\bn$ we let  $\bzi = {\bzi}^{\tau, (\ell)} (\bn)$ be defined by 
$\bzi := \bz^{\tau, (\ell)} \( \bn / |Q_\ell| \)$ where we recall the
function $\br \mapsto \bz^{\tau, (\ell)} (\br)$ has been defined in 
Section 4.1. We also recall the pressure has been defined in
\equ(pressure). We have the following Lemma.

\nproclaim Lemma [llclt]. 
Under the same hypotheses of the previous Theorem, 
there are constants $\e_0=\e_0(\bz,I) >0$, $C=C(\bz,I,\e_0)<\infty$
such that 
$$
\mu^\t_{\L,\bz} \(\bN=\bn \) =
\left[ (2\pi)^{|I|} \det v^{(\ell)}(\bzi) \right]^{-\12}
\exp\left\{ -I_\L (\bn) \right\}
\(1 + \hat R_\L^\t(\bn) \)
\Eq(ellclt)
$$
where
$$
I_\L (\bn) = I_{\L,\bz}^\t (\bn) :=
\sum_{i\in I} n_i \log {\z_i \over z_i} - 
\left[ P_\L^\t (\bzi) - P_\L^\t (\bz) \right]
$$
and
$$
\sup_{\t\in\O}\; \sup_{\bn \,: \atop 
\left| \bn - \br^{(\ell)} |Q_\ell| \right| \le \e_0 |Q_\ell| }
\; \left| \hat R_\L^\t(\bn) \right| 
\le C {1 \over |Q_\ell|}
\Eq(srllclt)
$$

\noi{\it Sketch of the proof.}\/
By definition of the multi--grancanonical state $\mu_{\L,\bz}^\t$ we
have
$$
\mu^\t_{\L,\bz} \(\bN=\bn \) = 
\prod_{i\in I} 
\( {z_i \over \z_i} \)^{n_i} \cdot { Z^\t(\bzi) \over Z^\t(\bz) } 
\cdot \mu^\t_{\L,\bzi} \(\bN=\bn \) 
= e^{- I_\L (\bn)}  {1\over (2\pi)^{|I|}} 
\int_{|\bt|\le \pi } \!d \bt \; e^{-i \< \bt,\bn\> }
 \mu^\t_{\L,\bzi} \( e^{i \< \bt,\bN \> } \)
$$
If we take $\e$ small enough,
$\left| \bn - \br^{(\ell)} |Q_\ell| \right| \le \e |Q_\ell|$ 
implies that 
$(\bzi, U)$ satisfy SM($\ell_0$) for some $\ell_0=\ell_0
(\bz,\e_0)$. In order to conclude the proof it is then 
enough to make the change of
variables $t_i = s_i / \sqrt{v^{(\ell)}_{i,i}}$, use Lemma
\thm[gaussbound] to estimate the tail and expand 
$\log  \mu^\t_{\L,\bzi} \( e^{i \< \bt,\bN \>} \)$ up to the third
order, using analyticity to estimate the remainder (see Section 4.2
for analogous computations). 
Note in fact that, by the definition of $\bzi$ we have 
$\mu^\t_{\L,\bzi}\( \bN\) =\bn$. \qed

\medskip
\noi{\it Sketch of the proof of Theorem \thm[lclt].}\/
By applying Proposition \thm[thtr] we have
$$
\sup_{\t\in\O}\; \sup_{\bn \,: \atop 
\left| \bn - \br^{(\ell)} |Q_\ell| \right| \le |Q_\ell|^{2/3 - \e}}
\; \left\| v^{(\ell)}(\bz) - v^{(\ell)}(\bzi) \right\| 
\le C |Q_\ell|^{2/3 - \e}
$$
which, together with the bound \equ(qfe), implies
$$
\( \det v^{(\ell)}(\bzi) \)^{-\12} = \( \det v^{(\ell)}(\bz) \)^{-\12}
\( 1 + R^{\t,(1)}_\L (\bn)\)
$$
where
$$
\sup_{\t\in\O}\; \sup_{\bn \,: \atop 
\left| \bn - \br^{(\ell)} |Q_\ell| \right| \le |Q_\ell|^{2/3 - \e}}
\left| R^{\t,(1)}_\L (\bn) \right| \le C { 1 \over |Q_\ell|^{1/3}}
$$

On the other hand, by the analyticity (uniform in $\ell$) of the
thermoduynamic functions, we have (see [DS4 Eq. 1.2.15])
$$
I_\L (\bn) = \12 \<  \( \bn - \br^{(\ell)} |Q_\ell| \),
 \( v^{(\ell)} (\bz) \)^{-1} \( \bn - \br^{(\ell)} |Q_\ell| \) \>
\( 1+ R^{\t,(2)}_\L (\bn)\)
$$
where
$$
\sup_{\t\in\O}\; \sup_{\bn \,: \atop 
\left| \bn - \br^{(\ell)} |Q_\ell| \right| \le |Q_\ell|^{2/3 - \e}}
\left| R^{\t,(2)}_\L (\bn) \right| \le C { 1 \over |Q_\ell|^{3\e}}
$$
in which we have used again that $\mu^\t_{\L,\bzi}\( \bN\) =\bn$. \qed

\newsection Gibbsianess and convergence

In this section we conclude the proof of the main results.
First, by applying the comparison of ensembles, we show the
constrained models satisfy a finite size effective condition
uniformly in the constraints.
Secondly, by applying the local central limit theorem, we prove 
the short range part of the renormalized potential converges to
the potential of independent harmonic oscillators.
Finally, when the global condition GMUSM holds, we verify that the
renormalized measure $\mu^{ (\ell) }_z$ (defined directly in infinite 
volume) is Gibbs w.r.t. the potential constructed in Section 3
(obtained via a thermodynamic limit).

\bigskip
\noindent {\it 5.1. Finite size condition for the constrained models}

\noi
We consider the BAT obtained by partitioning the original lattice 
$\cL$ into cubes of side $\ell$, $\cL = \cup_{i\in \cL_\ell} Q_\ell(i)$.
Let $\mu_z$ be the (infinite volume) Gibbs state of
the {\it original} system at activity $z$. We then introduce the
{\it constrained} system by fixing the total number of particles in
each cube; it is described by the conditional (multi--canonical) measure 
we introduced in the previous section.

We want to show that, provided Condition MUSM($\cA$) is satisfied, 
the local specification associated to the 
multi--canonical state $\nu_{\L,\bn}^\t$ satisfies \equ(CC) 
with $\delta(\ell)=C/\ell $.
We shall consider $\ell$ to be an integer multiple of $\ell_0$.
Recall that $\cB=\r (\cA)$, $L=d\ell$ and $\cD^{(\ell)}_{\hat\Lambda}=
(|Q_\ell|\cD)^{\hat\Lambda}\cap\Omega^{(\ell)}_{\hat\Lambda}$
(see Theorem \thm[summa]).

\nproclaim Proposition [fsbat]. 
Assume the interaction $U$ satisfies MUSM($\cA$). Then
for each closed set $\cD \subseteq \cB $ 
there is a constant $C$ such that for all $L$ the
following bound holds. 
$$
\sup_{i\in\cL_\ell} ~~
\sup_{k=1,\cdots,d} ~~
\sup_{\L\in P^{(k)}_{L}(i)}~~  
\sup_{\bn\in\cD^{(\ell)}_{\hat\Lambda}}~~ 
\sup_{\s,\z,\t}
\left| { Z_{\L,\bn}\(\s^{(k,+)},\s^{(k,-)},\t\) 
         Z_{\L,\bn}\(\z^{(k,+)},\z^{(k,-)},\t\) 
\over    Z_{\L,\bn}\(\s^{(k,+)},\z^{(k,-)},\t\)  
         Z_{\L,\bn}\(\z^{(k,+)},\s^{(k,-)},\t\)
}
-1\right|
\le {C\over \ell}
\Eq(efsbat)
$$

\nproclaim Lemma [fsbatin].
In the same setting and notation of the above theorem, there is a
constant $C$ such that for any $\D\su\L$ for which 
$d\(\D,\partial^{(k,-)}\Lambda\)\le r$, $\diam (\D) \le r$  
$$
\sup_{i\in\cL_\ell} ~~
\sup_{k=1,\cdots,d} ~~
\sup_{\L\in P^{(k)}_{L}(i)}~~  
\sup_{\bn\in\cD^{(\ell)}_{\hat\Lambda}}~~ 
\sup_{\s,\z,\t}
\Var\( \nu_{\L,\bn; \D}^{\s^{(k,+)},\t,\t} , 
\nu_{\L,\bn; \D}^{\z^{(k,+)},\t,\t} \) \le C { 1 \over |Q_\ell|}
\Eq(efsbatin)
$$

Postponing the proof of the Lemma, we show how it implies the main estimate.

\smallskip
\noi{\it Proof of Proposition \thm[fsbat].}\/
Let us first show that \equ(efsbatin) implies the following 
condition
$$
\sup_{i\in\cL_\ell} ~~
\sup_{k=1,\cdots,d} ~~
\sup_{\L\in P^{(k)}_{L}(i)}~~  
\sup_{x\in \partial^{(k,-)}\Lambda}~~
\sup_{\bn\in\cD^{(\ell)}_{\hat\Lambda}}~~ 
\sup_{\s,\z,\t}
\left| { Z_{\L,\bn}\(\s^{(k,+)},\t^x,\t \) 
Z_{\L,\bn}\(\z^{(k,+)},\t,\t \) 
\over  Z_{\L,\bn}\(\z^{(k,+)},\t^x,\t \) 
Z_{\L,\bn}\(\s^{(k,+)},\t,\t \) 
}
-1\right|
 \le C { 1 \over |Q_\ell|}
\Eq(fsbatin')
$$
We have in fact 
$$
\eqalign{
&{ Z_{\L,\bn}\(\s^{(k,+)},\t^x,\t \) 
Z_{\L,\bn}\(\z^{(k,+)},\t,\t \) 
\over  Z_{\L,\bn}\(\z^{(k,+)},\t^x,\t \) 
Z_{\L,\bn}\(\s^{(k,+)},\t,\t \) } 
-1
\cr
&~~~~=
{ Z_{\L,\bn}\(\s^{(k,+)},\t^x,\t \) 
\over  Z_{\L,\bn}\(\s^{(k,+)},\t,\t \) }
\left[ { Z_{\L,\bn}\(\z^{(k,+)},\t,\t \) 
\over  Z_{\L,\bn}\(\z^{(k,+)},\t^x,\t \) }
-
{ Z_{\L,\bn}\(\s^{(k,+)},\t,\t \) 
\over  Z_{\L,\bn}\(\s^{(k,+)},\t^x,\t \) 
}
\right]
\cr
&~~~~=
{ Z_{\L,\bn}\(\s^{(k,+)},\t^x,\t \) 
\over  Z_{\L,\bn}\(\s^{(k,+)},\t,\t \) }
\left[ \nu_{\L,\bn}^{\z^{(k,+)},\t^x,\t} \( h_x^\t \) -
\nu_{\L,\bn}^{\s^{(k,+)},\t^x,\t} \(h_x^\t\)
\right]
}
\Eq(r>d)
$$
where
$$
h_x^\t (\h) := e^{ - 
\left[ H_\L\( \h\circ_\L \t^x\) - H_\L\( \h \circ_\L \t\) \right] }
$$
is a local function with support contained in an $r$ neighborhood of
$x$. Since the first factor on the r.h.s. of \equ(r>d) is bounded
uniformly and the same holds for $\| h_x^\t \|$, \equ(fsbatin')
follows from \equ(efsbatin).
\par
An easy telescopic argument shows \equ(fsbatin') implies 
\equ(efsbat). Indeed, for any two
configurations $\z^{(k,-)}$,  $\s^{(k,-)}$, differing only on 
$\partial^{(k,-)}\Lambda$,
we can find a path $\{\h_l\}_{l=0,\cdots,M}$ of
length $M \le r\cdot (3L)^{d-1}$ such that $\h_0=\s^{(k,-)}$,
$\h_M=\z^{(k,-)}$ and $\h_{l}$ differs from  $\h_{l-1}$ at most
in one single site $x\in\partial^{(k,-)}\Lambda$. We
then write 
$$
\eqalign{
&{ Z_{\L,\bn} \(\s^{(k,+)},\s^{(k,-)},\t\)  
Z_{\L,\bn}\( \z^{(k,+)},\z^{(k,-)},\t\) 
\over  Z_{\L,\bn}\(\s^{(k,+)},\z^{(k,-)},\t\)  
Z_{\L,\bn}\(\z^{(k,+)},\s^{(k,-)},\t\)
}
\cr
&~~~~
= \prod_{l=1}^M 
{ Z_{\L,\bn} \(\s^{(k,+)},\h_{l-1},\t\)  
Z_{\L,\bn}\( \z^{(k,+)},\h_l,\t\) 
\over  Z_{\L,\bn}\(\s^{(k,+)},\h_{l},\t\)  
Z_{\L,\bn}\(\z^{(k,+)},\h_{l-1},\t\)
}
}
$$
and use \equ(fsbatin') to get  \equ(efsbat).
\qed

\medskip
\noi{\it Proof of Lemma \thm[fsbatin].}\/
Let us recall that $\Var\(\mu,\nu\)= \sup_{\|f\|=1} | \mu f - \nu f|$.
Let $f$ be a local function with support contained in $\D$.
By Theorem \thm[eqen'] we have
$$
\left| \nu_{\L,\bn}^{\z_1} f - \nu_{\L,\bn}^{\z_2} f
\right| \le C {1\over |Q_\ell|} + 
\left| \mu_{\L,\bz^1}^{\z_1} f - \mu_{\L,\bz^2}^{\z_2} f 
\right|
$$
where $\bz^{\alpha}=\bz^{\alpha}(\L,\bn,\z_{\alpha})$, 
$\alpha=1,2$ is chosen so that
$\mu_{\L,\bz^{\alpha}}^{\z_{\alpha}} 
\(\bN \)= \bn$. Since $\z_1$ differs from $\z_2$
only on  $\partial^{(k,+)}\Lambda$, by Condition MUSM($\cA$) we now have
$$
\left| \mu_{\L,\bz^1}^{\z_1} f - \mu_{\L,\bz^1}^{\z_2} f 
\right| \le C e^{-d\(\D,\partial^{(k,+)}\Lambda\)/C}
$$
On the other hand, by Lagrange theorem,
for a suitable $\bar\bz$,
$$
\left| \mu_{\L,\bz^1}^{\z_2} f - \mu_{\L,\bz^2}^{\z_2} f 
\right| \le 
\sum_{i\in\hat\Lambda} {1\over \bar z_i} 
\left| \mu_{\L,\bar\bz}^{\z_2} \(f;N_i\) \right|
\cdot \left| z^2_i -z^1_i \right|
$$

By the exponential decay of correlations we have
$$
\left| \mu_{\L,\bar\bz}^{\z_2} \(f;N_i\) \right|
\le C \bar z_i e^{- d\(\D, Q_\ell(i)\) / C}
$$
the bound \equ(efsbatin) is thus obtained by applying Proposition
\thm[thtr'] to estimate $\left| z^2_i -z^1_i \right|$. Note in fact
that $d\(\D,\partial^{(k,+)}\Lambda \) \ge d\ell-r$.
\qed

\bigskip
\noi{\it 5.2. Short range renormalized potential.}

\noi
In this section we consider the limit $\ell\rightarrow\infty$ of the
short range part of the renormalized potential. By applying  
Theorem \thm[lclt], we prove the necessary estimates.
This would also allow us to conclude the proof of Theorem \thm[teol].

\nproclaim Proposition [sril].
Recall that the short range part of the renormalized potential 
$\Phi^{(\ell),sr}_{ X}$ have been defined in \equ(2.73/7).
We introduce
$$
\Psi^{(\ell),sr}_{ X}(m_X):=
S(X) {1\over 2} \sum _{i\in X} m_i^2  
+\F^{(\ell),sr}_{ X} (m_X)  
\Eq (2.79)
$$
where 
$$
S(X):= 
\left\{\eqalign{ 
+1& {\rm \;if} \;X= \widetilde A_{k_1}, D_{k_4}\cr 
-1& {\rm \;if} \;X= \widetilde C_{k_3}, F_{k_2}\cr 
0&  {\rm \;otherwise}\cr 
}\right.
\Eq (2.80)
$$
Then the renormalized Hamiltonian can be written as
$$
H^{(\ell,\t)}_{I_p} (\underline {n}) = -{1\over 2} 
\sum _{i\in I_p} m_i^2 + \sum_{X\su I_p} \Psi^{(\ell),sr}_{ X} (m_X)
+ \sum_{X\su I_p} \F^{(\ell,\t),lr}_{ X} (m_X)
\Eq (2.83)
$$
Moreover there is a constant $a>0$ such that 
$$
\lim_{\ell\ten\infty} 
\sup_{m_X \in \bar \O^{(\ell)}_X \atop |m_X| \le \ell^a}
\left| \Psi^{(\ell),sr}_X (m_X) \right| =0 \quad  \quad
\quad\quad {\rm for ~any}~~ X\su\su \cL_\ell, ~|X|\ge 2
\Eq(e:sril)
$$

Note that Theorem \thm[teol] follows directly from Theorem \thm[summa] and 
Propositions \thm[fsbat] and \thm[sril].

\medskip
\noi{\it Proof of Proposition  \thm[sril].}\/
By using \equ(elclt), for each $V = \cup _{i\in X} Q_{\ell}(i)$, 
recalling that $m_i = ( n_i -\rho |Q_\ell| ) / \sqrt{ \chi |Q_\ell|}$,
we have
$$
\eqalign{
\log \m^{\t}_{V,z} \( {M}_i = {m}_i, \; i \in X \) 
= & 
{\rm const \;} -
{1\over 2} \sum _{i\in X} m_i^2
\cr
&-\left\{ 
\12 \< \( \bn - \br^{(\ell)} |Q_\ell|\) , 
\( v^{(\ell)} \)^{-1}\( \bn - \br^{(\ell)} |Q_\ell|\)\>
-{1\over 2} \sum _{i\in X} m_i^2 \right\}
\cr
&+ \log \left[ 1 + R^{\t}_{V} ({m}_X) \right]
}
\Eq (2.82)
$$
Therefore, by \equ(2.73/7) (where the boundary condition is $\t=0$ and
$d_{\ell}(X,I_p^c)>d$), we have
$$
\eqalign{
\Psi^{(\ell),sr}_{ X} (m_X) 
& = - S(X) 
\left \{ 
\12 \< \( \bn - \br^{(\ell)} |Q_\ell|\) , 
\( v^{(\ell)} \)^{-1}\( \bn - \br^{(\ell)} |Q_\ell|\)\>
-{1\over 2} \sum _{i\in X} m_i^2 \right\}
\cr
&+ \log \left[ 1 + R^{0}_{V} ({m}_X) \right]
}
\Eq (2.83')
$$
Indeed, to get \equ (2.83), it is sufficient to observe that 
\item{{\it i)}} given a block $Q_{\ell}(i)$ contained
in $\cA_p$, the corresponding one--body renormalized 
interaction $-{1\over 2} m_i^2$ appears only 
in one term $\F^{(\ell),sr}_{ X} (m_X)$ with $X = \widetilde A_{k_1}$ for one and only one 
$A_{k_1}\in \cA_p$ with $S(X) = +1$
\item{{\it ii)}}
given a block $Q_{\ell}(i)$ contained
in $\cB_p$, the corresponding one--body renormalized interaction $- {1\over 2} m_i^2$ appears 
in two terms $\F^{(\ell),sr}_{ X} (m_X)$ with $X = \widetilde A_{k_1}$ with $S(X) = +1$
and in one term with  
$X = F_{k_2}$ with $B_{k_2}\in \cB_p$ and $S(X) = -1$
\item{{\it iii)}}
given a block $Q_{\ell}(i)$ contained
in $\cC_p$, the corresponding one--body renormalized interaction $- {1\over 2} m_i^2$ appears 
in four terms $\F^{(\ell),sr}_{ X} (m_X)$ with $X = \widetilde A_{k_1}$ with $S(X) = +1$, 
in two
terms with  
$X = F_{k_2}$ with $B_{k_2}\in \cB_p$ and $S(X) = -1$ and in one term 
$X = \widetilde C_{k_3}$ with $C_{k_3}\in \cC_p$ and  $S(X) = -1$
\item{{\it iv)}}
given a block $Q_{\ell}(i)$ contained
in $\cD_p$, the corresponding one--body renormalized interaction 
$- {1\over 2} m_i^2$ appears  in two terms $\F^{(\ell),sr}_{ X} (m_X)$ 
with $X = \widetilde A_{k_1}$ with $S(X) = +1$,
 in two terms with  $X = \widetilde C_{k_3}$ with 
$C_{k_3}\in \cC_p$ and $S(X) = -1$ and in one term 
$X = D_{k_4}$ with $D_{k_4}\in \cD_p$ and $S(X) = +1$

\noindent
Performing the different cancellations in the four sub--lattices 
$\cA_p,\cB_p,\cC_p,\cD_p$ we easily get \equ (2.83).

Finally, to prove \equ(e:sril), we note that by 
Proposition \thm[thtr'] we have
$$
\left| |Q_\ell| \( v^{(\ell)} \)^{-1}_{i,j} - \d_{i,j} \right|
\le {C \over \ell}
$$
and, by strong mixing,  
$$
\left| \rho^{(\ell)}_i (\bz) -\rho (z_i)\right| \le C \ell^{-1}
$$
Hence the bound \equ(e:sril) follows from \equ(2.83') and 
Theorem \thm[lclt].
\qed

\bigskip
\noi{\it 5.3. Gibbsianess of renormalized potential}

We show here that, provided Condition GMUSM holds and $\ell$ is 
large enough, the renormalized measure $\mu^{(\ell)}$ is Gibbsian
w.r.t. the potential $\Phi^{(\ell)}$ which has been constructed in
Section 3. We have in fact the following result.

\nproclaim Proposition [gmul].
Assume Condition GMUSM holds and define 
the renormalized potential $\Phi^{(\ell)}$ as in Section 3. 
Then the renormalized measure $\mu^{(\ell)}_z$ is Gibbsian w.r.t. 
$\Phi^{(\ell)}$, i.e.
$$
\mu^{(\ell)}_{z} \( 
\left. m_I \right| m_{I^c} \)
= 
{
\exp\left\{
\sum_{X \cap I\neq \emptyset } 
\Phi^{(\ell)}_{X} \( m_I \circ m_{I^c} \) \right\}
\over
\sum_{m_I \in \bar \O^{(\ell)}_{I}}
\exp\left\{
\sum_{X \cap I\neq \emptyset } 
\Phi^{(\ell)}_{X} \( m_I \circ m_{I^c} \) \right\}
}
,\quad
\mu_z^{(\ell)}\; {\rm a.s.}  
\Eq(lsmul)
$$

Note that Theorem \thm[teog] follows directly from Theorem \thm[summa] and 
Propositions \thm[fsbat], \thm[sril] and \thm[gmul].
Indeed GMUSM implies $\cB = \rho ( [0,\infty) ) = [0,1]$

\medskip
\noi
{\it Proof of Proposition \thm[gmul].}\/
We recall the random variables $M_i=M_i(\h)$ have been defined in 
\equ(Centr). We introduce the two families of $\s$--algebras:  
$\cF_\L := \s\{ \h_x, \, x\in \L\}$, $\Lambda \su \cL$,
and $\cF^{(\ell)}_I := \s\{ M_i,\, i\in I\}$, $I \su \cL_\ell$.
For $I \su\su \cL_\ell $ and 
$F : \bar \O^{(\ell)}_I \mapsto \bR$ let us first prove that
$$
\mu^{(\ell)}_z \( \left. F(m_I) \right| \cF^{(\ell)}_{I^c} \)
=\mu_z \( \left. F(M_I) \right| \cF^{(\ell)}_{I^c} \)
,\quad\quad \mu_z^{(\ell)} ~ {\rm a.s.}
\Eq (mul=mu)
$$
let $G$ be a local function measurable w.r.t. 
$\cF^{(\ell)}_{I^c}$; by definition of the measure
$\mu^{(\ell)}_z$ we have
$$
\mu_z \( F(M_I) G(M_{I^c}) \) = 
\mu^{(\ell)}_z \( F(m_I) G(m_{I^c}) \)
= \int \! d \mu^{(\ell)}_z (\bm) \,\, G(m_{I^c}) 
\mu_z^{(\ell)}\( \left. F(m_I) \right| \cF^{(\ell)}_{I^c} \)
$$
on the other hand,
$$
\eqalign{
\mu_z \( F(M_I) G(M_{I^c}) \) 
& = \int\! d\mu_z (\h) \,\, G \( M_{I^c}(\h) \)
\mu_z\( \left. F\( M_I(\h) \) \right| \cF^{(\ell)}_{I^c} \)
\cr
& = \int\! d\mu_z^{(\ell)} (\bm) \,\, G \( m_{I^c}\)
\mu_z\( \left. F\( M_I(\h) \) \right| \cF^{(\ell)}_{I^c} \)
}
$$
which proves \equ(mul=mu).

Let $V =\cup_{i\in \hat V} Q_\ell(i) \su\su \cL$; 
we note that for $I\su \hat V$ we have
$$
\eqalign{
\mu_z \( \left. M_I=m_I \right| \cF^{(\ell)}_{I^c} \)
&=\mu_z \(  \left. 
\mu_z\( \left. M_I = m_I \right| \cF_{V^c} \vee \cF^{(\ell)}_{I^c}  \)
\right| \cF^{(\ell)}_{I^c} \)
\cr
&=\mu_z \(  \left. 
\mu_z\( \left. M_I=m_I \right| 
\cF^{(\ell)}_{\hat V \setminus I}\vee \cF_{V^c} \)
\right| \cF^{(\ell)}_{I^c} \)
}
\Eq (mul=muL)
$$
on the other hand, by definition of the renormalized 
Hamiltonian and the corresponding potential, see Section 3
$$
\mu_{V,z}^\t \( \left. M_I = m_I  \right| 
M_{\hat V\setminus I} =  m_{\hat V\setminus I}\)
=
{
\exp\left\{
\sum_{X\su \hat V \atop X \cap I\neq \emptyset } 
\Phi^{(\ell,\tau)}_{X} \( m_I \circ m_{\hat V \setminus I} \) \right\}
\over
\sum_{m_I \in \bar \O^{(\ell)}_{I}}
\exp\left\{
\sum_{X\su \hat V \atop X \cap I\neq \emptyset } 
\Phi^{(\ell,\tau)}_{X} \( m_I \circ m_{\hat V \setminus I} \) \right\}
}
\Eq(spevolfin) 
$$
Since Condition GMUSM holds, by Proposition \thm[fsbat],
\equ(CC) is satisfied with $\cD =[0,1]$ and therefore, 
by Theorem \thm[summa], the r.h.s. of \equ(spevolfin) converges, 
as $V\uparrow \cL$, to the r.h.s. of \equ(lsmul) uniformly in $\t$ and $\bm$. 
By using also \equ(mul=muL) and \equ(mul=mu) we thus conclude 
the proof.
\qed

\bigskip
\expandafter\ifx\csname sezioniseparate\endcsname\relax%
\input macro \fi
\numsec=-1              
\numfor=1\numtheo=1\pgn=1
\leftline{\bf A.1 Proof of USM(${\cal A}$)
$\Longrightarrow$ MUSM(${\cal A}$) in dimension 2.}
\par
Let $R_{L,  3 L}(i)$ be the rectangle with 
vertical and horizontal sides  $L,3L$,
respectively, and which is centered at $Q_{(L)}(i)$. 
\par
The fact that we only consider this rectangle  
with longer horizontal side  does not represent, of course, 
a loss of generality and is made only to fix notation.\par
For $M$ an even integer, $M/2$ and $L_0$ odd integers, we write: 
$$
\bar L = M L_0; \;\;\;\;\; 
R_{\bar L,  3 \bar L} =R_{\bar L,  3 \bar L}(({L_0 - 1 \over 2 }, {L_0 - 1\over 2 }));
$$
again the choice of the center is made to fix notation and does not constitute a loss of
generality. Recall that since $M$ is even and $L_0$ is odd the center of 
$Q_{\bar L}(({L_0 - 1 \over 2 }, {L_0 - 1\over 2 }))$ is in 
$({L_0  \over 2 }, {L_0 \over 2 })$.
\par
We set $R_{\bar L, 3 \bar L}$ = $Q^l_{\bar L}\cup Q^c_{\bar L}\cup Q^r_{\bar L}$ where by 
$Q^l_{\bar L}$, $Q^c_{\bar L}$, $Q^r_{\bar L}$ we denote the left, central and right $\bar L\times \bar L$
squares, respectively, contained in $R_{\bar L, 3 \bar L}$.\par
Consider a 2D lattice gas with an interaction satisfying USM ($\cA$) for some $\cA \subseteq
 [0,\infty)$.
 We start noticing that from the validity
of USM($\cA$) it is immediate to deduce  that for each $z \, \in\, \cA$ there exists an
integer $L_0$  such that the following condition 
$$
\sup_{\s,\t\in \O}
\;\sup_{i\in \{ 1,2\} }
\left |{  Z_V \( \s^{(i,+)},\s^{(i,-)},\t \)
Z_V\(\t^{(i,+)},\t^{(i,-)},\t\) \over
Z_V\( \s^{(i,+)},\t^{(i,-)},\t \) Z_V \( \t^{(i,+)},\s^{(i,-)},\t \) }
-1 \right|
<\e(2)
\Eq (C7)
$$
is verified for $V = $ 
$Q_{L_0}(i)$, $R_{L_0,  3 L_0}(i)$ in the homogeneous activity case. This, together with the
results of [O], [OP] establishes the equivalence of USM 
and C1 in the homogeneous activity
case; this result is valid in any dimension.
\par
Now, given a closed set $\cC \subseteq \cA$ suppose that we are able to prove the existence
of $\bar L$ such that: for all $z,z' \,\in\, \cC$, if we consider our lattice gas
enclosed in 
$V = R_{\bar L, 3 \bar L}$ with activity $z' $ in $Q^l_{\bar L}$ and  $z$ in 
$ Q^c_{\bar L}\cup Q^r_{\bar L}$ (i.e. we take the same activity both in 
$ Q^c_{\bar L}$ and $ Q^r_{\bar L}$), then,
 calling
$Z_{V,z,z'}(\t)$ the corresponding partition function with $\t$ boundary condition, we have:
$$
\sup_{\s,\t} \;
 \; \sup_{i\in \{ 1,2\} }
\; \sup _{ {y\in \de ^{(i,+)} V \atop  y' \in \de ^{(i,-)} V}}
\left |{  Z_{V,z,z'}(\s_y,\s_{y'},\t ) Z_{V,z,z'}(\t_{y},\t_{y'},\t)\over
Z_{V,z,z'}(\s_y,\t_{y'},\t )Z_{V,z,z'}(\t_{y},\s_{y'},\t )} -1\right|
<{\e(\bar L)\over \bar L^{2(d-1)}}
\Eq (C4)
$$
with $\e(\bar L)$ going to $0$ as $\bar L$ goes to infinity; 
then, using methods and results of [O], [OP] it is easy to get MUSM($\cA$).
Indeed in the two--dimensional, multi--grancanonical case, to get strong mixing condition using
effectiveness of some finite--size conditions for volumes of the form \equ (La=) with $\ell$
sufficiently large, it is sufficient to verify: 
\item{{\it (i)}} \equ (C4) for $V = Q_{\bar L} (i)$ and $V = R_{\bar L, 3 \bar L}(i)$
with {\it uniform} activity in $V$ arbitrarily chosen in $\cC$, and
\item{{\it (ii)}} \equ (C4) for $V = R_{\bar L, 3 \bar L}(i)$ and activity
$z' $ in $Q^l_{\bar L}$ and  $z$ in 
$ Q^c_{\bar L}\cup Q^r_{\bar L}$ uniformly for $z,z'$ in $\cC$.\par
In the homogeneous case {\it (i)}, as we noticed before, if, given $\cC$, $L_0$ is the size
for which SM($L_0$) holds uniformly in $\cC$, as prescribed by USM($\cA$), then, for $\bar
L$ sufficiently large \equ (C4) holds for $V = Q_{\bar L} (i)$ and $V = R_{\bar L, 3 \bar
L}(i)$ for each (constant in $V$) activity $z \in \cC$.
Then \thm[moslike] will follow from next Proposition A1.1

\nproclaim Proposition [1].
Suppose that  Condition
${\rm C1}^{(2)}(V)$
holds  for any $V= Q_{L_0} (i), R_{L_0,3L_0}(i)$ 
contained in one of the three squares
$Q^l_{\bar L}$, $Q^c_{\bar L}$ or $Q^r_{\bar L}$;
then, for $M\equiv {\bar L \over L_0}$ sufficiently large,
 \equ (C4) holds for $V=R_{\bar L, 3 \bar L}$.

\noi{\it Proof.}

We make a geometrical construction similar to the one introduced in
[O], [OP] and used in Section 3 to 
compute, via cluster expansion, the renormalized potential.
We recall that we denote by 
 $\cL$ our original lattice $\bZ^2$  whereas we denote
 by $ \cL_{L_0}$ the
$L_0$--rescaled lattice: 
we partition $\cL$ into cubes of side $L_0$. We write:
$$
\cL = \cup _{i\in  \cL_{L_0}} \;\;Q_{L_0}(i)
$$
 From now on we will mainly consider the $L_0$--rescaled lattice; our
unit length will be 
$L_0$. In other words we will use the distance $d_{L_0}$.  The
``bricks" of our construction will be  the blocks
$Q_{L_0}$ or  
$R_{ L_0, 3 L_0}$ and the original length--scale will enter only when
considering some 
properties of the partition functions in the regions $ Q_{L_0}$ or  
$R_{ L_0, 3 L_0}$ that we use as input of our perturbative theory.
\par
Let
$e_1,e_2$ denote, respectively, the horizontal and vertical lattice
unit vectors in $\cL_{L_0}$: 
$e_1= (1,0), e_2 =(0,1)$. Following definitions and notation of Section 3
we further partition $\cL_{L_0}$ into four
sub--lattices:
$$
\cL_{L_0} =\cL_{2L_0}^A\cup \cL_{2L_0}^B\cup\cL_{2L_0}^C\cup\cL_{2L_0}^D
$$
where:
$$
\eqalign{
\cL_{2L_0}^A & := \{ i= (i_1,i_2)\, \in \, \cL_{L_0} : i_1 = 2 j_1,x_2 = 2 j_2,\hbox 
{for some integers}\;\; y_1,y_2\}
\cr
\cL_{2L_0}^B &:= \cL_{2L_0}^A + e_2
\cr
\cL_{2L_0}^C &:= \cL_{2L_0}^A + e_1 + e_2 = \cL_{2L_0}^B + e_2 
\cr
\cL_{2L_0}^D &:= \cL_{2L_0}^A + e_1 = \cL_{2L_0}^C + e_2 = \cL_{2L_0}^B + e_1 + e_2
\cr
}
\Eq (4.2)
$$

We also set, for $i \in \cL _{L_0}$:
$$
A_i := Q_{L_0}(2i),\;\;\; 
B_i := Q_{L_0}(2i +e_2)\;\;\;
C_i := Q_{L_0}(2i+e_1 +e_2)\;\;\;
D_i := Q_{L_0}(2i+e_1).
\Eq (4.3)
$$

Then we can partition
$V\equiv R_{\bar L, 3
\bar L}$ into the union of the
$L_0$--blocks of the four types:
$A, B, C, D$:
$$
V = \cA_V \cup \cB_V \cup \cC_V \cup \cD_V
$$
where
$$
\cA_V := 
\{ A_i: i = (i_1,i_2) \,\in \cL_{L_0} \,:\, |i_2| \leq ({M/2 } -1)/2
\, ,\, |i_1| \leq  (3 {M/2 } -1)/2 
\}
$$
and similarly for $\cB_V,\cC_V,\cD_V$.\par
We have that the left  block on the bottom is an $A$--block whereas the right 
one on the top is a $C$--block.\par

We denote
by $\a_i$ a generic  spin configuration in $A_i$: $\a_i \in \{-1,+1\} ^{L_0^2}$. Similarly for
$\b_i,\g_i,
\d_i$. We simply denote by $\a,\b,\g,\d$ the configurations in $\cA_V,\cB_V,\cC_V,\cD_V$,
respectively.\par 
Notice that we have used the same notation (with a very similar meaning) as the one we used in
Section 3 to describe ``multi--canonical" block variables.

Consider the ``column" $V_l$ namely the rectangle with basis $L_0$ and height $\bar L$ placed
at the left--hand of 
$Q^c_{\bar L}$ ,  adjacent, from the exterior, to $Q^l_{\bar L}$:
$$
V_l = \{(x_1,x_2) \in \cL:  -{\bar L\over2} +{L_0+1\over2}\leq  x_1 \leq
-{\bar L\over2} +{L_0+1\over2} + L_0,\;\;
-{\bar L\over2} +{L_0+1\over2}\leq  x_2 \leq 
+{\bar L\over2} +{L_0-1\over2}
\}
$$
we decompose $V_l $ as disjoint union of $A$ and $B$ blocks:
$$
V_l = \cA_l \cup \cB_l
$$
where
$$
 \cA_l :=  \cA_V \cap V_l,\;\;\;\; \cB_l :=  \cB_V \cap V_l 
$$

We have: 
$$
\cA_l = \cup _{i \in I^l_A} A_i
$$
where 
$$
 I^l_A= \{ (i_1,i_2) \; : \; i_1 = -({M/2}-1)/2, \; |i_2| \leq
({M\over 2}-1)/2\}
$$
similarly for $\cB_l$
 
We write $\cA _V = \hat \cA \cup \cA _l$, $ \cB_V = \hat \cB \cup \cB _l$; in
other words $\hat \cA,\hat \cB$ denote the union of $A$ and $B$ blocks, respectively, which belong to 
$V = R_{\bar L, 3 \bar L}$ but not to $V _l $.\par

We will repeat almost the same computation that we made , in the multi--canonical framework, to compute the
renormalized potential. 
Namely we adopt the same strategy based on a block decimation procedure over the sequence of sub--lattices
$D,C,B,A$.\par
The main difference here is that we will treat in a different manner the region
in $V \equiv R_{\bar L, 3 \bar L}$ adjacent to the boundary between 
 $Q^c_{\bar L}$ and $Q^l_{\bar L}$. Here we will exploit the fact that this
boundary is one--dimensional.\par

Indeed we will see that the system of the surviving $\a$--variables in
$\cA_l$, after decimation 
on $\d,\g,\b$ and $\a$ in $\hat \cA$, gets an effective interaction which is
exponentially decaying with the distance and uniformly bounded in
norm. The resulting 
one--dimensional system, regarded on a sufficiently large
scale, is in the weak coupling region and from this it easily follows
a weak coupling between opposite horizontal sides of $V \equiv R_{\bar
L, 3 \bar L}$ so that condition C3 with an infinitesimal $\e_3$ is
satisfied for $V$.

We want to perturbatively treat, similarly to what we did in Section
3, the partition function: 
$$
Z_V^{\t} := \sum _{\h \in \O_V} \exp \left(H_V^{\t} (\h)\right )
$$
where 
$$
 H_V^{\t} (\h) := \sum _{\D :\D \cap V \neq 
\emptyset} \F_{\D} (\h \circ_V \t)
$$
and we recall that we are using the notation:
$$
V :=  R_{\bar L, 3 \bar L},\;\;\; \t \; \in \O_{V^c} \;\;\equiv \;\; {\rm  \; boundary \; condition\;
outside} \;V
$$

Given $\t,\t' \in \O_{V^c}$  and $ x$, $y$ belonging to the set of
conditioning sites above  the upper side and below the lower side of $V$,
respectively,  we want to consider the ratio
$$
{Z_{V,z,z'}(\t'_x,\t'_y,\t ) Z_{V,z,z'}(\t)\over
Z_{V,z,z'}(\t'_x,\t )Z_{V,z,z'}(\t'_y,\t )} 
\Eq (4.3')
$$
where $(\t,\t'_x,\t'_y)$, $(\t,\t'_x)$,$(\t,\t'_y)$ are the
configurations obtained from $\t$ by substituting 
$\t$ with $\t'$ in $\{x,y\}$, $\{x\}$ and  $\{y\}$, respectively.\par
The perturbative expression that we will obtain for $Z_V^{\t}$ will
show  an almost  factorized dependence on
boundary conditions in opposite horizontal faces so that we will be able
to show that the quantity
$$
 \left |{  Z_{V,z,z'}(\t'_x,\t'_y,\t ) Z_{V,z,z'}(\t)\over
Z_{V,z,z'}(\t'_x,\t )Z_{V,z,z'}(\t'_y,\t )}  -1 \right|
\Eq (4.3'')
$$
can be made arbitrarily  small for 
$\bar L$ sufficiently large so that condition C2 is satisfied.

It is easily seen, using DLR structure of the multi--grancanonical Gibbs
field, that the case when $x,y$ are close from the exterior to the two 
opposite vertical faces (at distance $3\bar L$) can be treated exactly
like in the homogeneous (constant activity) case; thus we will only
consider the above mentioned case of $x,y$ belonging to upper and
lower sets of conditioning spins.  
  
Sometimes, just for the sake of simplicity of notation, we will
actually drop the explicit dependence on the boundary condition $\t$
(even though this dependence is crucial). We express $H_V^{\t}$
exactly as we did in \equ (2.18): 
$$
\eqalign{
H_V(\h) &= \sum _{k_1:A_{k_1} \in  \cA_V}H_{A_{k_1}} (\a_{k_1})
+
\sum_ {k_2:B_{k_2} \in \cB_V} H_{B_{k_2}} (\b_{k_2}) + W_{B_{k_2},V\setminus
B_{k_2}} (\b_{k_2} |
\a) \cr
&~~~ +
\sum_ {k_3:C_{k_3} \in  \cC_V} H_{C_{k_3}} (\g_{k_3}) + 
W_{C_{k_3},V\setminus
C_{k_3}} (\g_{k_3} | \b
,\a) 
\cr
&~~~+ \sum_ {k_4:D_{k_4} \in  \cD_V} H_{D_{k_4}} (\d_{k_4}) +
W_{D_{k_4},V\setminus D_{k_4}} (\d_{k_4} |\g, \b,\a) 
\cr
}
\Eq (4.4)
$$
where, as in \equ (2.17),
$$
W_{\L_1,\L_2}(\h_{\L_1}|\h_{\L_2}) := W(\h_{\L_1}|\h_{\L_2})= 
H_{\L_1\cup\L_2}(\h_{\L_1},\h_{\L_2}) \,-\,
H_{\L_1}(\h_{\L_1})\, -\, H_{\L_2}(\h_{\L_2})
\Eq (4.5)
$$
We now proceed to the summation  over the $\d, \g, \b$ variables; we  repeat exactly the same operations
of splitting and gluing that we performed in section 3.
We get:
$$
\eqalign{
Z_V^{\t}  & =
\sum_{\a} \prod_ {k_1 : A_{k_1 } \in  \cA_V }
\exp \left\{ H(\a_{k_1}) \right\} 
\left[  
Z_{D_{k_1}} \left( (0),( \a_{k_1}), (0)  \right) 
Z_{D_{k_1-e_1}}  \left( (0),(0), (\a_{k_1}) \right) \right]^{-1}
\cr
&~~~\times
\prod _{k_2:B_{k_2} \in \cB_V}
Z_{\widetilde B_{k_2}} \left((0),(\a )^u, (\a )^d  \right)
\sum_{\b} \m^{\a}_2(\b)
\cr
&~~~\times
\prod _{k_3:C_{k_3} \in  \cC_V}
\left[ Z_{\widetilde C_{k_3}} \left((0)\right) \right]^{-1}
\prod_{k_3:C_{k_3} \in  \cC_V}
\left(1+\F^{(3)}_{C_{k_3}}(\alpha,\beta)\right)
\sum_{\g} \m^{\a,\b}_3 (\g) 
\cr
&~~~\times
\prod _{k_4:D_{k_4} \in  \cD_V}
\left ( 1 + \F^{(4)}_{D_{k_4}}(\a,\b,\g)\right)
\prod _{k_4:D_{k_4} \in  \cD_V}
\left ( 1 + \Psi^{(4)}_{D_{k_4}}(\a)\right)
\prod _{k_4:D_{k_4} \in  \cD_V}
\left [ Z_{D_{k_4}}\left ( (0)\right)\right ]^{-1}
\cr
}
\Eq (4.5')
$$
where the terms $[Z_{D_{k_1-e_1}} ( (0), (\a_{k_1}), (0))]^{-1}$,
$[Z_{D_{k_1}} ( (0), (0),(\a_{k_1} ))]^{-1}$ (defined in \equ (2.34))
come from the splitting described in \equ (2.35): in \equ(4.5'), by an 
abuse of notation, we still denote by $\widetilde C$, 
$\widetilde B$ and $\widetilde A$ their truncation in 
$R_{\bar L,3\bar L}$. Indeed, since we have generic and not periodic b.c., 
we have to introduce the modifications described in Section 3 
(below Proposition \thm[prop2.1]) in $\mu^{\alpha\beta}_{C_{k_3}}$,
$\mu^{\alpha}_{B_{k_2}}$ as well as in the $\Phi$ and $\Psi$ error
terms.      
Moreover notice that in the expression in
\equ (4.4) above,  we continue to
denote by $\a,\b,\g,\d$ also the 
configurations on the $A,B,C,D$ blocks outside $V$; in other words we
continue to denote by 
$\a,\b,\g,\d$ also the part of the $\t$ (exterior) configuration in
$A,B,C,D$  sub--lattices.  
We did not have them in \equ (2.39) since, there, we
were using periodic boundary conditions.
Now we continue with the same operations of splitting as in \equ
(2.44) (and gluing  as in \equ(2.54') ) only for the $B$  (and $A$)
blocks in $\hat \cA$, $\hat \cB$ namely outside the two vertical column
$V _l$. It is clear that we cannot perform the gluing operation described in
\equ (2.44) for the $B$ blocks in  $V _l$ and obtain  
a small value for the term $\F^{(2)}_{B_{k_2}}(\a)$. Indeed to get a
good upper bound for $\sup _{\a}|\F^{(2)}_{B_{k_2}}(\a)|$ we need the
validity of condition C1, with a sufficiently small $\e_1$ for horizontal
$R_{L_0,3L_0}$ rectangles and this condition is supposed to hold only
for $R_{L_0,3L_0}$ rectangles {\it completely contained} in one of the
three squares  $Q^l_{\bar L}$, $Q^c_{\bar L}$ or $Q^r_{\bar L}$.
For $R_{L_0,3L_0}$ rectangles centered at  $B$ block in $\cB_l$ we
cannot use condition C1 since these rectangles have simultaneously
non--empty overlap with two of the big squares namely 
$Q^l_{\bar L}$, $Q^c_{\bar L}$; the rectangles $R_{L_0,3L_0}$ having 
non--empty overlap with 
$Q^c_{\bar L}$, $Q^r_{\bar L}$ behave exactly like in the homogeneous
case since the activity in  $Q^c_{\bar L}\cup Q^r_{\bar L}$ is
supposed to be constant. 
\par
In this way we obtain the following
expression 
$$
\eqalign{
Z_V^{\t}  & =
\bar Z_V^{\t}
\sum_{\a}
\tilde Z_{V_l} (\a) 
\prod _{k_1 : A_{k_1} \in \hat \cA} \m_{A_{k_1}} (\a_{k_1})
\cr
&~~~\times
\prod _{k_1 : A_{k_1} \in \hat \cA}
\left (
1 + \Psi^{(1)}_{A_{k_1}}(\a_{k_1})
\right)
\prod _{k_2 : B_{k_2} \in \hat \cB}
\left ( 1 +\F^{(2)} _{B_{k_2}}(\a) \right)
\prod _{k_4 : D_{k_4} \in  \cD_V}
\left ( 1 + \Psi^{(4)}_{D_{k_4}}(\a)\right)
\cr
&~~~\times
\sum_{\b} \m^{\a}_{2}(\b)
\prod _{k_3 : C_{k_3} \in  \cC_V}
\left (1+\F^{(3)}_{C_{k_3}}(\a,\b)\right )
\cr
&~~~\times
\sum_{\g} \m^{\a,\b}_3 (\g)
\prod _{k_4 : D_{k_4} \in \cD_V}
\left ( 1 + \F^{(4)}_{D_{k_4}}(\a,\b,\g)\right) 
\cr
}
\Eq (4.6)
$$
where 
$ \m_{A_{k_1}} (\a_{k_1})$ is defined in \equ (2.56),
$\bar Z_V^{\t}$
is given by
$$
\bar Z_V^{\t}  =
\prod _{k_1 : A_{k_1} \in \hat \cA} 
Z_{\widetilde A_{k_1}}((0)) 
\prod _{k_2 : B_{k_2} \in \hat \cB}
\left[ Z_{ F_{k_2}} (0)\right ]^{-1}
\prod _{k_3 : C_{k_3} \in  \cC_V}
\left [ Z_{\widetilde C_{k_3}} \left((0)\right) \right]^{-1}
\prod _{k_4 : D_{k_4} \in  \cD_V}
 Z_{D_{k_4}}\left ( (0)\right)
$$
and
$$
\eqalign{
\tilde Z_{V_l} (\a) &
:= \prod _{\k_1: A_{k_1} \in V_l}\exp ( H (\a_{k_1} ) )
[Z_{D_{k_1}} ( (0), (\a_{k_1}), (0))]^{-1}  [Z_{D_{k_1+e_1}} ( (0),
(0), (\a_{k_1} ))]^{-1}
\cr
&~~~\times
\prod _{\k_2: B_{k_2} \in V_l} Z_{\tilde B_{k_2}} ((0),(\a _{k_2+
e_2}), (\a _{k_2}) ) 
\cr
}
\Eq (4.7)
$$

Let us call $\a_l$ the complex of $\a$ variable in $\cA_l$.
If we perform, in the r.h.s. of \equ (4.6) the sum over the $\g,\b$
variables  and over the $\a$ variables in $\hat \cA_V$, we get:
$$
Z_V^{\t} = \bar Z_V^{\t}
\sum_{\a_l \in \O_{\cA_l}} \tilde Z_{V_l} (\a_l)  \X_V^{\t}(\a_l)
\Eq (4.8)
$$
where, of course, 
$$
\eqalign{
\X_V^{\t}(\a_l) & =
\sum_{\a \in \O_{\hat\cA_V}}
\prod _{k_1 : A_{k_1} \in \hat \cA} \m_{A_{k_1}} (\a_{k_1})
\cr
&~~~\times
\prod _{k_1 : A_{k_1} \in \hat \cA}
\left (
1 + \Psi^{(1)}_{A_{k_1}}(\a_{k_1})
\right)
\prod _{k_2 : B_{k_2} \in \hat \cB}
\left ( 1 +\F^{(2)} _{B_{k_2}}(\a) \right)
\prod _{k_4 : D_{k_4} \in  \cD_V}
\left ( 1 + \Psi^{(4)}_{D_{k_4}}(\a)\right)
\cr
&~~~\times
\sum_{\b} \m^{\a}_{2}(\b)
\prod _{k_3 : C_{k_3} \in  \cC_V}
\left ( 1 +\F^{(3)}_{C_{k_3}}(\a, \b)\right )
\cr
&~~~\times
\sum_{\g} \m^{\a,\b}_3 (\g)
\prod _{k_4 : D_{k_4} \in \cD_V}
\left ( 1 + \F^{(4)}_{D_{k_4}}(\a,\b,\g)\right) 
}
\Eq (4.9)
$$
Like  in Section 3 we can write:
$$
\X_V^{\t}(\a_l)\; = \;
1 + \sum _{n\geq 1}\sum _{R_1, \dots
, R_n:\tilde R_i \subset V ,\atop \tilde R_i\cap \tilde R_j
=\emptyset, i<j=1,\dots,n}\;\;
\prod _{i=1}^n
\z^{\t}_{R_i} (\a_l)
\Eq (4.10)
$$
where the polymers $R_i$ are defined like in Section 2 with the
obvious changes. In this way we are 
reduced to one--dimensional system on $V_l$,  with finite norm,
rapidly decaying interaction.  Indeed we can write:
$$
Z^{\t}_V = \sum_{\a_l \in \O_{\cA_l}}\exp \left(\hat H(\a_l)\right)
\Eq (4.11)
$$
Where
$$
\eqalign{
\hat H(\a_l) & := \hbox{const.}\; + \sum_{k_1 \in V_l} H(\a_{k_1}) 
- \log Z^{\t}_{D_{k_1}}(\a_{k_1}) - \log Z^{\t}_{D_{k_1- e_1}}(\a_{k_1 - e_1})
\cr
&~~~
+\sum _{k_2 : B_{k_2}\in V_l \atop k_2\neq k_l^* } 
\log Z_{\widetilde B_{k_2}} (\a_{k_2},\a_{k_2 +e_2})
+
\log Z_{\widetilde B_{k^*_l}} (\a_{k^*_l}, \a^{(\t)}_{k^*_l +e_2}) 
+ \sum _{\G \subset\cA_l} 
\bar \F^{\t}_{\G} (\a_{\G})
\cr
}
\Eq (4.12)
$$
where\par \noindent
1)
$
\;\;\;\bar \F^{\t}_{\G} (\a_{\G}):= \sum_{R_1,\dots,R_n}^{\G}
\f_T(R_1,\dots,R_n) \prod _{i=1}^n \z^{\t} _{R_i}(\a_{\G}),
$
\par \noindent
2) the sum $\sum_{R_1,\dots,R_n}^{\G}$ runs over the clusters of (incompatible)
polymers ``touching" the whole set $\G$ of $A$--blocks in the sense that the product of
the activities of  the polymers $R_1,\dots,R_n$ explicitly depend on all the
$\a$--variables corresponding to the 
$A$--blocks in $\G$ and does not depend on any
other $\a$.
\par \noindent
3) we introduced 
$ Z_{D_{k}}(\a_{k}) = Z_{D_{k}} ( (0), (\a_{k}), (0))$,
$ Z_{\tilde B_{k}} (\a_{k},\a _{k+ e_2}) = Z_{\tilde B_{k}} ((0),(\a _{k+
e_2}),  (\a_{k}) )$
\par \noindent
4) $k^*_l$ is the index of the uppermost $B$--block in $V_l$:
$$
k^*_l := - \( M/2  -1 \) / 2 , \( M/2  -1 \) / 2 
$$
and$\a^{(\t)}_{k^*_l +e_2}$ is the configurations in the
$A$--blocks immediately outside (on the top) of $V_l$.\par

Notice that the dependence on the boundary condition $\t$ external to
$V$ is really present (beyond the term $Z_{\widetilde B_{k^*_l}}
(\a_{k^*_l}, \a^{(\t)}_{k^*_l+e_2})$),  only in $Z_{D_k}$ with $D_k$
adjacent to the boundary $\de V$ (upper and lower side).

  From \equ(4.11),\equ (4.12) and the general theory of
cluster expansion (see Proposition \thm[prop2.1]) it  follows  
that,for $M$ sufficiently large,
there exist positive constants $c_1, m_1, m_2, m_3$ such that:
 $$
 \sum _{\G \ni A_0}\| \F^{\t}_{\G}\|_{\infty} e^{m_1 |\G|} e^{m_2 \hbox {diam }\G}
 < \infty,
 \Eq (4.18)
 $$
 and, for any $y \in \de^ V, \G \subset \cA_l ,$:
 
$$
\sup _{\a_l} \sup _{\t,\t': \t_x= \t_{x'} \, \forall \, x \neq y}
|\bar \F^{\t}(\a_l) - \bar \F^{\t'}(\a_l)|\leq c_1 e^{- m_3 \;\hbox {dist}\,
(\G,y)}
\Eq (4.19)
$$
We are now reduced to a  one--dimensional system with finite
norm, rapidly decreasing potential.
We can then apply the theory developed in [CO2] and especially  in
[CCO] (see also [CEO],[CO1]).\par
Let us summarize the strategy of [CO2], [CCO] to find good mixing properties of the
Gibbs states for the one--dimensional systems like ours.
Consider the system of $M/2$  variables $\a_k$ on $V_l$. 
Suppose that the integers $p,n$ are such that $M/2$ is a multiple of $pn$.
We divide the interval $[1,\dots,M/2]$ into $m = {M\over2pn}$ intervals 
$I_1, \dots, I_m$
of length $pn$.
We call long range the contribution to the interaction coming from the terms with
range larger than $p$.
We decompose the potential $\F$ as:
$$
\F = \F^{sr} + \F^{lr} \;\;\;\;\; \hbox {with}\;\;
 \F^{sr}_{\G}=0 \;\; \hbox {if}\;\; \hbox {diam} \, \G \, >\, p;\;\;\;\;\;\;\;\;\;\;
 \F^{lr}_{\G}\neq 0 \;\; \hbox {only if}\;\; \hbox {diam} \, \G \, >\, p
\Eq (4.20)
$$
The idea is to treat $ \F^{lr}$ as a small perturbation. Indeed given a single block
$I_j$, a uniform upper bound on the sum of the absolute values of the contributions of the
long--range terms involving 
$I_j$ is of the order of $n \exp(-c p)$ for a suitable positive constant $c$.
On the other hand for the ``reduced" system with only short range interactions 
we can exploit the one--dimensionality and the uniform boundedness of the interaction.
Indeed the short range transfer matrix has a uniform positive gap in
its spectrum. This would 
imply an  exponential clustering of the short range Gibbs measure: the
truncated correlations at the extrema of an interval $I_j$ 
 would decay as
$
\exp(-c'n)$ with
$c'$ depending only on the gap of the transfer matrix. 
In the perturbative expansions  in [CO2],[CCO], the intervals $I_j$ involved in at least one
long range term are treated separately from the other ones and they happen to be very rare; on
the other ones, where only the short range terms are present the mechanism of 
strictly positive gap of the transfer matrix is active, inducing exponential decay of
correlations. We refer to [CO2],
[CCO] for more details; in these articles, (actually in a more complicated situation),
   analyticity of the free energy and decay of truncated correlations are proved. In
our case, as a consequence of the methods of [CO2], [CCO] we get exponential decay of
truncated correlations. This, together with \equ (4.19) allows to conclude the proof of
Proposition \thm[prop2.1] 
\QED

\bigskip
\expandafter\ifx\csname sezioniseparate\endcsname\relax%
     \input macro \fi
      \numsec=-2              
      \numfor=1\numtheo=1\pgn=1
\leftline{\bf A.2 A counterexample to USM $\Longrightarrow$ MUSM in
dimension 3.}
\par
We give here an example that, in general, the implication 
USM($\cA$) $\Longrightarrow$ MUSM($\cA$) does not
hold. We stress that our example is {\it ad hoc}, in particular the
interaction is translation invariant only by even shifts. We believe
however it sheds some light on the pathologies that may happen.
\midinsert
\vskip 17 truecm\noindent
\includegraphics{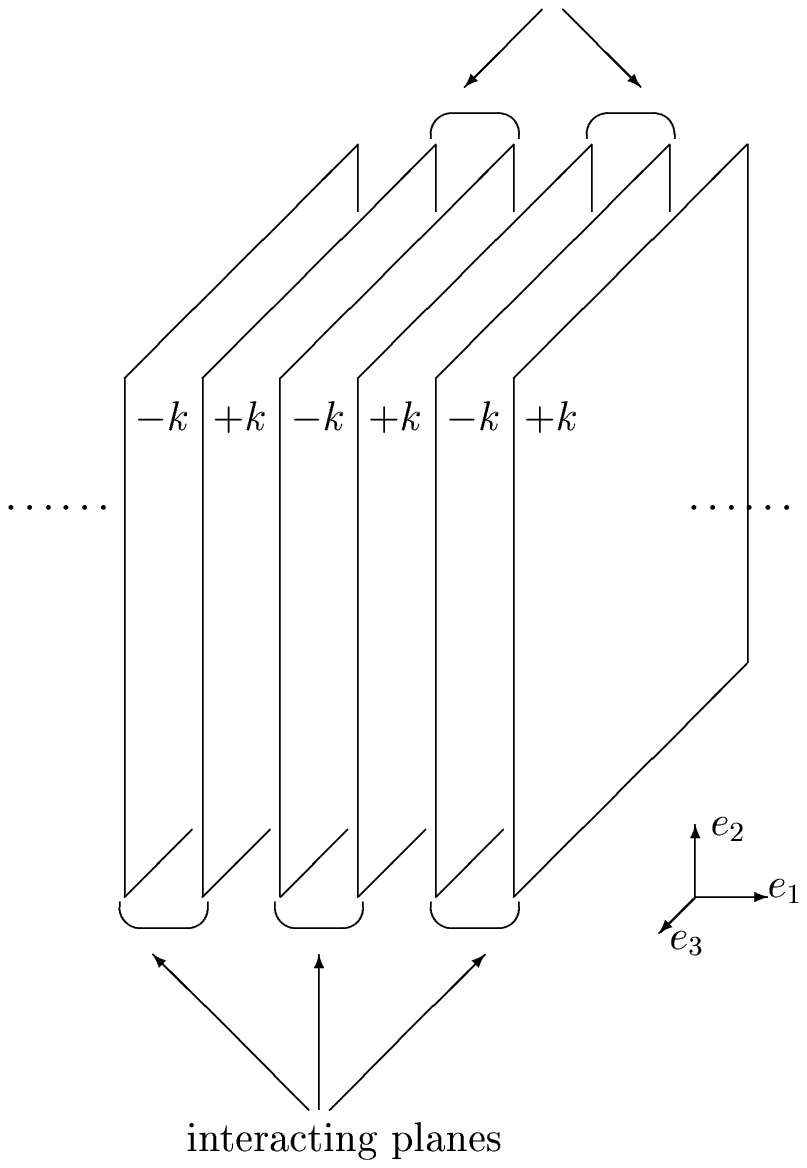} 
\vskip -1 truecm
\par\noindent 
{\centerline {\bf Fig. 9a}} 
\endinsert
\par

It is convenient to describe the example (see Fig. 9) 
by using spin variables, 
$\s\in \{-1,1\}^{\bZ^3}$. We denote by $\(e_1,e_2,e_3\)$
the canonical basis in $\bZ^3$. 
The one body potential (magnetic field) is
given as follows 
$$
\Phi_{\{x\}} (\s_x) =\cases
{ -k \s_x & if $x_1$ is even \cr
 k \s_x & if $x_1$ is odd \cr
}
$$
the two body potential is instead given by
$$
\Phi_{\{x,y\}} (\s_x,\s_y) =
\cases
{ 
-J \s_x \s_y & if $x_1$ is even, $y_1$ is odd
and $y=x+e_1+ a e_2 + b e_3$ \cr
&\phantom{if} for some $(a,b) \in\{ (0,0), (1,0), (0,1), (0,-1)\}$ \cr
0 & otherwise \cr
}
$$
where $J>0$. All the other potentials vanish, i.e. $\Phi_\L = 0$ for $|\L|>2$.

Note that the layer $\{x: x_1=a, a ~{\rm even} \}$ interacts only
with the layer $\{x: x_1=a+1\}$; in particular each double--layer
is independent of everything else.
Furthermore we claim that each double--layer is isomorphic to a
standard two dimensional Ising model with staggered magnetic field
(see Fig. 9b). 
\midinsert
\vskip 12 truecm\noindent
\includegraphics{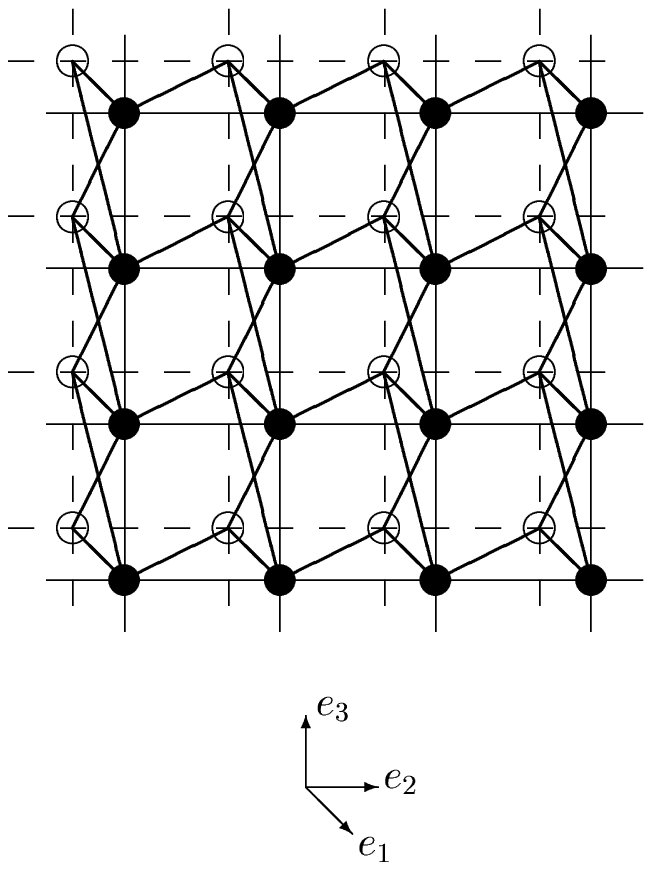} 
\par\noindent 
{\centerline {\bf Fig. 9b}} 
\endinsert
\par
We can in fact map the layer $\{ x\, : x_1=a, a ~{\rm even} \}$ to the even
sub--lattice of $\bZ^2$, as follows
$$
(a,x_2,x_3) \mapsto 
\cases
{
(2x_2,x_3) & if $x_3$ is even \cr
(2x_2-1,x_3) & if $x_3$ is odd \cr
}
$$ 
and the layer $\{ x\, : x_1=a+1\}$ to the odd sub--lattice of $\bZ^2$, 
as follows
$$
(a+1,x_2,x_3) \mapsto 
\cases
{
(2x_2-1,x_3) & if $x_3$ is even \cr
(2x_2,x_3) & if $x_3$ is odd \cr
}
$$
It is easy to verify that under the above mapping the double layer 
$\{ x\, : x_1=a, a ~{\rm even} \} \cup \{ x\, : x_1=a+1\}$ is mapped
onto the two dimensional Ising model with the following interaction
$$
\Phi_{\L}(\s) = 
\cases
{ 
- k \s_x & if $\L=\{x\}$ and $x_1+x_2$ is even \cr
k \s_x & if $\L=\{x\}$ and $x_1+x_2$ is odd \cr
-J\s_x \s_y & if  $\L=\{x,x+e_1\}$,  $\L=\{x,x+e_2\}$\cr
0 & otherwise \cr
}
\Eq(id2smf)
$$
and we are left with studying the strong mixing properties of such a
model. 

Let us denote by $\mu_\L^\t$ the Gibbs local specification
associated to the interaction \equ(id2smf) and by  $\mu_{\L,h}^\t$
the measure obtained from $\mu_\L^\t$ by adding a (constant) magnetic
field $h\in \bR$, i.e. $-k$ on the first line of \equ(id2smf) becames $-k-h$
whereas  $k$ on the second line of \equ(id2smf) becames $k-h$.
We claim that, if $k$ is chosen large enough (depending on $J$) such a
measure does satisfy condition GUSM. Roughly speaking, we have
a large magnetic field in either the odd or the even sub--lattice,
therefore the phase is determined on that sub--lattice; since the other
sub--lattice is conditionally independent (given the first sub--lattice)
we get the strong mixing condition. Indeed one can verify that the
finite size condition C1 holds on squares of side 2 with constants
uniform in $h$. 

On the other hand it is very easy to show that there is no $\ell_0$
such that the 3 dimensional model we started from satisfies
GMUSM. Let $\ell$ be an odd integer and consider
$\L = Q_\ell\( (- (\ell-1)/2,0,0) \) \cup  Q_\ell\( (\ell+1)/2,0,0) \)$; 
put a magnetic field $h_1=-k$ (resp. $h_2= +k$) on the first 
(resp. second) cube. The  image, under above mapping, of the
double--layer $\{ x\, : x_1=0 \} \cup \{ x\, : x_1=1 \}$ is now  
the standard two--dimensional Ising with zero magnetic field. If $J$
is chosen large enough we then have a long range order, hence
\equ(e:sm) fails to hold.

The pathology that has occurred is the following. Even if the local
specification does satisfy the strong mixing condition separately 
in each one of the two cubes $Q_\ell\( (- (\ell-1)/2,0,0) \)$,
$Q_\ell\( (\ell+1)/2,0,0)\)$, 
when we put them together we have a long range order which
propagates inside the double--layer which sits across the interface
between the two cubes.

\bigskip
\expandafter\ifx\csname sezioniseparate\endcsname\relax%
     \input macro \fi
      \numsec=-2              
      \numfor=1\numtheo=1\pgn=1
\leftline{\bf Acknowledgements.}

It is a pleasure to thank  N. Cancrini, A.C.D. van Enter,  
F. Martinelli and H.-T. Yau for fruitful discussions.
E.C. ackowledges the financial support of the european network 
``Stochastic Analysis and its Applications" ERB--FMRX--CT96--0075. 

\bigskip
\expandafter\ifx\csname sezioniseparate\endcsname\relax%
     \input macro \fi
      \numsec=-2              
      \numfor=1\numtheo=1\pgn=1
\leftline{\bf References.}

\item{[ABF]}
Aizenman M., Barsky D. J., Fern{\'a}ndez R. : 
{\it The phase transition in a general class of 
Ising-type models is sharp}.\ 
J. Statist. Phys. {\bf 47}, 343--374. (1987)

\item{[Ba]}
Basuev A.G. :
{\it Hamiltonian of the phase separation border and phase transition
of the first kind. I.}\
Theor. Math. Phys. {\bf 64}, 716--734. (1985)

\item{[BMO]}
Benfatto G., Marinari E., Olivieri E. : 
{\it Some numerical results on the block spin transformation 
for the 2D Ising 
model at the critical point.}
\ J. Stat. Phys. {\bf 78}, 731--757 (1995).

\item{[BKL]}
Bricmont J., Kupiainen A., Lefevere R. :
{\it Renormalization group pathologies and the definition of 
Gibbs states.}
\ Comm. Math. Phys. {\bf 194}, 359--388. (1998)

\item{[C]}
Cammarota C. : 
{\it The Large Block Spin Interaction.} 
\ Nuovo Cimento {\bf B(11) 96}, 1--16. (1986)

\item{[CCO]}
Campanino M., Capocaccia D., Olivieri E. :
{\it Analyticity for one--dimensional systems with long-range 
superstable interactions.}
\ J. Statist. Phys. {\bf 33}, 437--476. (1983)

\item{[CEO]} 
Campanino M., van Enter A. C. D.,  Olivieri E. :
{\it One-dimensional spin glasses with potential 
decay $1/r\sp {1+\epsilon}$. Absence of phase transitions and cluster
properties.}
\ Comm. Math. Phys. {\bf 108}, 241--255. (1987)

\item{[CO1]}
Campanino M., Olivieri E. :
{\it One--dimensional random Ising systems with
interaction decay $r\sp {-(1+\epsilon)}$: a convergent cluster
expansion.} 
\ Comm. Math. Phys. {\bf 111}, 555--57. (1987)

\item{[CM]}
Cancrini N.,  Martinelli F. :
{\it Comparison of finite volume canonical and gran canonical Gibbs
measures under a mixing condition.}
Preprint 1999.

\item{[CG]}
Cassandro M., Gallavotti G.:
{\it The Lavoisier law and the critical point.} \
Nuovo Cimento {\bf B 25}, 691. (1975)  

\item{[CO2]}
Cassandro M., Olivieri E. :
{\it Renormalization group and analyticity in one
dimension: a proof of Dobrushin's theorem.}
\ Comm. Math. Phys. {\bf 80}, 255--269. (1981)

\item{[CM]}
Cesi F., Martinelli F. : 
{\it On the layering transition of an SOS
surface interacting with a wall. I. Equilibrium results.} 
\ J. Statist. Phys. {\bf 82}, 823--913.  (1996)

\item{[CiO]}
Cirillo  E.N.M., Olivieri E. :
{\it Renormalization group at criticality and
complete analyticity of constrained models: a numerical study.}\
J. Statist. Phys. {\bf 86}, 1117--1151. (1997)

\item{[DM]}
Dinaburg E. I.,  Mazel A. E. :
{\it Layering transition in SOS model with external magnetic field.}\
J. Statist. Phys. {\bf 74}, 533--563. (1994)

\item{[D1]} 
Dobrushin R. L. : 
{\it Prescribing a system of random variables by
conditional distributions.}
\ Theory  Probab. Appl. {\bf 15}, 453--486. (1970)

\item{[D2]} 
Dobrushin R. L. : 
Lecture given at the workshop: ``Probability and Physics"
Renkum, Holland, 1995.

\item{[D3]}
Dobrushin R.L.:
{\it Perturbation methods of
    the theory of Gibbsian Fields}.  In Ecole d'Et\'e de Probabilit\'es de
    Saint-Flour XXIV - 1994,  1--66.  Springer-Verlag (Lecture Notes in
    Mathematics 1648), Berlin-Heildeberg-New York, 1996.

\item{[DS1]}
Dobrushin R. L., Shlosman S. B. :
{\it Constructive Criterion for the Uniqueness of Gibbs Fields.}\
Stat. Phys. and Dyn. Syst., Birkhauser, 347--370. (1985)

\item{[DS2]}
Dobrushin R. L., Shlosman S. B. :
{\it Completely Analytical Gibbs Fields.}\ 
Stat. Phys. and Dyn. Syst., Birkhauser, 371--403. (1985)

\item{[DS3]}
Dobrushin R. L., Shlosman S. B. :
{\it Completely Analytical Interactions Constructive Description.}\
J. Stat. Phys. {\bf 46}, 983--1014. (1987)

\item{[DS4]}
Dobrushin R. L., Shlosman S. B. :
{\it Large and moderate deviations in the Ising model.}\ 
Probability contributions to statistical mechanics, 91--219, Adv. Soviet
Math., 20, Amer. Math. Soc., Providence, RI, 1994

\item{[DT]}
Dobrushin R. L., Tirozzi B. :
{\it The central limit theorem and the problem of equivalence 
of ensembles.}\ 
Comm. Math. Phys. {\bf 54}, 173--192. (1977)

\item{[E1]}
van Enter A. C. D. : 
{\it Ill--defined block--spin transformations at arbitrarily 
high temperatures.}\
J. Statist. Phys. {\bf 83}, 761--765. (1996)

\item{[E2]}
van Enter A. C. D. : 
{\it On the possible failure of the Gibbs property for
measures on lattice systems. Disordered systems and statistical
physics: rigorous results.}\ 
Markov Process. Related Fields {\bf 2}, 209--224. (1996) 
 
\item{[EFK]}
van Enter A. C. D., Fern{\'a}ndez R.,  Kotecky R. :
{\it 
Pathological behavior of renormalization group maps at high field and above 
the transition temperature.}\
J. Stat. Phys., {\bf 79}, 969--992. (1995)

\item{[EFS]}
van Enter A. C. D.,  Fern{\'a}ndez R., Sokal A. D. :
{\it Regularity Properties and Pathologies of Position--Space
Renormalization--Group Transformations: Scope and Limitations of
Gibbsian Theory.}\ 
J. Stat. Phys. {\bf 72}, 879--1167. (1994)

\item{[ES]}
van Enter A. C. D., Shlosmann S.:
{\it (Almost) Gibbsian description of the sign field of SOS
field.}\
J. Stat. Phys. {\bf 92}, 353--368. (1998)


\item{[GaK]}
Gallavotti G., Knops H. J. F. :
{\it Block-spins interactions in the Ising model.}\
Comm. Math. Phys. 36, 171--184.  (1974)

\item{[GMM]}
Gallavotti G., Martin L\"of A., Miracle Sole S.,  in Battelle Seattle 
(1971) Rendecontres, A. Lenard, ed. (Lecture Notes in Phisics, Vol. 20, Springer, 
Berlin, 1973), pp.162-204.

\item{[GrK]}
Gruber C., Kunz H. :
{\it General properties of polymer systems.}\
Comm. Math. Phys. {\bf 22}, 133--161. (1971)

\item{[GP]}
Griffiths R. B., Pearce P. A. : 
{\it Mathematical Properties of Position--Space
Renormalization Group Transformations.}\
J. Stat. Phys. {\bf 20}, 499--545. (1979)

\item{[HK]}
Haller K., Kennedy T. :
{\it Absence of renormalization group pathologies near the critical 
temperature. Two examples.}\
J. Statist. Phys. {\bf 85}, 607--637. (1996)

\item{[H]}
Higuchi Y. :
{\it Coexistence of infinite $(*)$-clusters. II. Ising
percolation in two dimensions.}\ 
Probab. Theory Related Fields {\bf 97}, 1--33. (1993)

\item{[IS]} 
Iagolnitzer D., Souillard B. :
{\it Random fields and limit theorems.}\ 
Random fields, Vol. I, II (Esztergom, 1979), 573--591, 
Colloq. Math. Soc. János Bolyai, 27, 
North-Holland, Amsterdam-New York, 1981.

\item{[I]}
Israel R. B.  :
{\it Banach Algebras and Kadanoff Transformations
in Random Fields.}\ 
J. Fritz, J. L. Lebowitz and D. Szasz editors
(Esztergom 1979), Vol. II, 593--608 (North--Holland, Amsterdam 1981).

\item{[Ka]}
Kashapov I.A. :
{\it Justification of the renormalization group method.}\
Theor. Math. Phys. {\bf 42}, 184--186. (1980) 

\item{[KP]}
Koteck\'y R., Preiss D. :
{\it Cluster expansion for abstract polymer models.}\
Comm. Math. Phys. {\bf 103}, 491--498. (1986)

\item{[Ko]}
Kozlov O. K. :
{\it Gibbs Description of a System of Random Variables.}\
Probl. Inform. Transmission. {\bf 10}, 258--265. (1974)

\item{[LM]}
L{\"o}rinczi J.,  Maes C. :
{\it Weakly Gibbsian measures for lattice spin
systems.}\ 
J. Statist. Phys. {\bf 89}, 561--579. (1997)

\item{[LV]}
L{\"o}rinczi J., Vande Velde K. : 
{\it A Note on the Projection of Gibbs Measures.}\
J. Stat. Phys. {\bf 77}, 881--887. (1994)

\item{[MO1]}
Martinelli F., Olivieri E. :
{\it Finite Volume Mixing Conditions for Lattice
Spin Systems and Exponential Approach to Equilibrium of Glauber
Dynamics.}\
Proceedings of 1992 Les Houches Conference on  Cellular Automata and
Cooperative Systems, N. Boccara, E. Goles, S. Martinez and P. Picco
editors (Kluwer 1993). 

\item{[MO2]}
Martinelli F., Olivieri E. :
{\it Approach to Equilibrium of Glauber Dynamics
in the One Phase Region I. The Attractive Case.}\ 
Commun. Math. Phys. {\bf 161}, 447--486. (1994)

\item{[MO3]}
Martinelli F., Olivieri E. :
{\it Approach to Equilibrium of Glauber Dynamics
in the One Phase Region II. The General Case.}\
Commun. Math. Phys. {\bf 161}, 487--514. (1994)

\item{[MO4]}
Martinelli F., Olivieri E. :
{\it Some Remarks on Pathologies of
Renormalization Group Transformations for the Ising model.}\
J. Stat. Phys. {\bf 72}, 1169--1177. (1994)

\item{[MO5]}
Martinelli F., Olivieri E. :
{\it Instability of renormalization group pathologies under decimation.}\ 
J. Stat. Phys. {\bf 79}, 25--42. (1995)

\item{[MOS]}
Martinelli F., Olivieri E., Schonmann R. :
{\it For $2$--$D$ Lattice Spin Systems Weak Mixing Implies Strong Mixing.}\
Commun. Math. Phys. {\bf 165}, 33--47. (1994)

\item{[NOZ]}
Nardi F.R., Olivieri E., Zahradn\'{\i}k M. :
{\it On the Ising model with strongly anisotropic external 
field}.\  Preprint 1999.

\item{[N]}
Newman C. M. :
{\it Normal fluctuations and the FKG inequalities.}\ 
Comm. Math. Phys. {\bf 74}, 119--128. (1980)

\item{[NL]} 
Niemeijer Th., van Leeuwen M. J. :
{\it Renormalization theory for Ising--like spin systems.}\
In ``Phase Transitions and Critical Phenomena", 
vol. 6, Eds. C. Domb, M. S. Green  (Academic Press, 1976).

\item{[O]}
Olivieri E. :
{\it On a Cluster Expansion for Lattice Spin Systems: a Finite
Size Condition for the Convergence.}\
J. Stat. Phys. {\bf 50}, 1179--1200. (1988)

\item{[OP]}
Olivieri E., Picco P. : 
{\it Cluster Expansion for $D$--Dimensional Lattice
Systems and Finite Volume Factorization Properties.}\
J. Stat. Phys. {\bf 59}, 221--256. (1990)


\item{[SS]}
Schonmann R.H., Shlosman S.B. :
{\it Complete analyticity for $2$D Ising completed.}\ 
Comm. Math. Phys. 170, 453--482.  (1995)

\item{[Sh]}
Shlosman S.B. :
{\it Uniqueness and half--space non--uniqueness of Gibbs states in 
Czech models.}\ 
Teor. Math. Phys. {\bf 66}, 284--293. (1986) 

\item{[Y]} 
Yau H.-T. : 
{\it Logarithmic Sobolev inequality for lattice gases with
mixing conditions.}\ 
Comm. Math. Phys. 181, 367--408. (1996)

\end
\bye